\documentclass[twocolumn]{aastex62}
\usepackage{float,graphicx,amsmath,multirow,mathtools,comment}
\usepackage{color}
\usepackage[version=3]{mhchem}
\newcommand{\mjb}{mJy~beam$^{-1}$}

\newcommand{\kms}{\mbox{km~s$^{-1}$}}

\begin{document}

\title{An Automated Chemical Exploration of NGC 6334I at 340 au Resolution}
\author{Samer J. El-Abd}
\altaffiliation{Samer El-Abd is a Grote Reber Fellow  of the National\\ Radio Astronomy Observatory}
\affiliation{Department of Astronomy, University of Virginia, Charlottesville, VA 22904, USA}
\author{Crystal L. Brogan}
\affiliation{National Radio Astronomy Observatory, Charlottesville, VA 22903, USA}
\author{Todd R. Hunter}
\affiliation{National Radio Astronomy Observatory, Charlottesville, VA 22903, USA}
\affiliation{Center for Astrophysics $\mid$ Harvard \& Smithsonian, Cambridge, MA 02138, USA}
\author{Kin Long Kelvin Lee}
\affiliation{Department of Chemistry, Massachusetts Institute of Technology, Cambridge, MA 02139, USA}
\author{Ryan A. Loomis}
\affiliation{National Radio Astronomy Observatory, Charlottesville, VA 22903, USA}
\author{Brett A. McGuire}
\affiliation{Department of Chemistry, Massachusetts Institute of Technology, Cambridge, MA, MA 02139, USA}
\affiliation{National Radio Astronomy Observatory, Charlottesville, VA 22903, USA}
\correspondingauthor{Samer El-Abd, Brett A. McGuire}
\email{sje2tu@virginia.edu, brettmc@mit.edu}

\begin{abstract}

%\noindent Due to the time-consuming nature of fitting synthetic spectra to observations by hand for line rich sources, analysis of spectroscopic data from astronomical observations is generally limited to data extracted from one to a handful of pixels. Star-forming regions, however, show a great deal of internal structure that is difficult, if not impossible, to accurately characterize with such a limited number of spectra. We therefore present the results of an automated fitting routine that visits every pixel in the field of view of an ALMA data cube for two sites of massive star formation in NGC 6334I. This routine finds best-fit parameters including excitation temperature and column densities simultaneously for 21 distinct molecules by generating synthetic spectra across 7.48 GHz of spectral bandwidth between 280 and 351 GHz. An overview of the fitting routine as well as spatial maps of the derived parameters for each of the $>8000$ pixels will be presented with special attention paid to the \ce{C2H4O2} isomers. This work is intended as a proof-of-concept for the automated fitting technique and the unique methods of analysis it enables.

\noindent Much of the information gleaned from observations of star-forming regions comes from the analysis of their molecular emission spectra, particularly in the radio regime. The time-consuming nature of fitting synthetic spectra to observations interactively for such line-rich sources, however, often results in such analysis being limited to data extracted from a single-dish observation or a handful of pixels from an interferometric observation. Yet, star-forming regions display a wide variety of physical conditions that are difficult, if not impossible, to accurately characterize with such a limited number of spectra. We have developed an automated fitting routine that visits every pixel in the field of view of an ALMA data cube and determines the best-fit physical parameters, including excitation temperature and column densities, for a given list of molecules. In this proof-of-concept work, we provide an overview of the fitting routine and apply it to $0\farcs26$, 1.1\,\kms\ resolution ALMA observations of two sites of massive star-formation in NGC 6334I. 
Parameters were found for 21 distinct molecules by generating synthetic spectra across 7.48 GHz of spectral bandwidth between 280 and 351 GHz. Spatial images of the derived parameters for each of the $>8000$ pixels are presented with special attention paid to the \ce{C2H4O2} isomers and their relative variations. We highlight the greater scientific utility of the column density and velocity images of individual molecules compared to traditional moment maps of single transitions. 

%the routine itself will soon be made publicly available as part of the \texttt{molsim} Python package.

%Analysis of spectroscopic data from astronomical observations is generally limited to spectra extracted from a handful of pixels. This is due to the time-consuming nature of fitting synthetic spectra by hand, particularly when the analysis is done for a large number of molecules across a wide bandwidth.

\end{abstract}

\section{Introduction}
\label{intro}

%Observational astronomy research by and large is conducted in two distinct regimes - surveys consisting of a number of objects, and in-depth case studies of a single source. While large scale surveys have long been utilized in aspects of extragalactic and stellar astronomy \citep{}, the field of star formation has until recently mostly relied on extensive studies of single sources such as Orion-KL or Sagittarius B2 \citep{}. These studies are conducted by analyzing the spectral emission from these sources and constraining the physical conditions based on the properties of the molecular emission. 

The energetic events associated with star formation and the clustered nature of massive protostars result in a complicated picture with respect to the kinematics and excitation mechanisms of the surrounding gas \citep{Hunter:2021:L17,Rivilla:2013:A48}. As the nascent protostars continue to evolve they heat up this gas, enabling a plethora of chemical reactions that cannot efficiently occur elsewhere in the interstellar medium \citep{Jorgensen:2020:727}. The molecules that are formed in these unique environments are useful tools for understanding the physical conditions in which they are found. The characteristics of their rotational line emission can be used to measure the temperature and velocity of the gas (assuming local thermodynamic equilibrium), and the relative abundances of molecules constrain their formation pathways \citep{Herbst:2009:427}.

% Massive stars rarely, if ever, form in isolation. These stars instead form in ``protoclusters'' with other massive and lower mass stars \citep{Krumholz14,Motte22}. 

%Too much?
% Just as the deaths of massive stars are responsible for much of the atomic enrichment of the ISM, it's possible that their birth is in fact responsible for much of the molecular enrichment that we observe. 
The molecular emission that arises from star forming environments is a valuable tool for constraining the physical conditions of protostellar regions, \emph{but only in the case that this emission can be properly identified}. The reality is that spectroscopic observations in the millimeter regime are often a dense conglomeration of lines due to the high spectral line density of many molecules. This makes it a challenging task to isolate the emission from a single molecule; attaining an accurate picture of the chemical inventory for such regions requires simultaneously fitting a large number of molecules to the data to accurately model as much of the observed spectral bandwidth as possible. We are fortunate to have a number of tools such as \texttt{molsim} \citep{Lee:2023:}, \texttt{XCLASS} \citep{Moller:2017:A7}, \texttt{MADCUBA} \citep{Martin:2019:A159}, \texttt{CASSIS} \citep{Vastel:2015:313}, \texttt{WEEDS} \citep{Maret:2011:A47} and \texttt{pyspeckit} \citep{Ginsburg:2011:ascl:1109.001} which make it a relatively trivial process to overlay a simulated rotational spectrum of a molecule over our observations for a wide range of physical parameters. Given a wide enough bandwidth of observations, we can then be confident of a molecule's detection and characteristics if we can match the observed emission with reasonable parameters for a large number of emission lines. Individually fitting molecules to observations with a large bandwidth is a time-consuming process, however, especially if one wants to characterize the behavior of a large number of molecules. For this reason, the spectral analysis of star-forming regions is often limited to a small number of positions, if not a single one - thus leaving a wealth of information on the proverbial table. This approach is particularly problematic when applied to massive star-forming regions, as they are often physically and kinematically complex \citep[e.g.,]{Brogan2007,Cunningham2023}. Because the physical conditions vary significantly over the field of view of the observation, it is impossible to extrapolate information across the region from measurements taken over a handful of pixels with any confidence. 

To address this problem, we have developed an automated least-squares fitting routine that will fit the combined spectra of a given list of molecules to every pixel in an ALMA image cube. This technique allows for a broader exploration of the physical conditions and molecular abundances surrounding massive star-forming regions and other protostellar environments by both accelerating the production of the measurements of interstellar molecular abundances and increasing the number of positions for which these measurements can be made. We have chosen ALMA data toward NGC 6334I spanning a total bandwidth of 7.48\,GHz in the frequency range from 280 to 351\,GHz for our initial proof of concept study. This massive Galactic protocluster hosts two extremely rich hot core line sources, each with distinct physical conditions and kinematics (see \S\,\ref{source} for additional details), providing ample fodder to test our fitting technique on a challenging use case. The goal of this work should be stated clearly: we are not claiming that the final fit parameters for each pixel are necessarily as accurate as those that would be derived were one to manually fit each of the molecules by hand in one of the pixels. That said, we do believe that the derived parameters are reasonably accurate for the vast majority of the pixels in our field of view. These fits - taking into account their uncertainties - provide a unique point of view with regards to the spatial morphology of complex interstellar molecules that is not biased by an {\it a priori} choice in the extraction location of the spectra. Massive star-forming regions are physically and kinematically complicated; the ability to take a holistic view of such regions will be a crucial step forward for understanding their ongoing physical processes. 

In this paper, the data on which the fitting routine was tested are described in \S\,\ref{data}, while the fitting technique itself is described in \S\,\ref{routine}. Images of the excitation temperature, linewidths, and velocity measured for each of $>$8000 pixels across two distinct regions in NGC 6334I are presented in \S\,\ref{results}, along with images of the physical column densities of the \ce{C2H4O2} isomers. \S\,\ref{discussion} expands on some of the methods of analysis this technique enables, as well as how the velocity and column density images compare with their moment map counterparts.

\section{ALMA Data Characteristics}
\label{data}

\subsection{Test Case: NGC 6334I}
\label{source}

\begin{figure*}[htb!]    \centering
   \includegraphics[width=\linewidth]{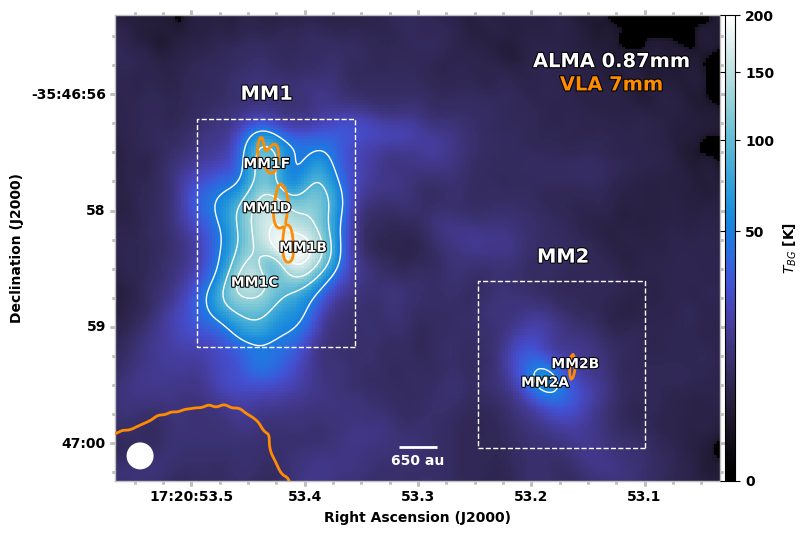}
    \caption{An ALMA 0.87\,mm continuum image of the central portion of NGC 6334I showing the proximity of MM1 and MM2. The white contours denote continuum brightness temperatures of 50, 90, 130, and 170\,K. The orange contours mark VLA 7\,mm emission of 11\,K \citep[the contour at the southwest edge of the frame arises from MM3,][]{Brogan2016}. The dashed boxes highlight the regions over which we ran the automated fitting routine.}
    \label{cont_fig}
\end{figure*}

NGC~6334I contains two prodigious hot core spectral line sources, MM1 and MM2 \citep{Beuther:2007:989,Zernickel:2012:A87,Bogelund:2018:A88}. At a distance of 1.3\,kpc \citep{Reid2014,Chibueze2014}, these two sources are separated by only $\sim$4000 au (see Figure \ref{cont_fig}, making any chemical differentiation between the two of them particularly diagnostic as it is reasonable to assume they likely formed from similar primordial material. MM1 (the brighter of the two sources) hosts at least two young protostars, MM1B and MM1D. Complicated bulk motions are evident in the source, with at least two outflows: one on a larger scale oriented NE-SW \citep{Qiu2011} and another more compact, dynamically young outflow in the N-S direction \citep{Brogan:2018:87}. It is currently unclear which source within MM1 is driving these outflows. MM2 (129\,K continuum peak at 350 GHz) is a source with significantly lower brightness temperature than MM1 (222\,K continuum peak), but MM2 is still rich with molecular line emission and exhibits much narrower linewidths (see Figures \ref{3panel_mm1_univ} and \ref{3panel_mm2_univ}).

%How much of this paragraph is really necessary?
NGC 6334I-MM1 has recently undergone an accretion outburst in 2015 \citep{Macleod2018,Hunter:2017:L29}. Accretion onto young stellar objects is an avenue by which stars are believed to gain a significant portion of their final mass \citep{Fischer:2019:183,Fischer:2022:}. While this mechanism has long been associated with low-mass star formation, recent observational and theoretical evidence indicates this mechanism occurs in massive star formation as well \citep{Garatti:2017:276, Hunter:2021:L17, Meyer:2021:4448}. Such accretion may occur rapidly in discrete episodes as opposed to a steady process that takes place over a longer stretch of time. These episodes are also responsible for energetic outflows driven through the surrounding cloud, raising the level of continuum emission as well as driving maser emission. In NGC 6334I-MM1, the continuum emission quadrupled in intensity and coincided with a rapid increase in the maser emission observed from the source. Both effects are still visible six years after the event \citep{Hunter:2017:L29,Hunter:2021:L17}. 
It is highly likely that these energetic events serve as the impetus for a variety of unique chemical reactions across the region in question. Attaining a better understanding of this unique, complex environment is not feasible through the analysis of a small handful of positions. A new approach is required to be able to understand the entire breadth of the physical conditions and their associated molecular products.

\subsection{Observations}
\label{obs}

The ALMA data toward NGC~6334I used for this study were observed during Cycle 3 in 2016, under project code 2015.A.00022.T. These are the same data used in \citet{El-Abd:2019:129} but have since been reprocessed with the ALMA Cycle 8 pipeline \citep[version 6.2.1-7-pipeline-2021.2.0.128,][]{Hunter2023} to account for an ALMA renormalization issue that can affect the flux scaling of strong spectral lines\footnote{see https://help.almascience.org/kb/articles/what-errors-could-originate-from-the-correlator-spectral-normalization-and-tsys-calibration}; these corrections were found to be of order 10\% for the most affected transition, CS (J=6-5), and much lower to undetectable for the majority of transitions arising from complex organic molecules.

The observations consisted of two tunings, each with four spectral windows, with a bandwidth of 1.87 GHz per pair of windows. The first set of spectral windows were centered at 280.1, 282.0, 292.1, and 294.0 GHz. The second set of spectral windows were centered at 337.1, 339.0, 349.1, and 351.0 GHz. Both tunings were taken with a factor of 2 online channel averaging producing a channel width of 976.6\,kHz; the spectral cubes were made with a channel width of 1.1\,\kms\/, comparable to the effective spectral resolution.
The observations were centered at $\alpha$(J2000) = 17$^h$:20$^m$:53$^s$.36, $\delta$(J2000) = -35$^\circ$:47$^\prime$:00$^{\prime\prime}$.0 and the images were created with CASA \citep{CASA2022} using {\tt robust} image weighting values of 0.2 and 0.5 for the lower and higher frequency tunings, respectively. The resulting angular resolution of the images was a bit less than $0\farcs26$
(the beam was oversampled by a factor of 5 relative to the minor axis of the synthesized beam). The data were self-calibrated using the bright continuum emission, and the solutions were also applied to the continuum-subtracted line data; the continuum subtraction was performed as described in \citet{Brogan:2018:87}.  The full width at half-power (FWHP) of the primary beam is $\sim 20''$ and the images were corrected for the primary beam response. The cubes were smoothed to a uniform circular angular resolution of $0\farcs26$ before analysis with an rms noise per channel of 1.8\,\mjb\/ for the lower tuning and 3.0\,\mjb\/ for the upper tuning.\footnote{The rms noise measurements were made in line-free channels of the data cube at several positions between the 0.8 and 0.6 primary beam response annulus; a further distance from the phase center than either MM1 or MM2. These are conservative estimates because the noise is not uniform in the maps corrected for the primary beam response; the noise level increases with distance from the phase center.} Further data processing details are given in \citet{McGuire:2017:L46}, \citet{Hunter:2017:L29}, and \citet{Brogan:2018:87}.

%These data have been reprocessed since the work of \citet{El-Abd19}, which explains in part why derived abundances may differ between the two works. 

\section{The Fitting Routine}
\label{routine}

\subsection{Technique}
\label{technique}

\texttt{molsim} is a publicly-available Python package tailored for the analysis of high spectral resolution spectroscopic observations of astronomical sources. The use of this package, or others like it, enables astronomers to match the rotational spectra of molecules measured in the lab to observations made of the ISM and in turn derive physical parameters from those observations. This package has already been used to great effect across a number of different studies ranging from single-dish observations of a dark molecular cloud \citep{McGuire:2020:L10} to interferometric surveys of massive star-forming regions \citep{Schuessler:2022:102,Remijan:2022:A85}. In tests comparing similar Python packages, \citet{Ginsburg:2022:291} found that the simulations produced by \texttt{molsim} were in good agreement with those produced by \texttt{XCLASS} and \texttt{pyspeckit}. 

%Due to the ``line forests" that are a common feature of star-forming regions observed at radio wavelengths (a consequence of many molecules having a high spectral line density), an accurate picture of the chemical inventory for such regions requires simultaneously fitting a large number of molecules to the data to accurately model as much of the observed spectral bandwidth as possible. With the capabilities of modern day interferometers such as ALMA, individual sites of star formation can be spatially resolved; the result is that we want to fit a large number of molecules across a wide spectral range to multiple positions for a single observation - a time-consuming process. In order to both accelerate the production of the measurements of interstellar molecular abundances and increase the number of positions for which these measurements can be made, we aimed to produce an automated fitting routine that performs a fit for a number of rotational molecular spectra for each pixel across a chosen field of view for an ALMA image cube. 

In addition to telescope- and source-specific parameters, for each molecule we derive values for excitation temperature ($T_{ex}$), linewidth ($\Delta V$), velocity ($v_{LSR}$), and column density ($N_T$) by forward-modeling the spectra of each species and performing a least-squares fit of the combined spectra across the entire frequency range of the observations. For our treatment of NGC 6334I, we made the assumption that all of the molecules in the model share a single excitation temperature, linewidth, and velocity for each pixel while we allow the column density to vary on a molecule-by-molecule basis. We discuss the robustness of each of these assumptions in detail later. In the case of $T_{ex}$ specifically, this assumption is required in order to properly account for optical depth corrections when lines of different species overlap with one another. As many of the spectral lines in our observations are at least modestly optically thick, a full radiative transfer analysis would be needed in order to simulate contributions from molecules at more than one excitation temperature, which is not possible in the current version of \texttt{molsim}. 
%This is not currently feasible because the collisional coefficients necessary for such an analysis have not yet been measured or calculated for the vast majority of complex species.

%Decide how much of this text replaces something above or add it in
%For each pixel we derive a unique excitation temperature and linewidth that is applied to all of the molecules; we also derive a unique velocity and column density for each molecule resulting in a total of 2+(\emph{N}*2) parameters for each pixel when fitting \emph{N} molecules. In order to better manage the computational resources required for this fitting, we had to be somewhat selective with the molecules included in our model. Because we are applying the fitting routine to the same source, we first used the simulations of \citet{El-Abd19} to sort the molecules by the flux contributed to the final simulated spectrum in order to identify the most important molecules to include. We tested a number of additional molecules of interest before settling on the final list of 21 molecules (see Table \ref{mol_table} for the complete list of included molecules). The molecules that we chose to exclude from the final simulation in this work invariably contained few lines $(<20)$ in the frequency range of our observations, making an automated fit of their physical parameters extremely difficult.

While our initial goals were to use a least-squares minimizer - specifically, the bounded limited-memory BFGS algorithm \citep{Zhu:1997:550} - to find all of the relevant parameters for simulating the molecules, the nature of the spectra we were attempting to fit meant we were left with less than satisfactory results in pixels that had a particularly bright continuum or showed substantial absorption signals. For our specific test data on NGC 6334I near 300 GHz, which exhibits both very high dust continuum and high line opacity for abundant molecules, as well as a fairly extreme level of line blending, we found that it would be better to make independent measurements of the excitation temperature and linewidth for each pixel and rely on the minimizer to fit the velocity and column densities. 
%Upon public release there will be an option to allow the excitation temperature to be fit with the least-squares minimization which will be reasonable for certain environments.

\begin{figure*}[htb!] 
    \centering
    \includegraphics[width=0.98\textwidth]{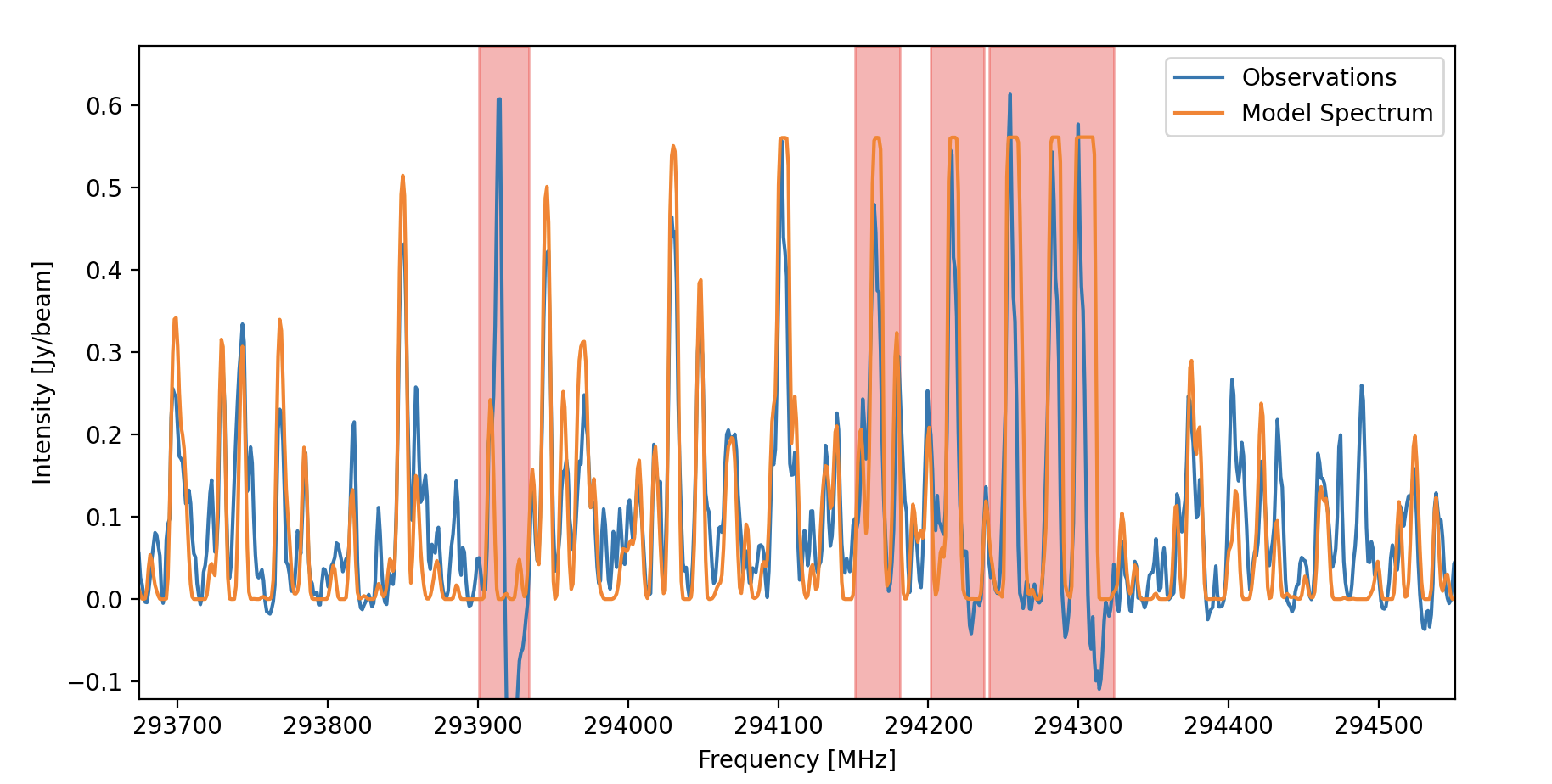}
    \caption{A sample spectrum extracted from MM1 (blue) with a simulated emission spectrum using fitted parameters from the routine (orange) overlaid. The channels highlighted in red either contain emission from a molecule used exclusively for masking purposes (such as at 293900 MHz), or contain emission from a molecule included in the model where the particular transition falls below the upper state energy threshold criteria - such lines are likely both optically thick and significantly impacted by absorption. These highlighted channels are excluded from the data used by the fitting routine in any capacity. See Section \ref{channel_exclusion} for a more detailed explanation.}
    \label{masking_example}
\end{figure*}

The excitation temperature is measured by assuming the brightest lines in our spectral range are optically thick (see Figure \ref{masking_example}). From this, it follows that we can directly solve for the excitation temperature adopting the formalism of \citet{Turner:1991:617}:

\begin{equation}
    \Delta T_B = [J_\nu(T_{ul}) - J_\nu(T_{bg}) ] [1 - \text{exp}(-\tau)]
    \label{Turner1}
\end{equation}
where the source is assumed to fill the beam, $\Delta T_B$ is the intensity of a single transition, $T_{ul}$ is the excitation temperature, $T_{bg}$ is the background temperature (measured from the continuum image of each spectral window), $\tau$ is the opacity, and 
\begin{equation}
    J_\nu(T) = (h\nu/k)[\text{exp}(h\nu/kT)-1]^{-1}.
    \label{Turner2}
\end{equation}

Assuming $\tau>>1$ and rearranging the equations to solve for $T_{ul}$ per pixel, hereafter referred to as $T_{ex}$, we have

\begin{equation}
    T_{ex} = \frac{h\nu/k}{\text{ln}\bigl(\frac{h\nu}{k(J_\nu(T_{bg})+\Delta T_B)}  + 1\bigr)}.
    \label{Turner3}
\end{equation}
The adopted value for $\Delta T_B$ is found by computing the mean intensity, $I$, of all $N$ channels within $5\%$ of the peak intensity (from any line) for each pair of spectral windows (there are two spectral windows per sideband, two sidebands per tuning, and two different tunings) using the following equations:

\begin{equation}
\begin{aligned}
    \Delta T_B & = \overline{\Delta T}_{Bspw}\\
    \Delta T_{B spw} & = \frac{\sum_{j=1}^{N} I_j}{N}
\end{aligned}
\end{equation}

%and averaging this value for each pair of spectral windows in the observations; the observations in this work were comprised of four pairs of spectral windows.

% There are, however, certain molecules that appear to consistently lie at a higher excitation temperature than the other molecules included in our model - methanol and methyl cyanide. 
%Other molecules like formaldehyde and sulfur monoxide share this trait as well, but such molecules are only used to mask the relevant parts of the observations.
%As we are finding only a single excitation temperature for all of the included molecules, the channels with emission from these aberrant molecules are excluded when calculating $\Delta T_B$. 
% As we are simulating the spectra from all of these molecules with a single excitation temperature, the channels with emission from these aberrant molecules are excluded when calculating $\Delta T_B$. 

\begin{figure}
    \centering
    \includegraphics[width=0.5\textwidth]{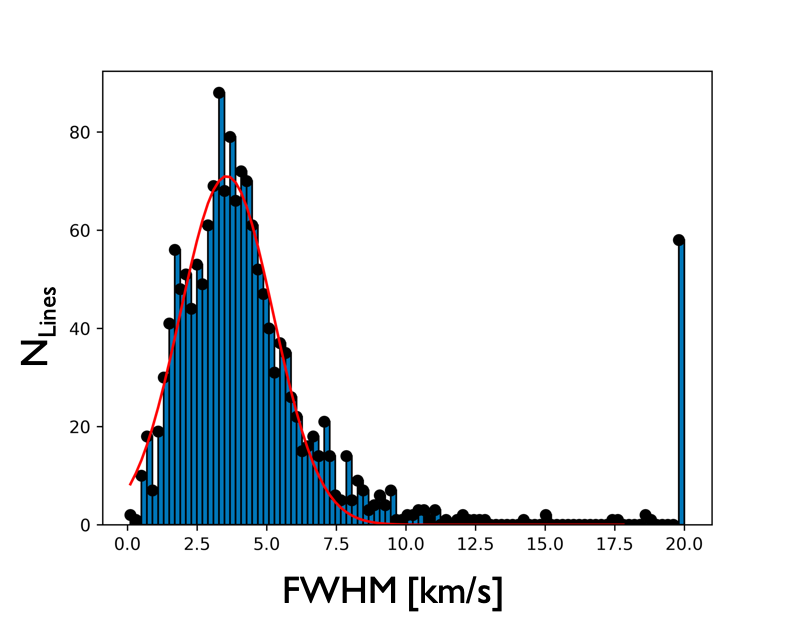}
    \caption{An example of the Gaussian fit to a histogram of FWHM values fitted for all transitions $>10\sigma$ from a single pixel (blue). The adopted value for the linewidth of a given pixel is taken from the central value of the fitted Gaussian (red line). The large value of $N_{Lines}$ for the highest velocity bin ($\sim 20$\, \kms\/ for this example) is a consequence of the peak-finding algorithm fitting a single Gaussian to the many blended line features in the data and includes any fitted FWHM $\geq 20$ \kms\/.}
    \label{dv_hist}
\end{figure}

In the case of deviations from our assumption of a well-mixed, homogeneous gas parcel, the temperature derived from the optically thick lines may not perfectly represent those of the optically thin lines for a given species.  We expect any such deviations to be small, and indeed find that the excitation temperatures derived from our optically thick lines result in fits to the optically thin lines well within expected uncertainties.
The linewidth for each pixel is found by first running a spectral peak-finding algorithm across the entire frequency range of our observations, excluding any peaks below $10\sigma$ (see Section \ref{obs} for a description of the noise measurement). A Gaussian is fit to each of these spectral peaks and the resulting FWHM values are binned in a histogram. Another Gaussian is fit to the histogram itself, with the central value of this final Gaussian adopted as the canonical linewidth for the pixel in question. Although the linewidth of any particular transition may be heavily affected by spectral line confusion, the ensemble is largely insensitive to this effect given the large number of fitted transitions  - see Figure~\ref{dv_hist}. Visual inspection of the fits and examination of residuals in numerous spectra show that the values extracted from this procedure indeed robustly describe the vast majority of spectral lines.

\begin{figure}
    \centering
    \includegraphics[width=0.4\textwidth]{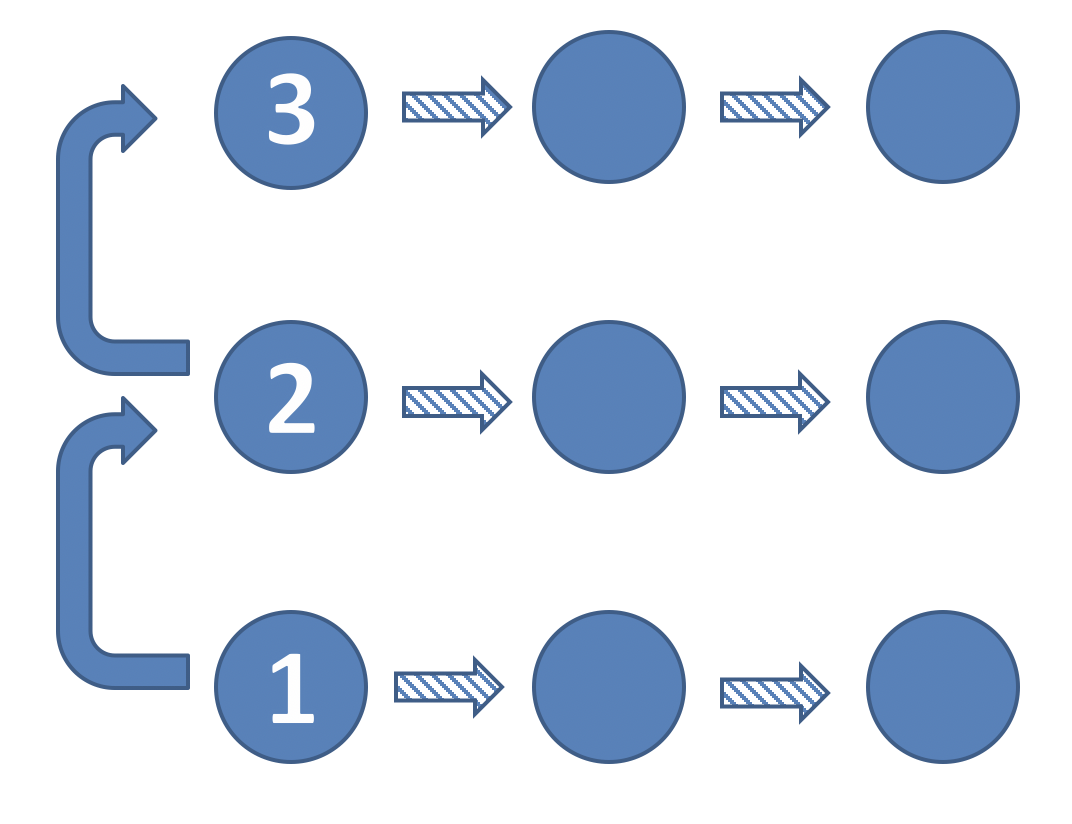}
    \caption{Diagram showing the path of the fitting routine in MM1 on a sample 3x3 image. The final fit parameters for one pixel are used as the starting parameters for the next pixel as denoted by the arrows. The dashed arrows denote the subsequent stage of fitting.}
    \label{pixels_mm1}
\end{figure}

Once the excitation temperature and linewidth have been measured, the fitting itself begins with a single pixel on the image with every molecule initialized to the same column density. Once the velocity and column densities have been measured, the same process is carried out for the pixel directly to the north, with the velocity and column densities from the previous pixel used as the starting parameters. This process is continued until the entire column has been fit (see Figure \ref{pixels_mm1}). Scripts are then launched in parallel that each use one of the pixels in this first column as a starting point and then moves across the rest of the corresponding row in the image. Using scripts in this fashion allows us to maintain the workflow of using the parameters from previous pixels as the starting point for the next fit while greatly decreasing the necessary computation time.

The minimization process is dramatically accelerated by the use of boundary conditions on the allowed values explored by the minimizer for each parameter. The flexibility on the bounds were carefully selected while balancing two factors: capturing the spread in physical parameters across NGC 6334I and a desire to make things more computationally efficient.  In this case, we are aided by the assumption that the physical parameters should not vary dramatically (orders of magnitude) between pixels.  Thus, we are able to set boundary conditions such that neighboring pixels should not differ significantly from one another. 

For the results presented in this work, the velocity was allowed to vary by $\pm$ 1.0 \kms\/ from pixel-to-pixel, while the column density for each molecule was allowed to vary by $\pm$ 0.4 dex - these bounds were centered on the fitted value from the previous pixel in the fit. While this methodology for setting the bounds works for the vast majority of pixels in our images, a poor fit in one pixel has the potential to adversely affect the fit in the subsequent pixel - see Section \ref{areasofconcern} for a discussion of where this might have impacted our results. 

\subsection{Molecules Included in the Model}
\label{molecules}

The results presented in this work are the product of a best-fit synthetic emission model containing rotational spectra from 21 different molecules. Regarding which species to simulate, the molecules included in the models of \citet{El-Abd:2019:129} served as the starting point, with molecules added and removed as we explored the capability of the automated fitting method (see Table \ref{mol_table} for the complete list). Several of the molecules initially included with the model had a relatively small number of transitions in the frequency range of our observations. Due to the small number of data points that the minimizer was working with - and more importantly, the overall number of available \emph{unblended} transitions - the fit parameters for these molecules were poorly constrained relative to the other molecules. 

Due to the poor fits in some regions for these molecules, we conducted another run of the fitting routine that was identical in every way except for the exclusion of these molecules. This run produced best-fit parameters for the remaining molecules that were well-matched to the previous results, indicating that the inclusion of a handful of poorly-constrained molecules does not negatively impact the fitting of the other molecules in our model. However we continued to exclude these molecules from future runs of the fitting routine as their inclusion drastically increased the required computation time. This increase was likely due to their minimal contributions to the final emission model over a wide range of parameters, meaning that a larger parameter space was being explored for each of these molecules. As these molecules did still have emission in a significant number of pixels in our field of view, channels that contained any of their emission were excluded from the fit. 

While the 21 molecules that we included in the fitting routine cover a significant amount of the spectral emission, there are still a number of lines that can be attributed to molecules that are not included, or are otherwise unidentified. This result is within expectations - the goal of this initial pass was to fit the bulk of the emission in as efficient a manner as possible with molecules of interest. 

%A useful aspect of fitting the emission in this manner is that it is possible to now fit additional molecules around this initial pass of established parameters. To test the feasibility of this technique, we selected two molecules to incorporate into the model after running the initial fit of 21 molecules and performed another run where they were included from the beginning (i.e. the initial run was done with all 23 molecules). The molecules in question were $^{13}$CH$_3$CN and t-H$^{13}$COOH, both of which are isotopologues of molecules already included in the model. Whether they were included in the initial set of molecules or only included in the fit afterwards, the resulting fit parameters were consistently withing the uncertainties. The ability to fit additional molecules around an established emission model has the potential to be especially useful as a starting point for searches of new interstellar molecules within complicated spectra.

\subsection{Exclusion of Channels}
\label{channel_exclusion}

We quickly found that we could not simply start the code and leave it with nothing but the minimization function to guide it. NGC 6334I is a kinematically complicated source with a number of outflows and a bright continuum \citep{Brogan:2018:87}; these properties cause effects in the spectra that make their simulations much more challenging. The key was figuring out how to identify the channels in our observations that exhibited significant absorption from molecular transitions with $E_{up} \ll T_{bg}$ or other such effects that we are currently incapable of modeling appropriately and excluding them from the fit. 
%Outflows serve to complicate the velocity structure, in some cases making multiple velocity components of a single molecule observable; this is an issue in the case where we seek to fit a single velocity for a particular molecule. A bright continuum will also manifest as significant absorption features in the observations that we cannot model. We experimented with a number of tweaks to accommodate such effects before settling on the final adjustments.

The primary solution was to use an upper state energy exclusion threshold that is applied to problematic molecules included in the model - methyl cyanide and formamide in our case. These molecules are extremely abundant and have many intense lines in the frequency range of our observations. Many of these transitions have low upper state energies - such transitions are easy to excite and are prone to displaying absorption effects due to an intervening colder layer of molecules along our line of sight. There is also the matter of the absorption of these low-lying transitions against the bright continuum  - this effect is more pronounced as the continuum level increases. These phenomena allow us to use the background temperature as a measure of which lines we can safely include in our simulations to fit. The exclusion threshold is dependent on the mean background temperature at that pixel (measured for each of four spectral windows), and excludes any channels of the observation from being fit if they are contaminated with emission from transitions with an upper state energy below the threshold (see Figure \ref{masking_example} for a demonstration). While the implementation of this threshold alleviated the opacity issue, there are still instances where the results of the fitting routine appear to be unphysical - see Figures \ref{b3} and \ref{b8} in Appendix B. The depression around MM1D is a likely consequence of the complex kinematics in that particular region, with visible evidence for multiple velocity components for some molecules. This effect appears to be more pronounced for molecules with higher column densities.

Methanol is an especially problematic molecule to fit in NGC 6334I. In the frequency range used for the study, the vibrational ground state of methanol ($v_t$=0) is dominated by relatively low $E_{up}$ transitions that are optically thick; this is true to a large degree even for the $v_t$=1 transitions. Thus, methanol required a separate solution - splitting the catalog into separate vibrational states such that we only attempted to fit the $v_t = 2$ transitions, as fitting the $v_t = 0,1$ transitions was an impossible task for most of the pixels in this source due to their significant opacity and consequent absorption effects. All of the $v_t = 0,1$ transitions were used to exclude channels from the observations as above regardless of their upper state energy. As with methyl cyanide and formamide, there appear to be regions that are difficult to get a handle on the methanol parameters, with MM1D again highlighted (Figure \ref{fig:b1}). For each of these molecules, and even less abundant molecules like methyl formate, the 13C isotopologues are certainly a more reliable representation of their spatial morphologies. 

Additionally, a handful of molecules such as \ce{H2CO} and \ce{CS} were present in our observations but had few to no unblended transitions that were also unaffected by significant absorption effects. The simulated emission from all of the transitions of these molecules was used to exclude additional channels from the observations (see Section \ref{molecules} for the criteria for including a molecule in the model and Table \ref{mol_table} for a complete list of molecules). All of these methods of treating the emission from various molecules not only improved the fits for the affected molecules like methyl cyanide and formamide, but also improved the fits for many of the other molecules included in the model. 

\subsection{Uncertainties}

%The uncertainties in the values from the fitting routine are adopted from the covariance matrix that is calculated during the minimization process with the \texttt{scipy.optimize.minimize} routine \citep{}. Due to the nature of the problem we're trying to solve, however, these calculated uncertainties should be treated with caution. Because the spectra we are attempting to fit are so dense with spectral emission, small variations in the fit parameters for certain molecules may have minimal impact on that molecule's net contribution to the minimization function. The result is that the some of the parameters we are attempting to fit are poorly constrained, and the associated errors should be treated accordingly. This is especially true in regions that we ended up masking, as we will discuss in the next section. We are actively working on an updated version of the code that is capable of more robust estimations of the fit parameter uncertainties.

The uncertainties in the values produced by the fitting are estimated through the propagation of uncertainties in the fundamental measurements with the exception of the linewidth, $\Delta V$. Spatial maps of the uncertainties for each parameter can be found in the Appendix.

\subsubsection{Excitation Temperature}

% The two relevant quantities in Eq. \ref{Turner3} for propagating the uncertainty in the excitation temperature are $\Delta T_B$ and $T_{bg}$; the uncertainty in both of these values follow from the flux calibration uncertainty, assumed here to be 10\%. The uncertainty for each channel averaged for $\Delta T_B$ are added in quadrature in the case of multiple channels. Once the uncertainty in the excitation temperature has been calculated for each spectral window (Eq. \ref{Tex_spw_unc}), these values are added in quadrature to get the final uncertainty in our excitation temperature.

The two relevant quantities in Eq. \ref{Turner3} for measuring the uncertainty in the excitation temperature are $\Delta T_B$ and $T_{bg}$. The rms noise for both of these quantities were measured and this uncertainty was propagated through the equation for excitation temperature for each spectral window; the values for each spectral window were added in quadrature. An additional factor of 10\% of the final fitted value was added in quadrature with the final rms noise value to account for the absolute flux calibration uncertainty \citep{CortesPaulo:2023:}. 

% \begin{equation}
%     \sigma_{spw} = \frac{(h\nu/k)^2 \sqrt{\sigma_{\Delta T}^2 + \sigma_{J(T bg)}^2}}{T_{ex}[(h\nu/k)+T_{ex}] \text{ln}^2(\frac{(h\nu/k)+T_{ex}}{T_{ex}})}
%     \label{Tex_spw_unc}
% \end{equation}

% \begin{equation}
%     \sigma_{spw} = \frac{(h\nu/k)^2 \text{exp}(h\nu/kT_{bg})}{T_{bg}^2[\text{exp}(h\nu/kT_{bg})-1]^2} \sqrt{\sigma_{\Delta T}^2 + \sigma_{J(T bg)}^2}
% \end{equation}

\subsubsection{Linewidth}

The uncertainty in the linewidths is taken directly from the uncertainty automatically calculated by the \texttt{lmfit} package used for the Gaussian fit to the histogram in Figure \ref{dv_hist}. 

\subsubsection{$V_{LSR}$}

The uncertainty in the central value of a Gaussian can be approximated with the following equation \citep{Campbell:2018:}

\begin{equation}
    \sigma_c = \frac{\sqrt{FWHM \Delta\nu}}{SNR}
\end{equation}

where $\Delta\nu$ is the channel width of the observation. From this, the uncertainty in our fitted velocity is the root mean square of $\sigma_c$ scaled by the number of transitions used to constrain the velocity. 

\begin{equation}
    \sigma_{V_{LSR}} = \frac{\sqrt{\sum_{j=1}^N\frac{\sigma_{cj}^2}{N}}}{\sqrt{N}}
\end{equation}

It should be stated that the uncertainty for the velocity given in this work is \emph{not the uncertainty in the velocity for any single molecule in our model.} Rather, it is the uncertainty of the single best-fit velocity for the \emph{ensemble} of included molecules. 

\subsubsection{Column Density}

%The biggest contributor to the column density uncertainty is the contribution from optically thin transitions, as small changes in the column density will not significantly affect the intensity of an optically thick line. 
The column density can be related to the intensity of a single optically thin transition with the following equation \citep{Hollis:2004:L45}

% \begin{equation}
%     N_T = \frac{4\pi^{\frac{3}{2}} \tau \nu^3 \Delta V Q}{\sqrt{\text{ln(2)}} A_{ij} g_u c^3 \text{exp}(\frac{-E_{up}}{kT_{ex}}) (\text{exp}(\frac{h\nu}{kT_{ex}}) - 1)}
% \end{equation}

% \begin{equation}
%     N_T = \frac{1}{2}\frac{3k}{8\pi^3}\sqrt{\frac{\pi}{\text{ln2}}}\frac{Q\text{exp}(E_u/T_{ex})\Delta T_B \Delta V}{B \nu S\mu^2 \eta_B} \frac{1}{1 - \frac{\text{exp}(h\nu / kT_{ex})-1}{\text{exp}(h\nu / kT_{bg})-1}}
% \end{equation}

\begin{equation}
    N_T = \frac{1}{2}\frac{3k}{8\pi^3}\sqrt{\frac{\pi}{\text{ln2}}}\frac{Q\text{exp}(E_u/T_{ex})\Delta T_B \Delta V}{B \nu S\mu^2 \eta_B \bigl(1 - \frac{\text{exp}(h\nu / kT_{ex})-1}{\text{exp}(h\nu / kT_{bg})-1}\bigr)}
\end{equation}
% prevent indent
where the relevant quantities for propagating the uncertainty are $\Delta T_B$, $\Delta V$, $T_{ex}$, and $T_{bg}$ for a single transition. However, our calculation of the column density is dependent on \emph{all} of the transitions included in the molecule's catalog. Due to the nature of performing a least-squares minimization the stronger transitions will be weighted more heavily in this calculation. The uncertainty in our final value for the column density can then be found by taking an intensity-weighted mean of the column density uncertainty calculated for individual optically thin transitions of a molecule

\begin{equation}
    \sigma_{N_T} = \sqrt{ \left(\sum_{j=1}^{N} \frac{\sigma_j I_j^2}{\sum_{k=1}^{N} I_k^2} \right)^2 + \sigma_{flux}^2 } 
\end{equation}
% prevent indent
where $N$ is the number of transitions, $\sigma_j$ is the propagated uncertainty of the column density for a single transition, $I_j$ is the intensity of the transition, and $\sigma_{flux}$ represents 10\% of the column density value to account for the absolute flux calibration uncertainty \citep{CortesPaulo:2023:}. Note that $\sigma_j$ does not take into account the absolute flux calibration uncertainty as this quantity is correlated across transitions and incorporating it before the final step would result in an underestimate of the uncertainty. As the intensity of optically thick transitions does not change with small changes to the column density, such transitions do not contribute to the uncertainty calculation and are thus excluded from the calculation. 

\subsection{Identifying Areas of Concern}
\label{areasofconcern}

While a visual inspection indicates that the simulated spectra generated by the automated fitting routine appear to match the observations within the uncertainties for the vast majority of the pixels in our field of view, there are a number of pixels for which the derived parameters were a relatively poor representation of the observations. The principle area of concern was a block of pixels toward MM1C, which displayed a large number of absorbed channels. While the derived parameters for many of the molecules were reasonable, the visible absorption negatively affected enough of the molecules that we decided to mask the region in its entirety for the column density maps. We have highlighted the pixels here as both a cautionary example and as a demonstration that the routine is capable of extricating itself from an unrealistic parameter space after fitting a problematic region. 

%While the results of the automated fitting routine are reasonable for the vast majority of the pixels in our field of view, there are some regions that the absorption from the continuum is so strong that it adversely affects the spectra in a way that can't be modeled. This effect is especially pronounced in the \ce{CH3OH} emission. Unfortunately, there are so many lines attributed to \ce{CH3OH} in our spectra that if the fit for that molecule is poor, it will negatively affect the fit of the other molecules in our list. For this reason we have identified regions in our field of view for which the \ce{CH3OH} is poor, and masked them from the presented maps. 
%Notably, these most troublesome regions do not necessarily track the known protostars. The brightest source, MM1B - which is the source that has recently undergone an accretion outburst - shows a reasonable fit for \ce{CH3OH}, while the fit around MM1C in the south necessitated a mask.

Since the rows are each fit independently in our pseudo-parallel fitting routine, we had to be careful about introducing artificial structure. This was apparent in early versions of the routine where two rows would diverge wildly in velocity and/or column density for the same molecule. There appeared to be two separate causes for such an effect:

\begin{enumerate}
    \item High opacity: When moving over areas of high opacity, the fitting routine could not match any of the strongly absorbed lines for molecules like methanol or methyl cyanide.
    \item Low initial signal: Starting the fitting routine in a region of the image with low signal would cause the column density for weak molecules to go to unrealistic values, which could propagate through the rest of the image due to the bounds being dependent on the fit from the previous pixel.
\end{enumerate}

\begin{figure}
    \centering
    \includegraphics[width=0.4\textwidth]{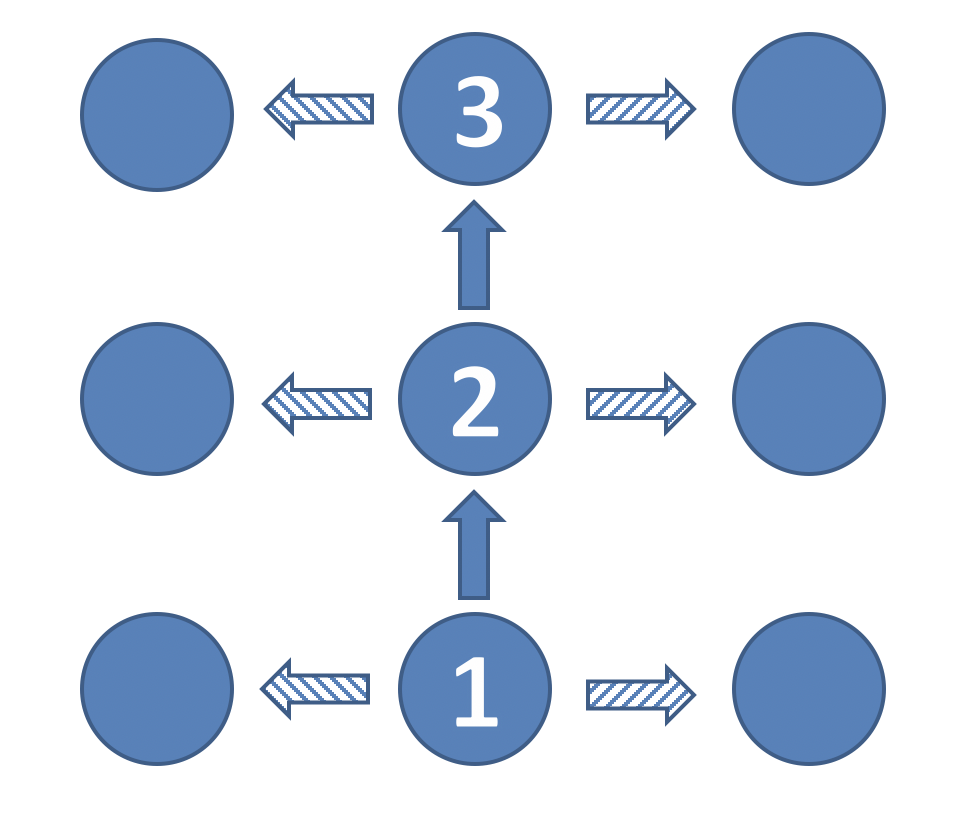}
    \caption{Diagram showing the path of the fitting routine in MM2 on a sample 3x3 image. This path was better suited to MM2 due to the low signal at the edges of our field of view for many molecules. The dashed arrows denote the subsequent stage of fitting.}
    \label{pixels_mm2}
\end{figure}

These two separate causes were evident in MM1 and MM2, respectively. The opacity issue in MM1 and efforts to mitigate it have already been discussed with the upper state energy threshold in Section \ref{channel_exclusion}, though this did not solve the issue for every pixel as shown with the additional mask we put in place around MM1C. In MM2, we found that simply starting the routine in an image column with higher signal and then branching out in either direction alleviated the issue. This fix necessitated a change in how the fitting routine hopped between pixels (see Figure \ref{pixels_mm2}). 

%Should I also mention how I tested various inputs for the first pixel and they had little impact on the results there?
% This issue, however, is symptomatic of a broader question - how much of an impact do the starting parameters have on the final parameters for a pixel? During the development of the routine we tested moving across the image from right-to-left, as well as bottom-to-top before settling on the current left-to-right configuration for MM1. During this testing we noted that the speed with which the routine was able to ``recover'' after attempting to fit pixels in the masked regions was variable. The reasons for this are not currently clear, but it should be noted that it may take a couple of attempts to find an orientation that works best for a specific source; the capability to set the starting row/column to best suit the observations of the user is functionality built in to the routine.
% With these adjustments in place, there now appears to be very little structure that is being introduced by the fitting routine itself; the fact that each of the rows are independently converging into a single smoothly varying image is an excellent sign for the efficacy and accuracy of the minimization routine. 

\section{Results}
\label{results}

\subsection{The Shared Parameters - $T_{ex}$, $\Delta V$, and $V_{LSR}$}

\begin{figure*}[htb!] 
    \centering
    \includegraphics[width=0.9\linewidth]{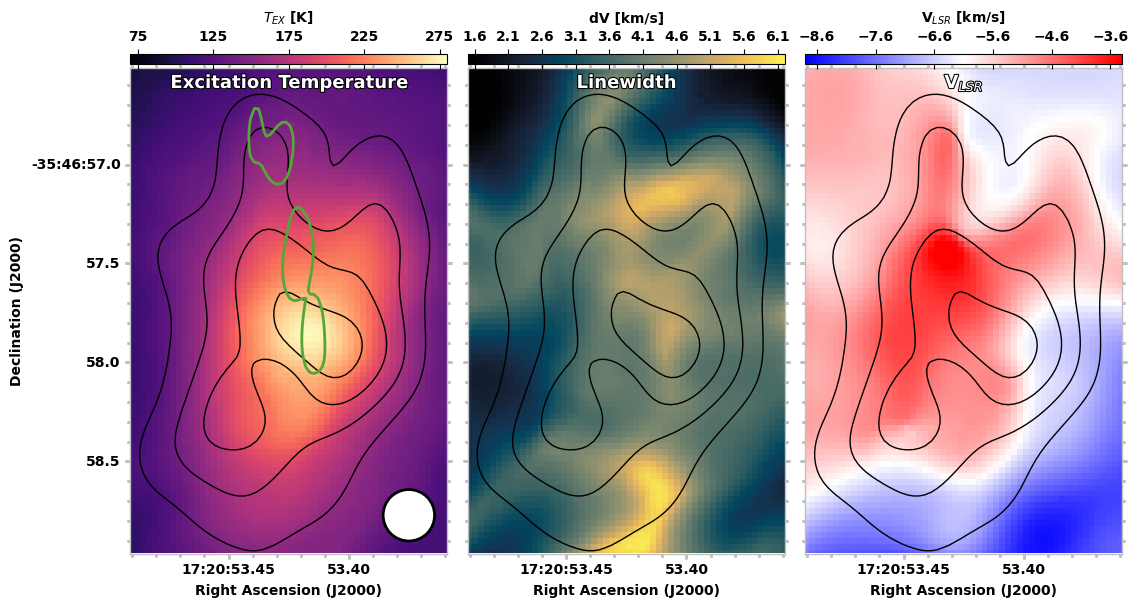}
    \caption{The excitation temperature (left) linewidth (middle), and velocity (right) maps for MM1 that were produced by the fitting routine. These parameters were shared for all of the molecules in our model. The contours mark continuum levels of 50, 90, 130, and 170 K. The green contours mark the same VLA 7mm emission as in Figure \ref{cont_fig}.}
    \label{3panel_mm1_univ}
\end{figure*}

\begin{figure*}[htb!]
    \centering
    \includegraphics[width=0.9\textwidth]{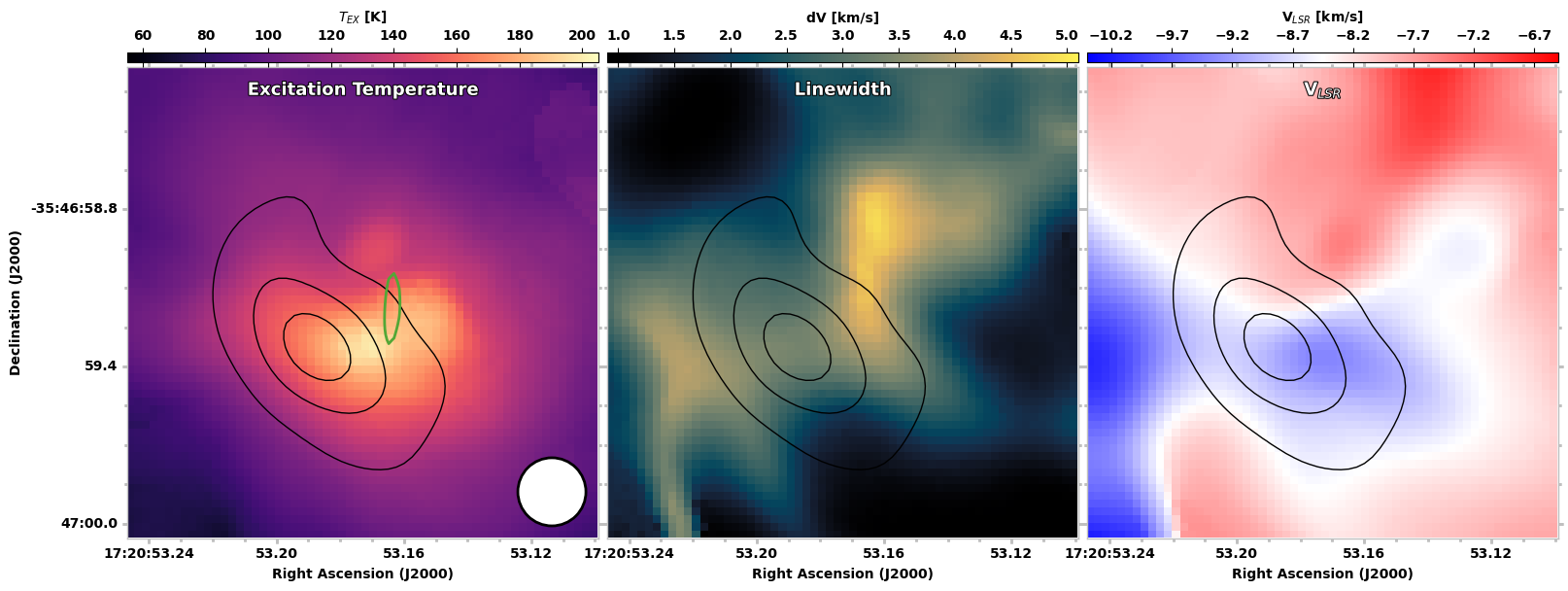}
    \caption{The excitation temperature (left) linewidth (middle), and velocity (right) maps for MM2 that were produced by the fitting routine. These parameters were shared for all of the molecules in our model. The contours mark continuum levels of 20, 35, and 50 K. The green contours mark the same VLA 7mm emission as in Figure \ref{cont_fig}.}
    \label{3panel_mm2_univ}
\end{figure*}

Figure \ref{3panel_mm1_univ} shows the final values for the excitation temperatures, linewidths, and velocities across MM1 while Figure \ref{3panel_mm2_univ} shows the same for MM2. For both of these sources, the presented linewidth images have been deconvolved from the channel width. In MM1, the excitation temperature tracks rather well with the continuum emission with little to no variation otherwise. The linewidth image, on the other hand, shows a significant amount of internal structure with broadening clearly visible in a column to the north and south. This broadening is consistent with our understanding of the kinematics of MM1 as the existence of a N-S outflow has previously been traced by maser emission and thermal lines of CS and HDO \citep{Brogan:2018:87,McGuire2018}. The line broadening is a likely consequence of the walls of the outflow impacting the more quiescent gas. This technique gives us a concrete visual demonstration on how such large scale effects have a measurable impact on the spectral emission. 

In MM2, neither the excitation temperature nor the linewidth images correlate particularly well with the continuum emission. The excitation temperature increases rapidly to the west of the continuum peak, whereas the linewidths display a clear enhancement to the northwest. 

%For both regions we used the methanol fit parameters as a gauge for the quality of the fit results taken holistically. If the methanol parameters hit the boundary conditions for at least 2 of the 4 parameters, that pixel was masked in the maps presented for this work. This masking method highlighted two specific areas of concern in MM1 around MM1C and MM1D. This criterion for masking pixels was never satisfied in MM2.

\subsection{Column Densities}
\label{column_densities}

\begin{figure*}[htb!] 
    \centering
    \includegraphics[width=0.9\linewidth]{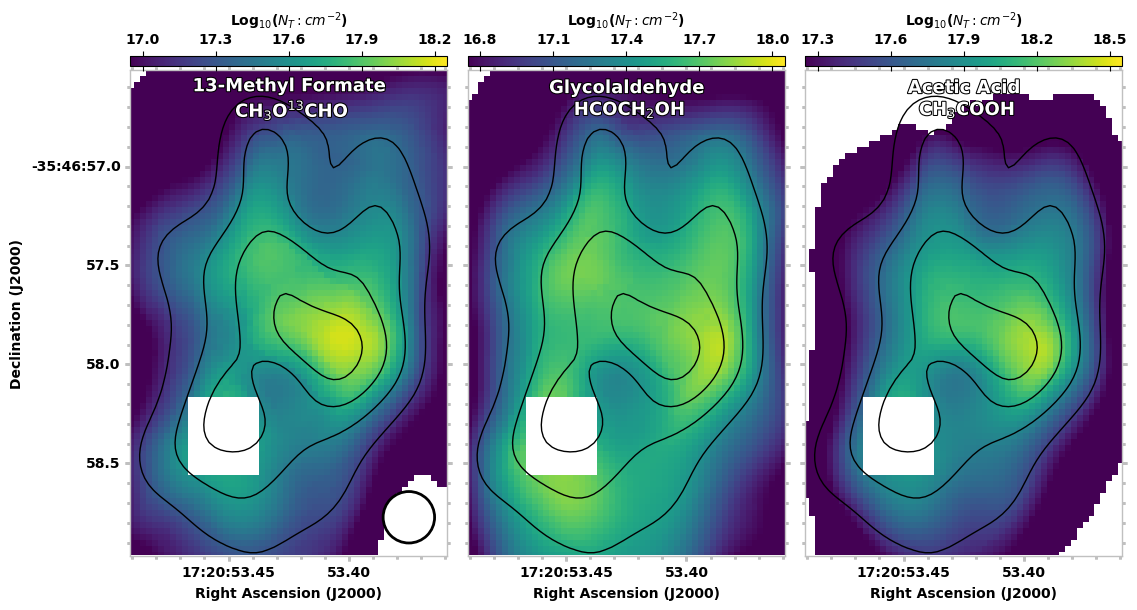}
    \caption{MM1 column density maps for the three \ce{C2H4O2} isomers: 13-methyl formate (left), glycolaldehyde (middle), and acetic acid (right) produced by the fitting routine. Pixels with a low signal-to-noise ratio have been masked on a per-molecule basis and appear white (mainly visible in the acetic acid map) with an additional mask applied around MM1C for all molecules.}
    \label{3panel_mm1}
\end{figure*}

The column density is uniquely derived for each of the 21 molecules that are included in the fitting routine's emission model. In this discussion, we will primarily be focusing on images for the three \ce{C2H4O2} isomers: methyl formate, glycolaldehyde, and acetic acid, with images for the other molecules available in the Appendix. The methyl formate image will, however, be that of the $^{13}$C isotopologue (CH$_3$O$^{13}$CHO) due to the fact that it is optically thin over our entire field of view as opposed to the standard isotopologue. Each of the images in the following sections (Figures~\ref{3panel_mm1} and \ref{3panel_mm2}) has had two masks applied. Using the final fit parameters for each molecule, any pixel with less than 3 transitions with intensities greater than 5x the rms noise was masked. This mask manifests largely in the outermost pixels of MM1 and also results in a significant portion of the MM2 field of view being masked for certain molecules. The previously-mentioned block of pixels around MM1C (\S\,\ref{areasofconcern}) was masked for all molecules. 

\subsubsection{MM1 Results}

% \begin{figure}[H]
%     \centering
%     \includegraphics[width=0.5\textwidth]{2_Panels/2panel_mm1_ch3ocho.png}
%     \caption{The column density (left) and velocity (right) spatial maps for methyl formate in MM1 that were produced by the fitting routine. }
% \end{figure}

The spatial morphology of 13-methyl formate follows the continuum emission in a fairly straightforward manner - peaks and troughs in the continuum emission are generally mirrored by increases and decreases in the 13-methyl formate column density, respectively.

%As shown in Figure~\ref{3panel_mm1}, methyl formate is the most readily detectable among the three \ce{C2H4O2} isomers across the entire field of view in MM1 and has a distinct structure with multiple locations of enhanced column density. The strongest enhancement lies to the west of MM1B, with two weaker enhancements to the southeast of MM1C and to the east of MM1D. Beyond that, notable dips in the continuum are also accompanied by drops in methyl formate's column density, but the column density in general does not necessarily reflect the continuum level otherwise.

% \begin{figure}[H]
%     \centering
%     \includegraphics[width=0.5\textwidth]{2_Panels/2panel_mm1_hcoch2oh.png}
%     \caption{The column density (left) and velocity (right) spatial maps for glycolaldehyde in MM1 that were produced by the fitting routine. }
% \end{figure}

The glycolaldehyde column density does not follow the continuum quite as closely as 13-methyl formate with several locations across MM1 presenting column densities close to the maximum; the strongest peak lies to the west of MM1B, which is offset from the peak of 13-methyl formate.

%While generally less abundant than methyl formate, glycolaldehyde is nonetheless clearly seen toward the inner regions of MM1. The column density peaks in the same position as methyl formate near MM1B but the peak does not stand out as much against the column density across the rest of the field of view. 

% \begin{figure}[H]
%     \centering
%     \includegraphics[width=0.5\textwidth]{2_Panels/2panel_mm1_ch3cooh.png}
%     \caption{The column density (left) and velocity (right) spatial maps for acetic acid in MM1 that were produced by the fitting routine. }
% \end{figure}

Acetic acid exhibits the sharpest increases and fall-offs in column density in our field of view for MM1, with a peak column density exceeding the other two molecules (its peak column density approaches that of 12-methyl formate) while also having a significant amount of masking visible due to the lack of detectable transitions in the north and south. While acetic acid also peaks to the west of MM1B, there is little apparent structure in the column density map besides also following the continuum emission fairly closely.

\subsubsection{MM2 Results}

\begin{figure*}[htb!]    
    \centering
    \includegraphics[width=\linewidth]{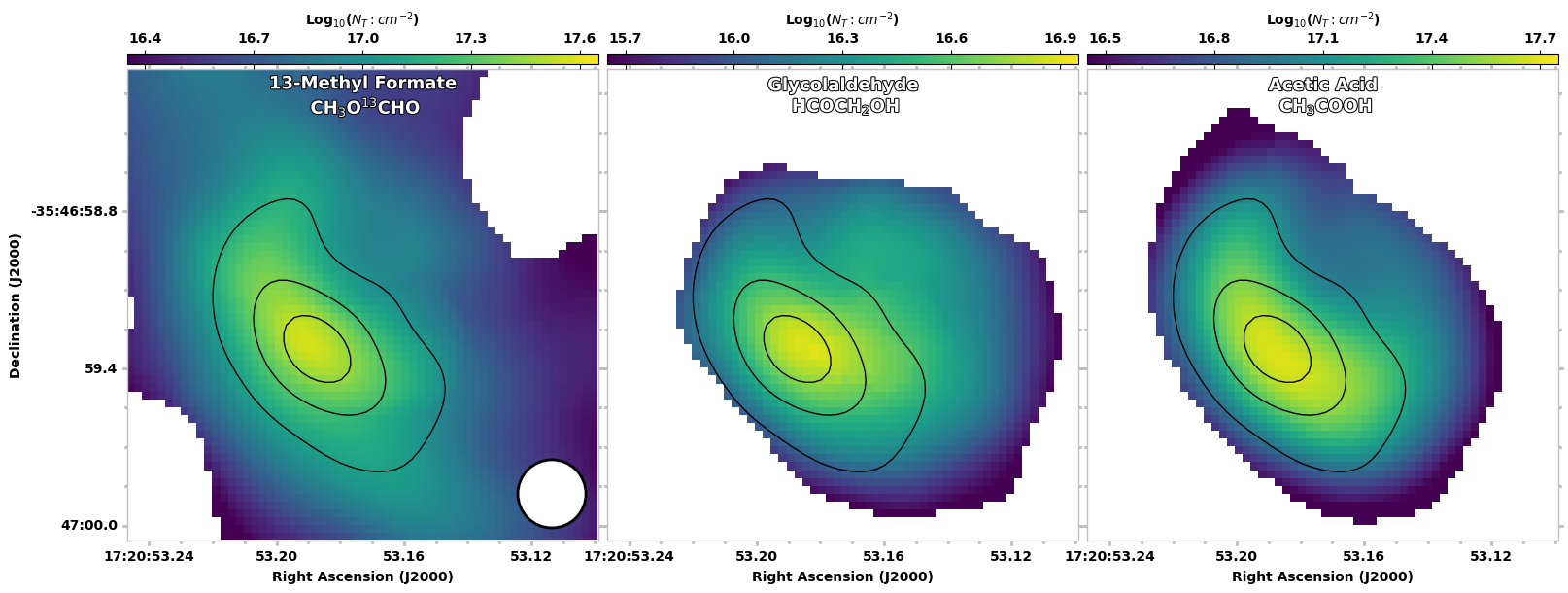}
    \caption{MM2 column density maps for the three \ce{C2H4O2} isomers: 13-methyl formate (left), glycolaldehyde (middle), and acetic acid (right) produced by the fitting routine. Pixels with a low signal-to-noise ratio have been masked on a per-molecule basis and appear white, which has a noticeable impact on all three molecules in MM2.}
    \label{3panel_mm2}
\end{figure*}

% \begin{figure}[H]
%     \centering
%     \includegraphics[width=0.5\textwidth]{2_Panels/2panel_mm2_ch3ocho.png}
%     \caption{The column density (left) and velocity (right) spatial maps for methyl formate in MM2 that were produced by the fitting routine. }
% \end{figure}

As shown in Figure~\ref{3panel_mm2}, 13-methyl formate is the easiest isomer to detect in MM2, with relatively few pixels meeting the criteria for masking. The principal structure in the column density is a narrow band that closely follows the continuum emission. A slight decrease in the column density is also visible to the northwest of the continuum peak, around MM2B.

% \begin{figure}[H]
%     \centering
%     \includegraphics[width=0.5\textwidth]{2_Panels/2panel_mm2_hcoch2oh.png}
%     \caption{The column density (left) and velocity (right) spatial maps for glycolaldehyde in MM2 that were produced by the fitting routine. }
% \end{figure}

In MM2, glycolaldehyde is largely seen towards the innermost regions, with much of the image masked due to the lack of readily detectable transitions. The previous work of \citet{El-Abd:2019:129} had established the difficulty in detecting glycolaldehyde toward this source; this is reinforced by the significant masking in the field of view due to the small number of visible transitions. The presented column densities for glycolaldehyde in MM2 should be treated as upper limits.

% \begin{figure}[H]
%     \centering
%     \includegraphics[width=0.5\textwidth]{2_Panels/2panel_mm2_ch3cooh.png}
%     \caption{The column density (left) and velocity (right) spatial maps for acetic acid in MM2 that were produced by the fitting routine. }
% \end{figure}

The acetic acid distribution in MM2 is compact around the continuum emission, with much of the field of view again being masked due to the low intensity of the strongest transitions. In pixels that haven't been masked, however, the acetic acid morphology closely resembles that of 13-methyl formate with the same narrow band cutting through the continuum contours.

\subsection{Column Density Ratios}

\begin{figure}[h!]
    \centering
    \includegraphics[width=0.5\textwidth]{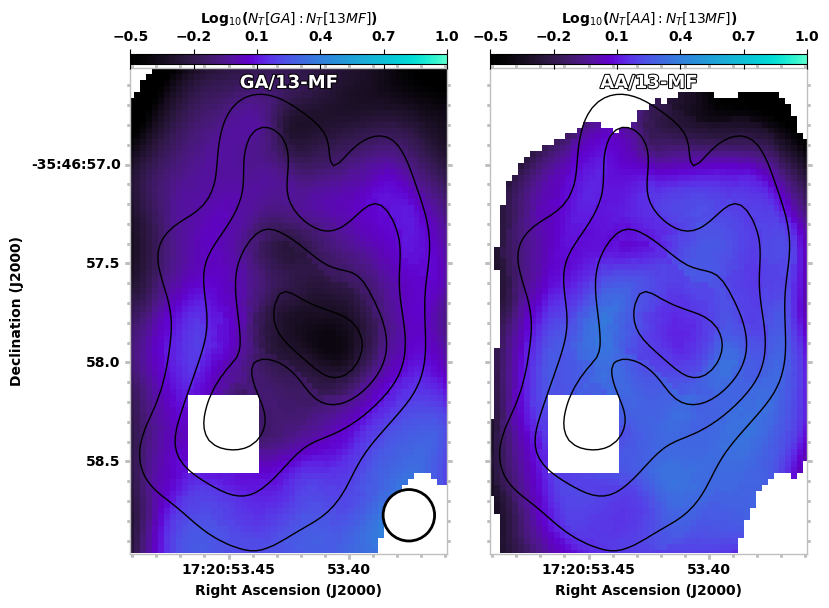}
    \caption{Images of the ratios of the glycolaldehyde \emph{(left)} and acetic acid \emph{(right)} column densities with those of 13-methyl formate in MM1.}
    \label{ratio_map}
\end{figure}

A unique benefit of this method of spectral analysis is the ability to produce images that directly display the column density ratios between molecules, shown in Figure \ref{ratio_map} for the three \ce{C2H4O2} isomers. 
There is a coherent structure to the map of the glycolaldehyde column density ratio with 13-methyl formate. The stand-out feature is the relative depletion of glycolaldehyde towards the continuum peak; it is conceivable, however, that the glycolaldehyde column density is getting underestimated in the region due to the optical depth of the spectra. The ratio with acetic acid, on the other hand, is remarkably constant over the field of view. 

%It is immediately apparent how the acetic acid column density increases rapidly compared to methyl formate as one looks at the higher continuum contours, with the two column densities approaching parity in the central regions. The glycolaldehyde column density also increases relative to methyl formate as one moves in the same direction, albeit at a much slower rate. Chemical modelling in \citet{Garrod:2022:1} suggested that methyl formate is able to desorb from dust grains more efficiently at lower temperatures than glycolaldehyde; it is possible that the shift in the glycolaldehyde to methyl formate ratio we observe is due in part to glycolaldehyde desorbing more efficiently as the temperature increases.

% Dimethyl ether is a molecule that has been commonly linked to the formation of methyl formate in the ISM \citep{}. When comparing the ratio of dimethyl ether to methyl formate column densities across our field of view in MM1, we identified a decrease in the abundance of dimethyl ether relative to methyl formate that was unique among all of the molecules included in our model. This decrease spatially coincides with the walls of the broadened linewidths in MM1, perhaps indicating that the kinematics responsible are somehow impacting the abundance of dimethyl ether. Examining the relative abundances of these molecules in this fashion is only possible through the techniques presented in this work, as one must be able to measure the physical column densities of these molecules over thousands of pixels.

%Is dme anticorrelated with mf in regions of 'interest'?

\section{Discussion}
\label{discussion}

\subsection{Previous Findings}

Analysis of the \ce{C2H4O2} isomers has previously been carried out in NGC 6334I by \citet{El-Abd:2019:129} where the relative abundances of the three molecules were compared across select locations toward both MM1 and MM2 along with measurements for other star-forming regions that were found in the literature. While methyl formate and acetic acid were found to track linearly across all of the star-forming regions in the sample, methyl formate and glycolaldehyde instead displayed a bimodal trend with their abundances. Interestingly, the abundances of these molecules in MM1 and MM2 followed separate trends, an indication of some factor(s) preferentially affecting the production of glycolaldehyde despite the regions' proximity. As these distributions were identified with a relatively small number of data points, we could now test whether the observed trends hold up with a much larger sample size across MM1 and MM2.

\subsection{Collating the Column Density Results}

The key finding in \citet{El-Abd:2019:129} was in how ratios of molecular column densities appeared to display distinct trends across multiple sources. This pointed to some physical or chemical factor affecting the production of the selected molecules in a way that had not previously been considered. In that work, the ratio of methyl formate to glycolaldehyde was shown to have two distinct linear trends among a number of star-forming regions. Crucially, NGC 6334I-MM1 and -MM2 lay in separate trends, despite their close proximity. As previously discussed however, these trends were derived from a small number of spectra extracted from each source. The values for other star-forming regions were limited to a single point each. Reproducing such a plot with the plethora of new data points we have produced in this work would serve as an excellent indicator of whether the bimodal trend previously observed was due to the small sample size, or whether it was a true distinction in the data. It should be noted that we are instead using the $^{13}$C isotopologue of methyl formate for this work, as the standard isotopologue was too optically thick for many of the positions in our field of view.

\begin{figure}[h!]
    \centering
    \includegraphics[width=0.5\textwidth]{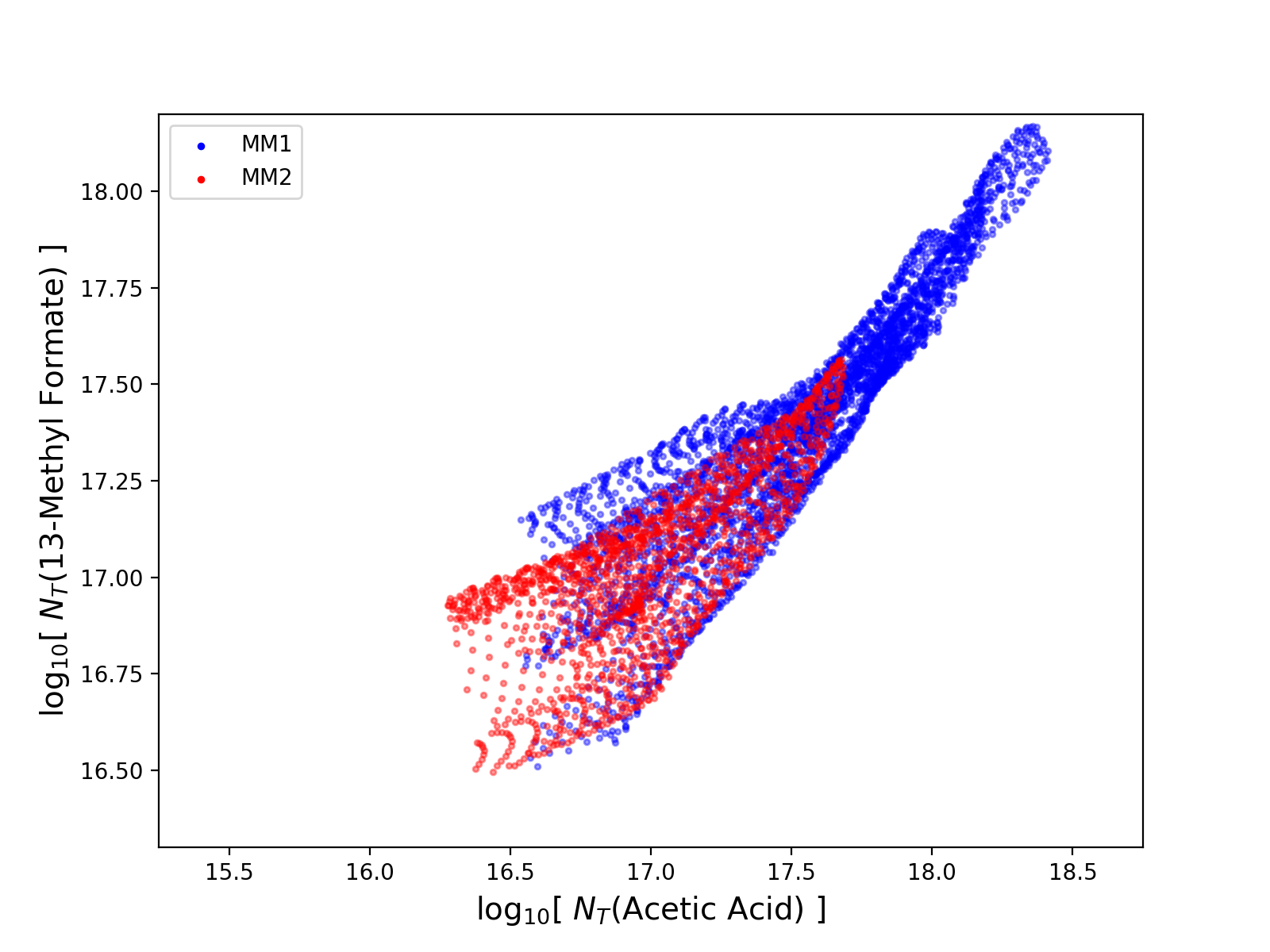}
    \includegraphics[width=0.5\textwidth]{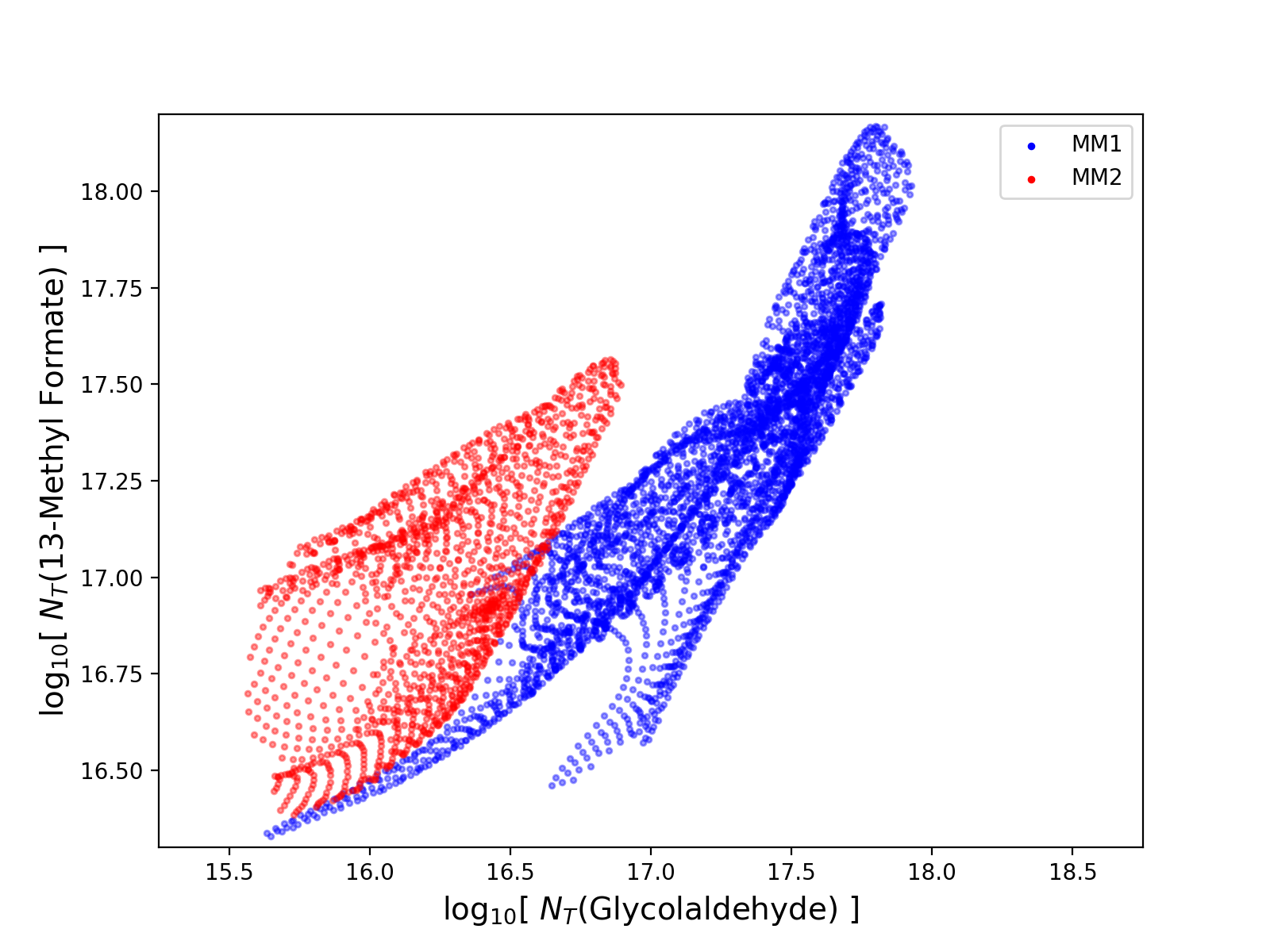}
    \caption{The column density of 13-methyl formate plotted against acetic acid (top), and glycolaldehyde (bottom). The 13-methyl formate and acetic acid column densities follow a single linear trend across MM1 and MM2, while the glycolaldehyde column density points to two distinct trends between MM1 and MM2. Masked pixels in Figures \ref{3panel_mm1} and \ref{3panel_mm2} have been excluded from these plots. The trends observed in \citet{El-Abd:2019:129} hold up with the additional data measured in this work.}
    \label{scatter_fig}
\end{figure}

%As can be seen in Figure \ref{scatter_fig}, the point clouds of methyl formate and acetic acid column density in MM1 and MM2 exhibit similar shapes, extents, and slopes, indicating that these species track linearly in a similar fashion in both regions. That is to say, as the methyl formate abundance increases, acetic acid increases at a roughly similar rate, regardless of how much methyl formate is present. The plot of methyl formate versus glycolaldehyde, however, tells a familiar story. Taking MM1 and MM2 separately, we again see that methyl formate and glycolaldehyde track with each other in a roughly linear fashion. Plotting them together, however, we see that the rate at which glycolaldehyde is increasing relative to methyl formate is demonstrably different in each source. 

As can be seen in Figure \ref{scatter_fig}, the addition of this new data has broadened the previously-observed trends of the \ce{C2H4O2} isomers; it also appears that this data now encompasses several distinct regimes in which the production of these molecules may differ from one another within the same source. Despite this, the conclusions of \citet{El-Abd:2019:129} would appear to be upheld; the acetic acid points in MM2 comprise a single trend when combined with data from MM1, while there still appears to be two wholly separate trends in the production of glycolaldehyde relative to 13-methyl formate between the two sources.

%As previously discussed, the column densities derived for glycolaldehyde in MM2 should be treated as upper limits; should the column densities in fact be significantly lower than the presented values, this would drive the MM2 trend \emph{further} away from the trend for MM1.

\begin{figure}[h!]
    \centering
    \includegraphics[width=0.5\textwidth]{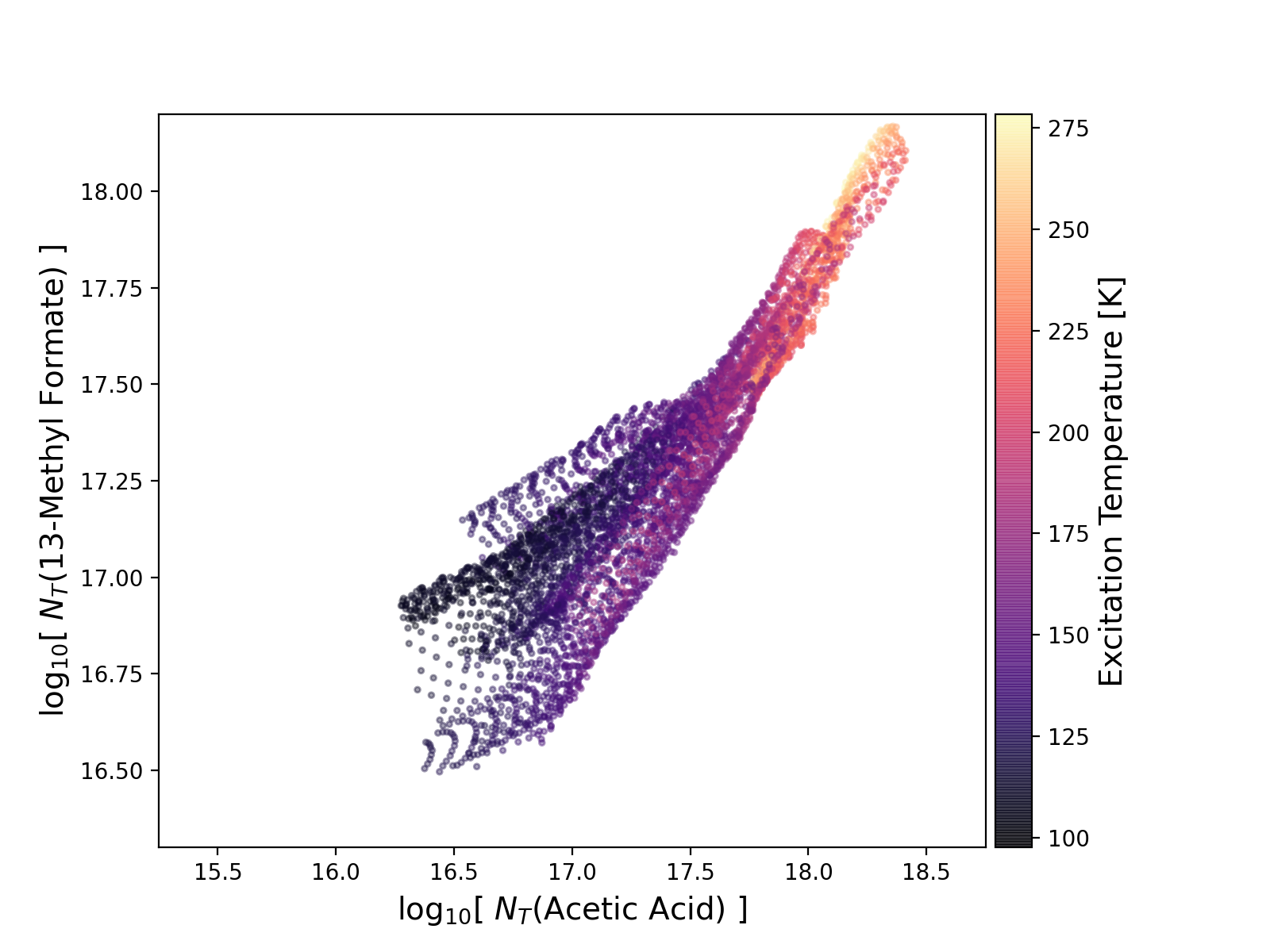}
    \includegraphics[width=0.5\textwidth]{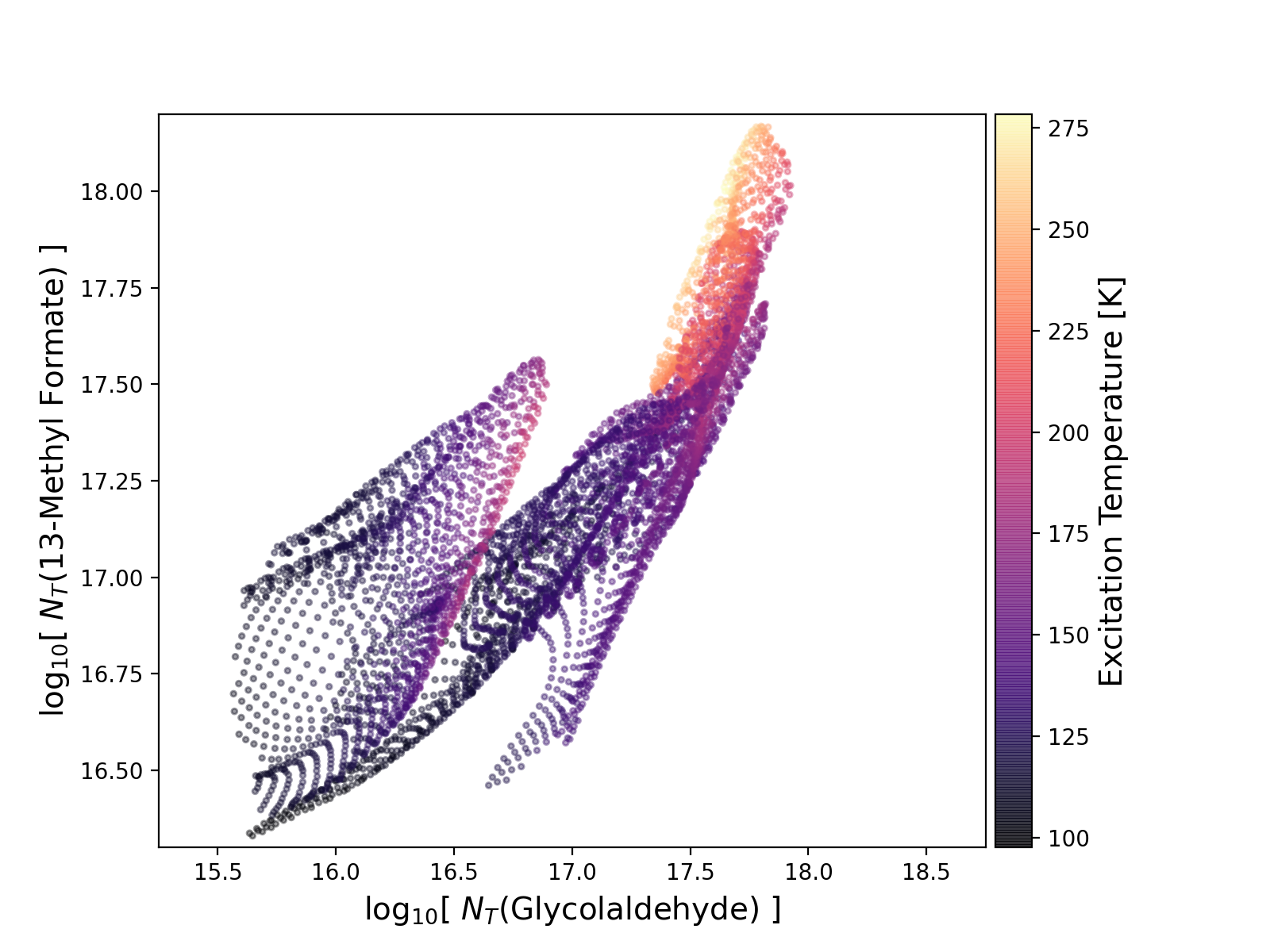}
    \caption{The same column density data in Figure \ref{scatter_fig}, this time colored by the excitation temperature of the fit.}
    \label{tex_scatter}
\end{figure}

Besides the large-scale linear trends in Figure \ref{scatter_fig} there is some finer structure that is particularly visible when the data is colored by the excitation temperature (Figure \ref{tex_scatter}). \citet{Wilkins:2022:4} took this approach when comparing the $^{13}$CH$_3$OH and CH$_3$OD column densities in Orion-KL. In the case of acetic acid, the column densities and excitation temperatures match up extremely well in the low excitation temperature regime (below $\sim150$ K). There is a distinct trend that is populated by the higher excitation temperature data in MM1, but there is no comparable MM2 data to compare. 
%The acetic acid column density increases relatively smoothly with the excitation temperature gradient across both sources, indicating that the temperature is likely impacting the gas-phase column densities for these two molecules in a similar way.
%Formed on grain surfaces and released when warm enough?

The glycolaldehyde plot in Figure \ref{tex_scatter} on the other hand, tells a much more complicated story. Taking each source separately, the glycolaldehyde column density generally increases with the excitation temperature in MM2. In MM1, the same thing happens at lower excitation temperatures (albeit at a faster rate) but the highest excitation temperature data carves out its own trend in the middle of the column density data. Interestingly, it is this high excitation temperature trend that matches the slope of the MM2 data. Overall, however, the column density and excitation temperature data are poorly correlated in the glycolaldehyde data, possibly indicating that there are other factors that play a more prominent role in the interstellar formation of glycolaldehyde. 
%Gas-phase formation?

%Figure \ref{tex_scatter} parallels work done by \citet{Wilkins:2022:4} in which the column densities derived for $^{13}\text{CH$_3$OH}$ and \ce{CH3OD} through a similar methodology were plotted against one another and colored by the pixel's rotational temperature. As in that work, the plots have distinct morphologies, but the overall shapes of the point clouds for Orion-KL and NGC 6334I do not resemble one another in a meaningful way.

%As in that work, several distinct features are visible in these plots, but the features visible in the NGC 6334I data do not correspond to any of the features in the Orion-KL observations. 

\subsection{Comparison with Traditional Moment Maps}
\label{mom_map_comparison}

\begin{figure}[h!]
    \centering
    \includegraphics[width=0.48\textwidth]{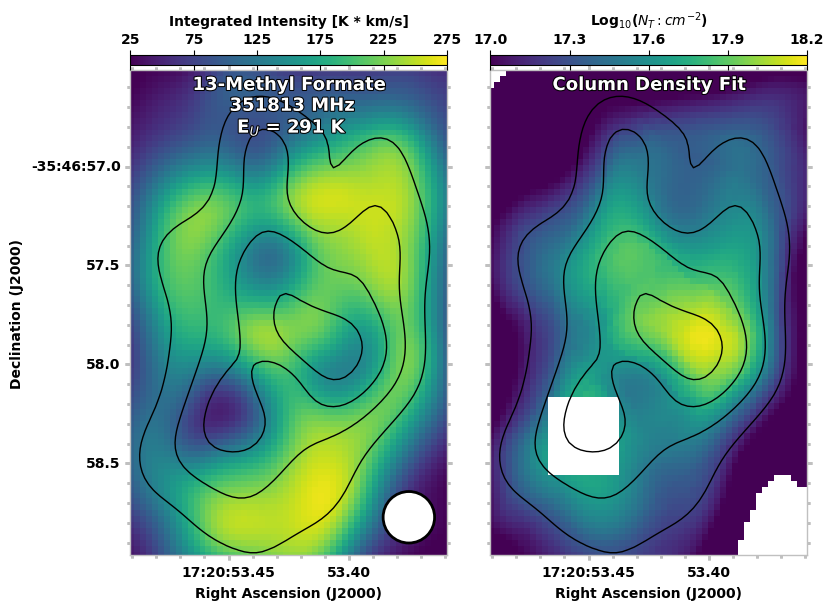}
    \caption{A side-by-side comparison of an integrated intensity map (left) using a single 13-methyl formate transition and the 13-methyl formate column density image produced by the fitting routine (right) in MM1. The integrated intensity map was produced by integrating over a velocity range of -10.4 to -2.7 \kms.}
    \label{mm1_mom0}
\end{figure}

%Substituting 13-MF for the figures in this section
% \begin{figure}[h!]
%     \centering
%     \includegraphics[width=0.48\textwidth]{2panel_mm1_ch3ocho_mom0_comparison.png}
%     \caption{A side-by-side comparison of an integrated intensity map (left) using a single methyl formate transition and the methyl formate column density image produced by the fitting routine (right) in MM1. The integrated intensity map was produced by integrating over a velocity range of -10.4 to -2.7 \kms.}
%     \label{mm1_mom0}
% \end{figure}

The standard methods for analyzing the spatial distribution and kinematics of interstellar molecules over a large field of view make use of a variety of moment maps. Moment 0 maps analyze the integrated intensity of a single transition of a molecule across a given region and are often used as a proxy for the spatial distribution of the molecule. It is impossible, however, to disentangle the excitation conditions of a single transition from the physical abundance of its associated molecule using such an analysis. In regions that span a range of temperatures, this raises the question of whether we are observing a variation in the abundance of the molecule or whether we are highlighting regions where the physical conditions more readily excite a particular transition. Moment 1 maps use the intensity-weighted velocity of a transition to generate a velocity-field and are useful for studying the kinematics of a molecule. A velocity or linewidth gradient across the field of view, however, significantly hampers the applicability of these techniques. Spectral contamination from other molecules may easily be introduced, or the emission that one is attempting to integrate may shift out of the selected channel range, though efforts have been made to mitigate these issues such as the VINE maps introduced by \citet{Calcutt:2018:A90}. In quiescent regions, these concerns can be alleviated by carefully selecting a transition and channel range to isolate only the emission of interest; however, in star-forming regions with complex kinematics and molecular inventories that vary from pixel to pixel, such a task is monumentally more difficult. 

\begin{figure}[h!]
    \centering
    \includegraphics[width=0.48\textwidth]{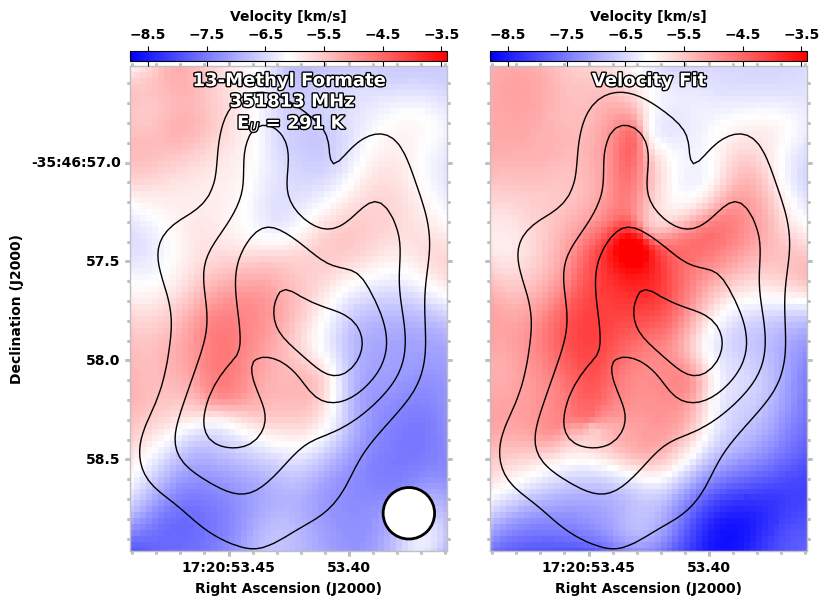}
    \caption{A side-by-side comparison of a velocity moment map (left) using a single 13-methyl formate transition and the velocity image produced by the fitting routine (right) in MM1.}
    \label{mm1_mom1}
\end{figure}

In contrast with moment map analyses, the images that are presented as part of this work -- for both the physical column densities and the velocity -- are derived using the entire set of transitions for each included molecule in our observed bandwidth. Thus, we have removed any ambiguity that comes from the excitation of a single transition - provided there are enough transitions for said molecule in the scope of our observations to appropriately model the column density - as well as alleviated any concerns about the velocity shifting out of a relevant range. The comparison of the two methods in this work was conducted with 13-methyl formate. There are a number of 13-methyl formate transitions in our frequency range; unfortunately, while the transition used to create these moment maps appeared to be unblended for a number of positions, it also appears to be noticeably optically thick in several positions in MM1. It remains, however, the best of the available options in our data, and serves to illustrate the potential pitfalls when attempting to apply the moment map methodology to such a complex dataset. Even the $^{13}$C isotopologue of an abundant molecule is capable of apparent opacity issues, along with the difficult task of identifying a transition that is perpetually unblended across thousands of pixels in a region with an evolving molecular inventory and varying physical parameters.

%The moment maps in Figures \ref{mm1_mom0}, \ref{mm1_mom1}, \ref{mm2_mom0}, and \ref{mm2_mom1} were produced using the same methyl formate transition used to create the moment maps in \citet{El-Abd:2019:129} though the maps were made using reprocessed data as described in Section \ref{obs}; this transition was carefully chosen as it was unblended and appeared to be optically thin for the positions analyzed in that work. Note that we are using 12-methyl formate in these images, not the $^{13}$C isotopologue.

\subsubsection{Comparison in MM1}

There are several differences in the morphologies of the integrated intensity maps and column density images of methyl formate (Figure \ref{mm1_mom0}) that are immediately apparent. Many of the regions that were highlighted in Figure \ref{cont_fig} coincide with dips in the integrated intensity map; this is most likely a consequence of this particular transition having a high optical depth at these positions. The column density image, as previously discussed, instead varies smoothly across our field of view in a manner that loosely follows the continuum emission. This highlights the challenges of using moment maps as proxies for molecular distributions, as there are clear morphological differences from a true map of the column density.

%As the depressions in intensity have such a drastic effect on the moment map, it is difficult to compare the spatial morphology of the molecule in the rest of the field of view between the two maps. In any case, direct comparisons of these maps should be taken with a grain of salt as they are representations of two distinct quantities in the integrated emission and the column density, respectively. That said, as these are both tools to explore the spatial distribution of molecules, it is a useful exercise to discuss them together.}

In contrast, the moment 1 map from this transition and the velocity image from the routine are qualitatively much more similar (Figure \ref{mm1_mom1}). There still exist discrepancies between the two, however, with the greatest divergence occurring around MM1F on the order of several kilometers per second. Delving into the spectra in some of these pixels, it was clear that the kinematics had not been adequately captured by the channels selected for the moment map (see Appendix C for a visual demonstration). This reinforces the difficulty in finding a single range of channels that adequately treats an entire star-forming region for the purposes of generating a velocity field from a moment map. Simply increasing the number of channels to capture the emission in one position would introduce unwanted emission in other positions. 
In both presented moment maps there are myriad effects which may negatively impact our ability to gain accurate information and are alleviated by the images produced by the fitting routine. The opacity of an individual transition is mitigated by leveraging all of the available transitions for a particular molecule; spectral contamination of an individual transition from molecular variation in different positions also ceases to be an issue. In addition, the velocity image is not skewed by improperly accounting for the velocity structure in a source - an impossible task for a source as complicated as NGC 6334I.

\subsubsection{Comparison in MM2}

\begin{figure}[h!]
    \centering
    \includegraphics[width=0.5\textwidth]{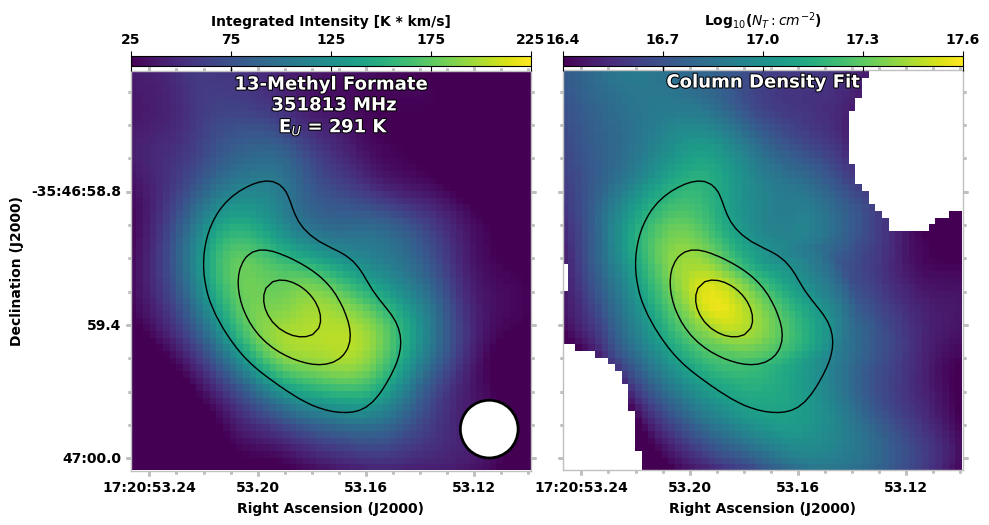}
    \caption{A side-by-side comparison of an integrated intensity map (left) using a single 13-methyl formate transition and the 13-methyl formate column density image produced by the fitting routine (right) in MM2.}
    \label{mm2_mom0}
\end{figure}

\begin{figure}[h!]
    \centering
    \includegraphics[width=0.5\textwidth]{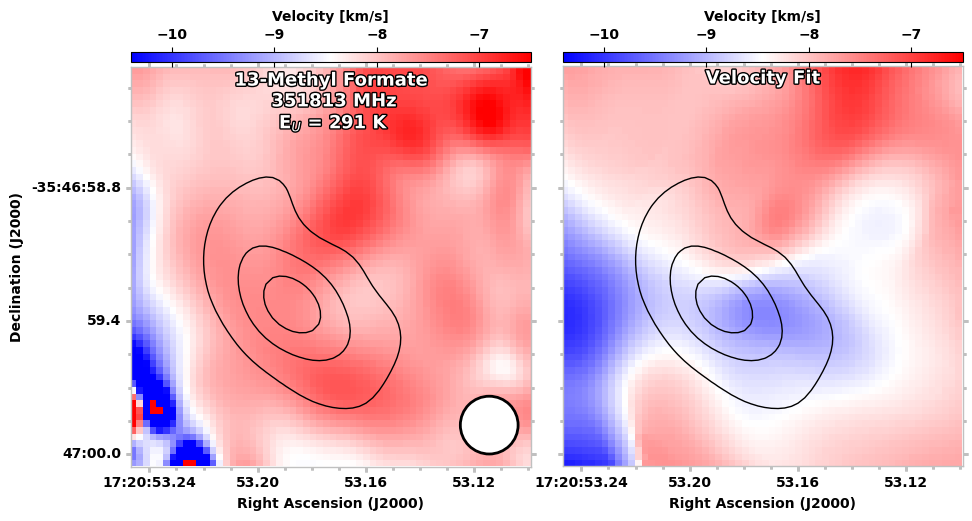}
    \caption{A side-by-side comparison of a velocity moment map (left) using a single 13-methyl formate transition and the velocity image produced by the fitting routine (right) in MM2. Note that the speckled portion in the bottom left of the moment map corresponds to a region with very weak 13-methyl formate emission (see Figure \ref{mm2_mom0}).}
    \label{mm2_mom1}
\end{figure}

The morphologies of the integrated intensity maps and column density images are in much better agreement in MM2 than MM1 (Figure \ref{mm2_mom0}). The molecule is fairly well matched to the continuum emission in both cases, and there are no regions where the two plots significantly differ, likely due to the generally lower opacity of MM2. 
The moment 1 map is in fairly good agreement with the velocity image outside of the displayed contours (Figure \ref{mm2_mom1}), but diverges fairly significantly (again on the order of a few kilometers per second) towards the warmer region. See Appendix C for a discussion on how such a discrepancy may arise.

% The morphologies of the integrated intensity maps and column density images are in much better agreement in MM2 than MM1 (Figures \ref{mm2_mom0}, \ref{mm2_mom1}). There are still notable differences; the column density extends more evenly over the field of view while the integrated intensity is more compact around the continuum, for instance. However, the same features can be picked out of both panels, perhaps most significant of which is the gap present to the northwest of the continuum peak. This gap lines up with MM2B, which is perhaps having an impact on the opacity which is manifesting in both images.
% The pair of velocity fields are again more similar in MM2 than in MM1, though there remain differences, mostly constrained to the blueshifted swathe of pixels cutting through the continuum peak of MM2. 

%Sort out the comments at the bottom
\begin{table*}
\centering
\footnotesize
\caption{Molecules Included in the Fitting Routine}
\label{mol_table}
\begin{tabular}{ccccccc}
\hline\hline
Molecule & Catalogs & Vib. States $^{1,2}$ & No. of Transitions $^3$ & $E_{U}$ Range [K] & $E_{U}$ Threshold & Database $^4$ \\
\hline

\ce{CH3OH} & 1 & $\mathbf{v_t = 2}$ & 49 & 573 - 2524 & None & CDMS \\
$^{13}$CH$_3$OH & 1 & $\mathbf{v_t = 0-1}$ & 57 & 17 - 891 & None & CDMS \\
\ce{CH3CN} & 2 & $\mathbf{v = 0, v_8 = 1}$ & 112 & 120 - 2969 & $3*T_{BG}$ & CDMS \\
CH$_3$ $^{13}$CN & 1 & $v = 0$ & 29 & 120 - 1721 & None & CDMS \\
\ce{NH2CN} & 1 & $v = 0$ & 76 & 101 - 1569 & None & JPL \\
\ce{H2CCO} & 1 & $v = 0$ & 29 & 102 - 2323 & None & CDMS \\
\ce{CH3CHO} & 1 & $\mathbf{v_t = 0-2}$ & 626 & 26 - 2939 & None & JPL \\
\ce{NH2CHO} & 2 & $v = 0, v_{12} = 1$ & 145 & 58 - 2472 & $3*T_{BG}$ & CDMS \\
t-HCOOH & 1 & $v = 0$ & 98 & 24 - 2048 & None & CDMS \\
\ce{CH3OCH3} & 1 & $\mathbf{v = 0}$ & 267 & 32 - 1302 & None & CDMS \\
\ce{C2H5OH} & 1 & $v = 0$ & 1143 & 49 - 2688 & None & CDMS \\
\ce{C2H5CN} & 1 & $v = 0$ & 348 & 25 - 2886 & None & CDMS \\
\ce{CH3COCH3} & 1 & $\mathbf{v_t = 0-2}$ & 1486 & 46 - 2241 & None & JPL \\
\ce{CH3OCHO} & 1 & $\mathbf{v_t = 0-1}$ & 993 & 38 - 1749 & None & JPL \\
\ce{HCOCH2OH} & 1 & $v_t = 0-3$ & 2142 & 27 - 4916 & None & CDMS \\
\ce{CH3COOH} & 2 & $\mathbf{v_t = 0, v_t = 1}$ & 2790 & 36 - 2398 & None & CDMS \\
CH$_3$O$^{13}$CHO & 1 & $\mathbf{v_t = 0-1}$ & 1435 & 25 - 1485 & None & CDMS \\
\ce{a-(CH2OH)2} & 1 & $v = 0$ & 951 & 60 - 1696 & None & CDMS \\
\ce{g-(CH2OH)2} & 1 & $v = 0$ & 1104 & 59 - 1912 & None & CDMS \\
\ce{CH3OCH2OH} & 1 & $v = 0$ & 421 & 67 - 1105 & None & CDMS \\
\ce{SO2} & 2 & $\mathbf{v = 0, v_2 = 1}$ & 90 & 29 - 5355 & None & CDMS \\
\hline
\ce{CH3OH}$^{\ddag}$ & 1 & $\mathbf{v_t = 0-1}$ & 216 & 17 - 2640 & N/A & CDMS \\
CH$_3$ $^{18}$OH$^{\ddag}$ & 1 & $\mathbf{v_t = 0-2}$ & 160 & 29 - 2620 & N/A & CDMS \\
\ce{HNCO}$^{\ddag}$ & 1 & $v = 0$ & 14 & 143 - 1536 & N/A & CDMS \\
\ce{SO}$^{\ddag}$ & 2 & $v = 0, v = 1$ & 7 & 26 - 1718 & N/A & CDMS \\
\ce{HC3N}$^{\ddag}$ & 1 & $\mathbf{v = 0}$ & 2 & 217 - 307 & N/A & CDMS \\
\ce{OCS}$^{\ddag}$ & 2 & $v = 0, v_2 = 1$ & 6 & 161 - 924 & N/A & CDMS \\

% \ce{} &  & $3.4 (7) \times 10^{13}$ & $< 2.1 \times 10^{13}$ & 1.6 (0.4) & Small \\
% \ce{} &  & $9 (3) \times 10^{15}$ & $< 8.0 \times 10^{14}$ & 6 (4) & Small \\
% \ce{} &  & $8 (2) \times 10^{15}$ & ... & 8 (3) & Small \\
\hline\hline
\multicolumn{7}{l}{$^1$Vibrational states in separate catalogs are denoted by commas.}\\
\multicolumn{7}{l}{$^2$Bolded text indicates entries where vibrational contributions to the partition function are included.}\\
\multicolumn{7}{l}{ Column density estimates for the non-bolded entries are potentially slightly underestimated in the case}\\
\multicolumn{7}{l}{ an updated partition function is calculated. All entries use the most up-to-date partition functions as of publication.}\\
\multicolumn{7}{l}{$^3$This number refers to all of the transitions in the catalog in our frequency range, not all detected transitions.}\\
\multicolumn{7}{l}{$^4$The catalogs were obtained from both the CDMS \citep{Muller:2005:215} and JPL \citep{Pickett:1998:883} spectroscopic databases.}\\
\multicolumn{7}{l}{$^{\ddag}$Channels containing emission from these molecules are excluded from the fit.}\\

\end{tabular}
\end{table*}

% \subsection{Comparing MM1B and MM1D}

% Despite the proximity of the two star-forming cores in MM1, their respective spectra display distinct properties. In addition to varying in typical parameters such as temperature, velocity, and linewidth, MM1D appears to display multiple velocity components in a large number of molecules. The possibility that an absorption effect was the cause for the apparent velocity components also had to be considered. The fact that i: this effect was observed in only one of the continuum sources and ii: this effect was evident in positions far removed from the continuum sources lends credence to the possibility that multiple velocity components are being observed.

% \begin{figure*}
%     \centering
%     \includegraphics[width=0.5\textwidth]{mm1b.png}
%     \includegraphics[width=0.5\textwidth]{mm1d.png}
%     \caption{The spectrum of vibrationally-excited \ce{CH3CN} in MM1B and MM1D, respectively. MM1B is modeled very well at a single velocity, whereas the same transitions in MM1D appear to display two distinct velocity components. }
%     \label{}
% \end{figure*}

\section{Conclusions}

Star-forming regions are physically complicated environments; if we are to better understand the effects that evolving protostars have on their environments, and how those environments in turn affect the formation of the protostars, we must first be able to understand the wide range of physical conditions present in these sources. While this problem is of a scale that would be intractable to solve by treating individual pixels, we have demonstrated the feasibility of an automated approach to the measurement of the physical conditions and molecular column densities in NGC 6334I-MM1 and -MM2. The observed trends with respect to the bifurcation of the relative column densities of the \ce{C2H4O2} isomers were in agreement with the work of \citet{El-Abd:2019:129}. Comparisons were conducted with the results of the fitting routine and the closest moment map analogues. In comparing the moment 1 map and velocity image the fitting routine more accurately represented the velocity of the emission. Comparing the moment 0 maps and column density images demonstrated shortcomings in attempting to use the moment 0 map as a proxy for the column density.

%While comparing the moment 0 map and column density image is not ideal as they are distinct quantities, the difficulty in selecting an appropriate channel range over which to integrate the emission for turbulent regions was highlighted; this issue is not present when creating the column density map.

\clearpage
%\begin{comment}
    
\section{Appendix A}
\label{appendix_a}
The spectra from NGC 6334I-MM1 and -MM2 are remarkably varied in both the level of spectral crowding and the overall intensity of the molecular emission. We have selected a number of positions from which to extract spectra to demonstrate this fact in Figures \ref{mm1spectra}--\ref{mm2spectra}, with emission from the \ce{C2H4O2} isomers highlighted. 

\begin{figure*}
    \centering
    \includegraphics[width=0.75\textwidth]{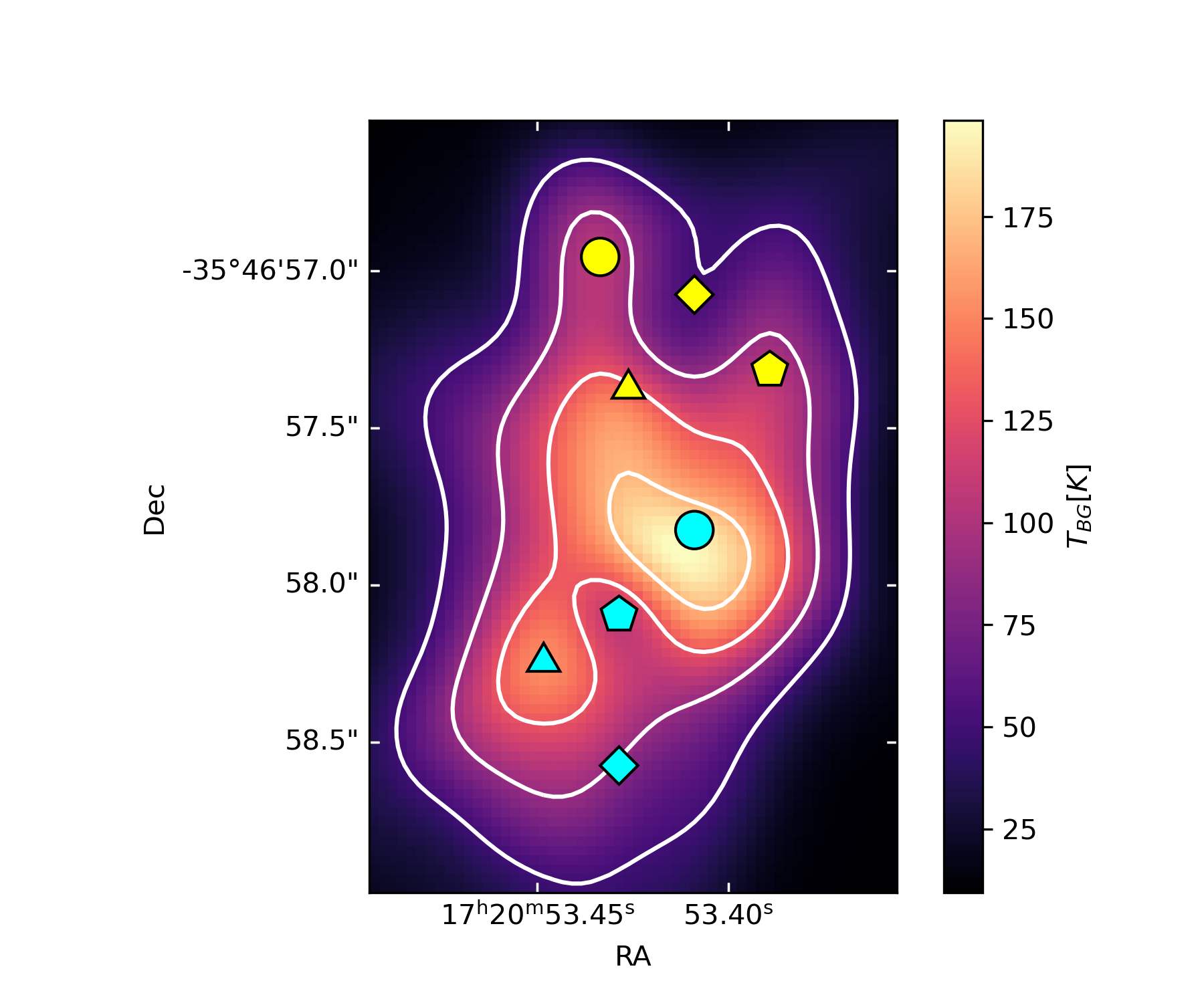}
    \caption{A sample of spectra were extracted from the marked positions to display the variety in physical conditions across MM1 in Figures~\ref{mm1spectra} and \ref{mm1spectra2}.}
    \label{mm1_finder}
\end{figure*}

\begin{figure*}
    \centering
    \includegraphics[width=0.75\textwidth]{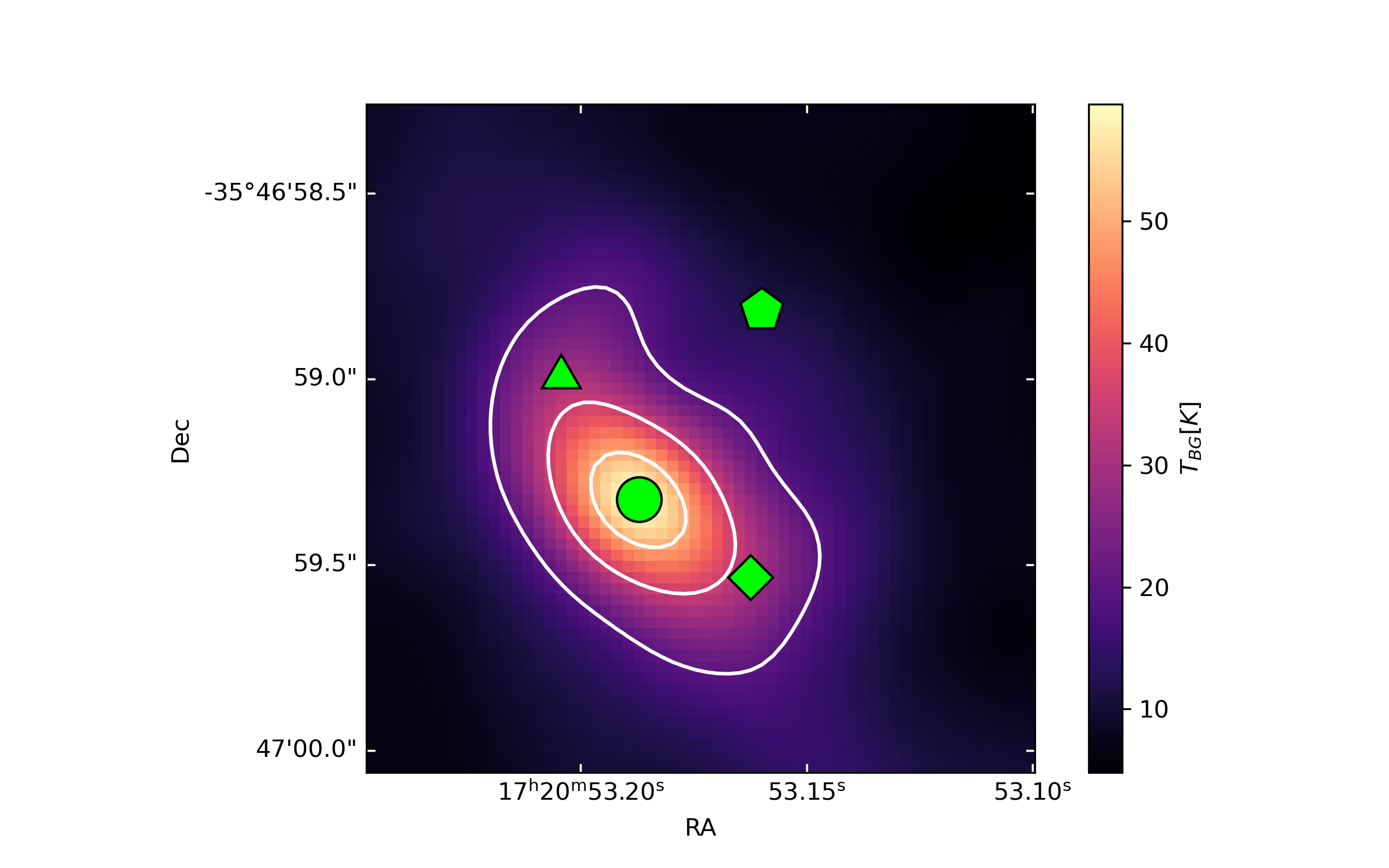}
    \caption{A sample of spectra were extracted from the marked positions to display the variety in physical conditions across MM2 in Figure~\ref{mm2spectra}.}
    \label{mm2_finder}
\end{figure*}

\clearpage

\begin{figure*}
    \centering
    \includegraphics[width=0.85\textwidth]{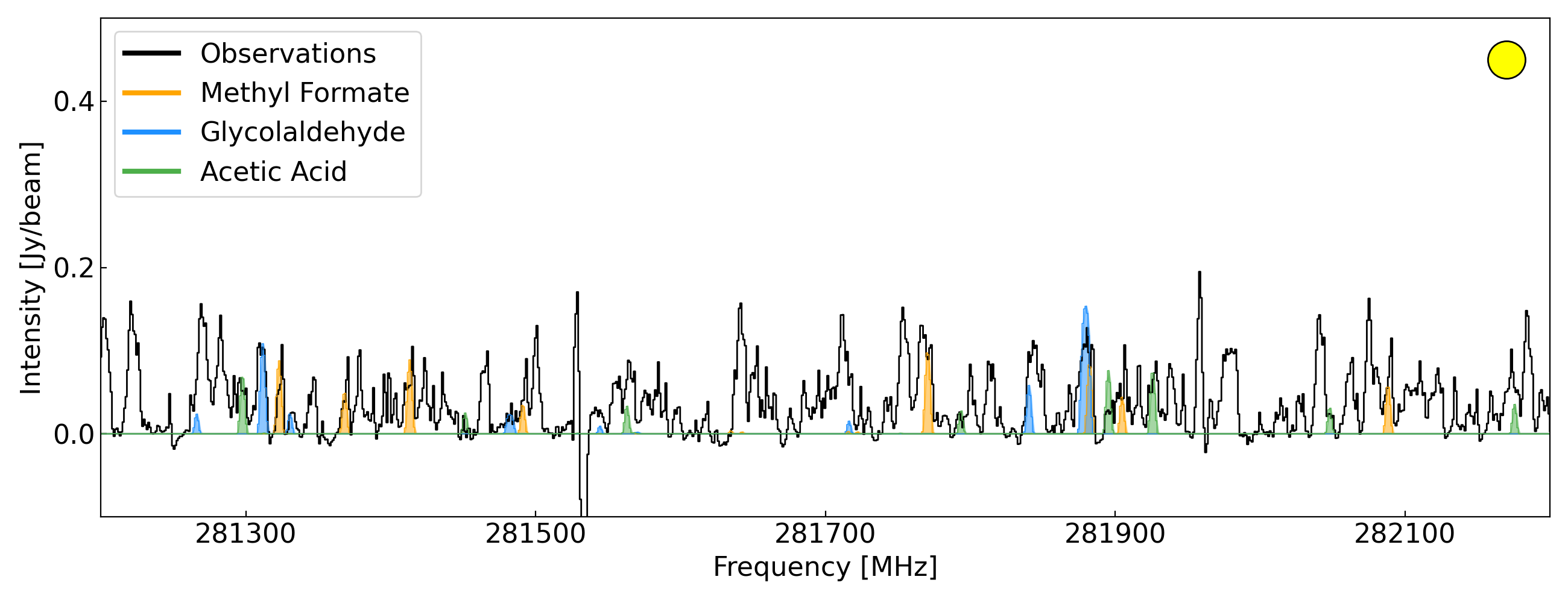}
    \includegraphics[width=0.85\textwidth]{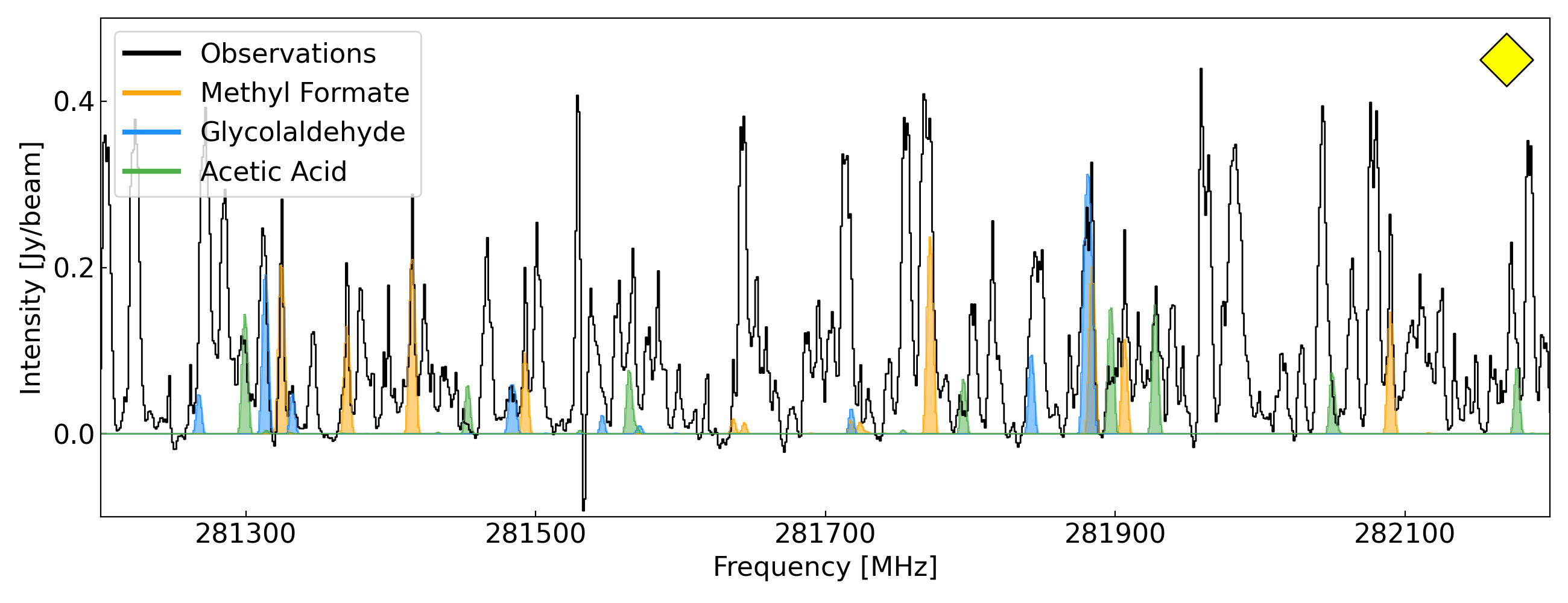}
    \includegraphics[width=0.85\textwidth]{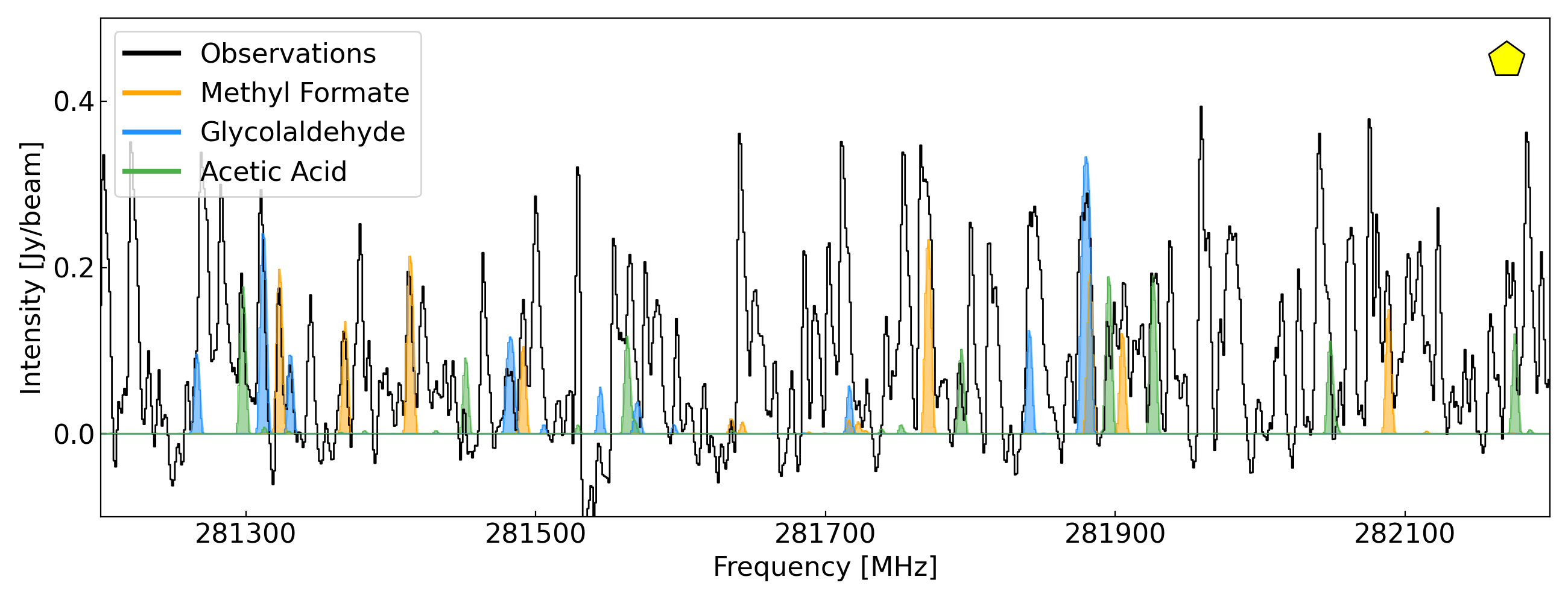}
    \includegraphics[width=0.85\textwidth]{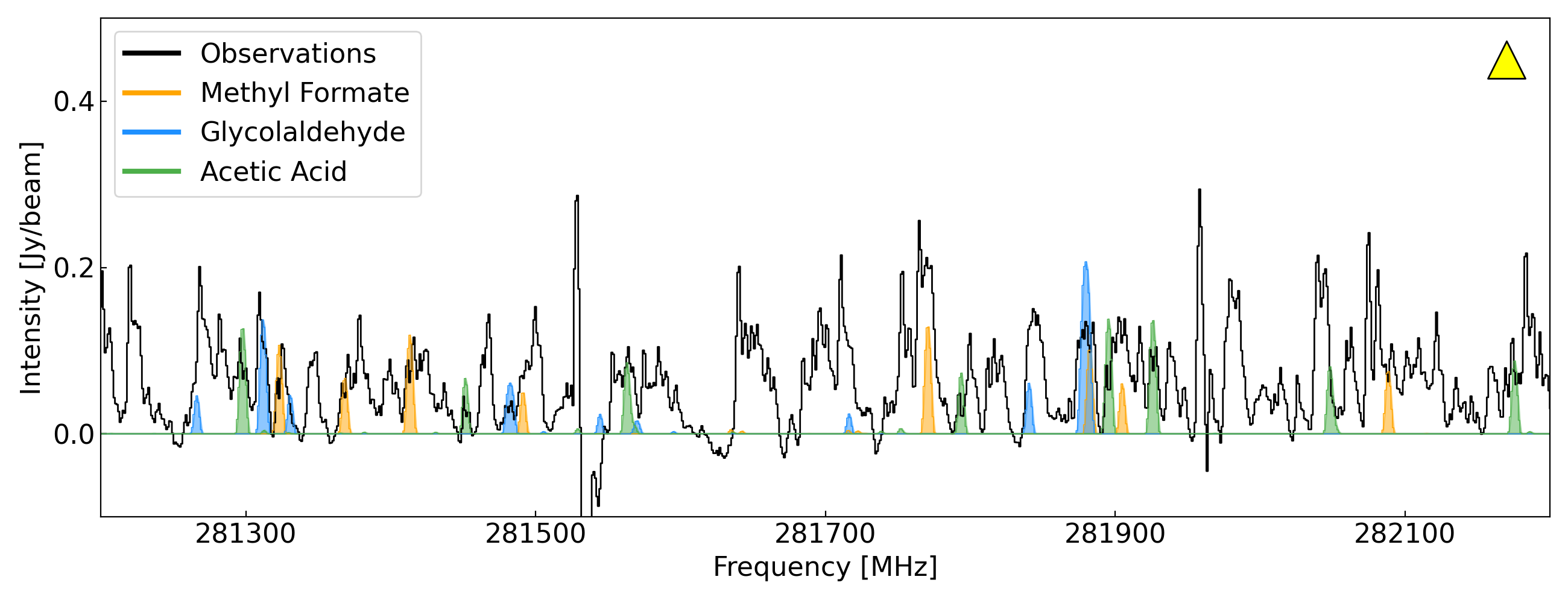}
    \caption{Spectra of MM1 extracted from the positions marked in Figure~\ref{mm1_finder}.}
    \label{mm1spectra}
\end{figure*}

\clearpage

\begin{figure*}
    \centering
    \includegraphics[width=0.85\textwidth]{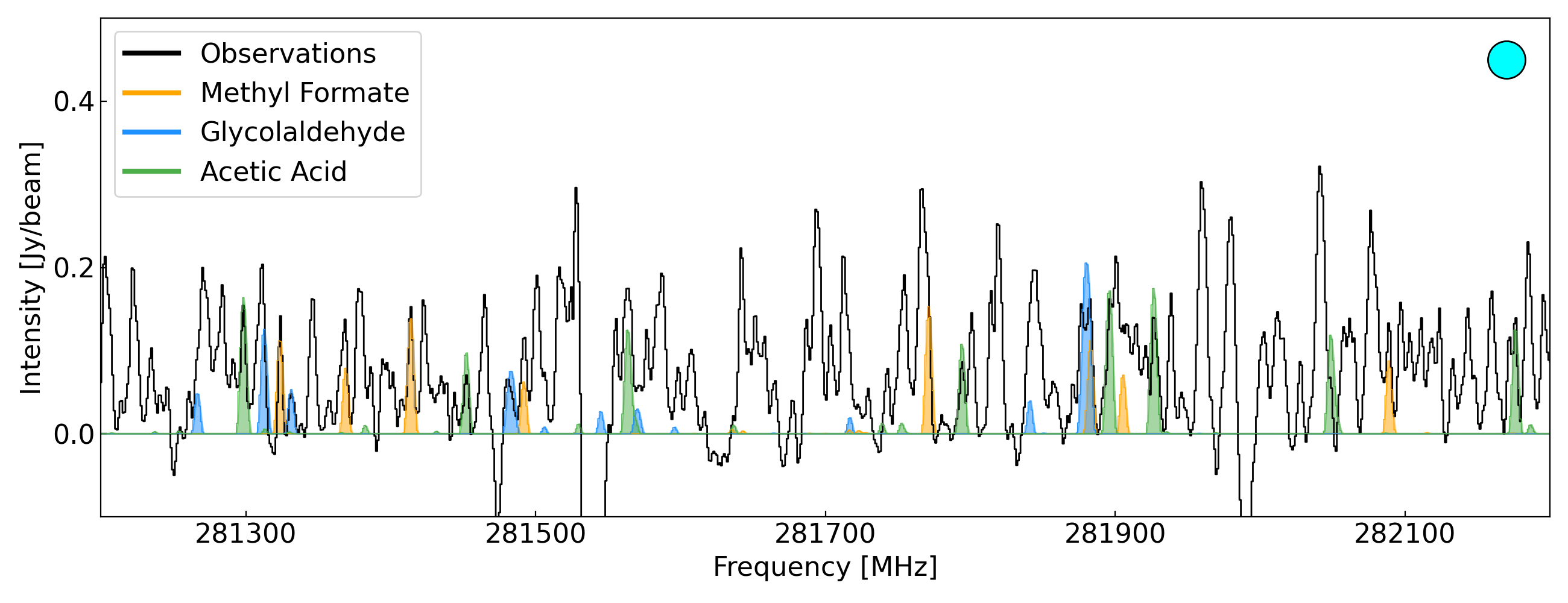}
    \includegraphics[width=0.85\textwidth]{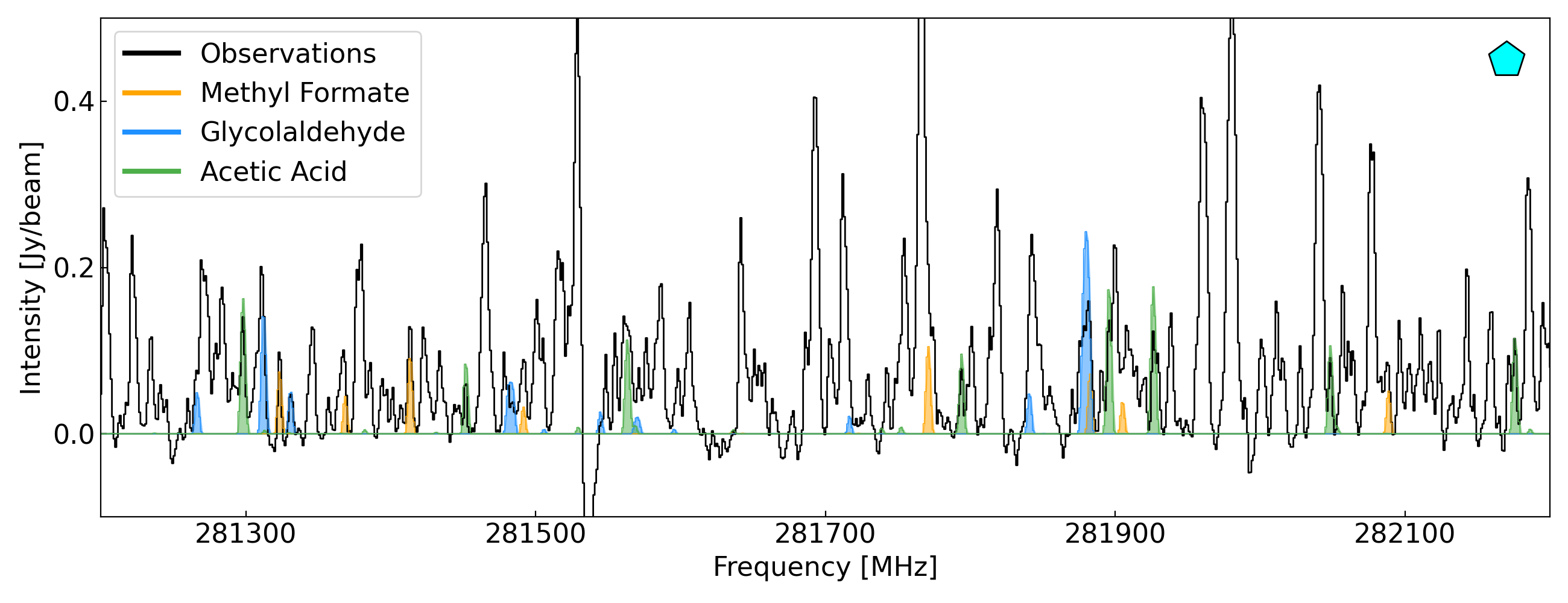}
    \includegraphics[width=0.85\textwidth]{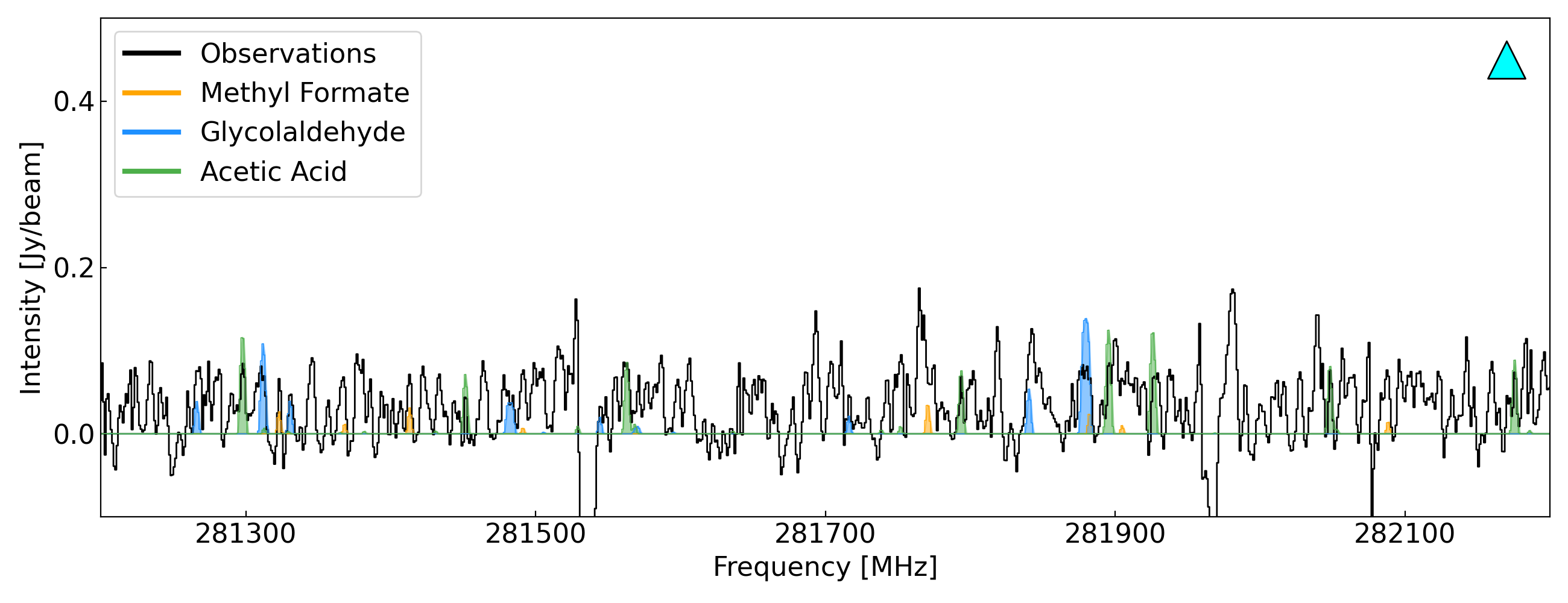}
    \includegraphics[width=0.85\textwidth]{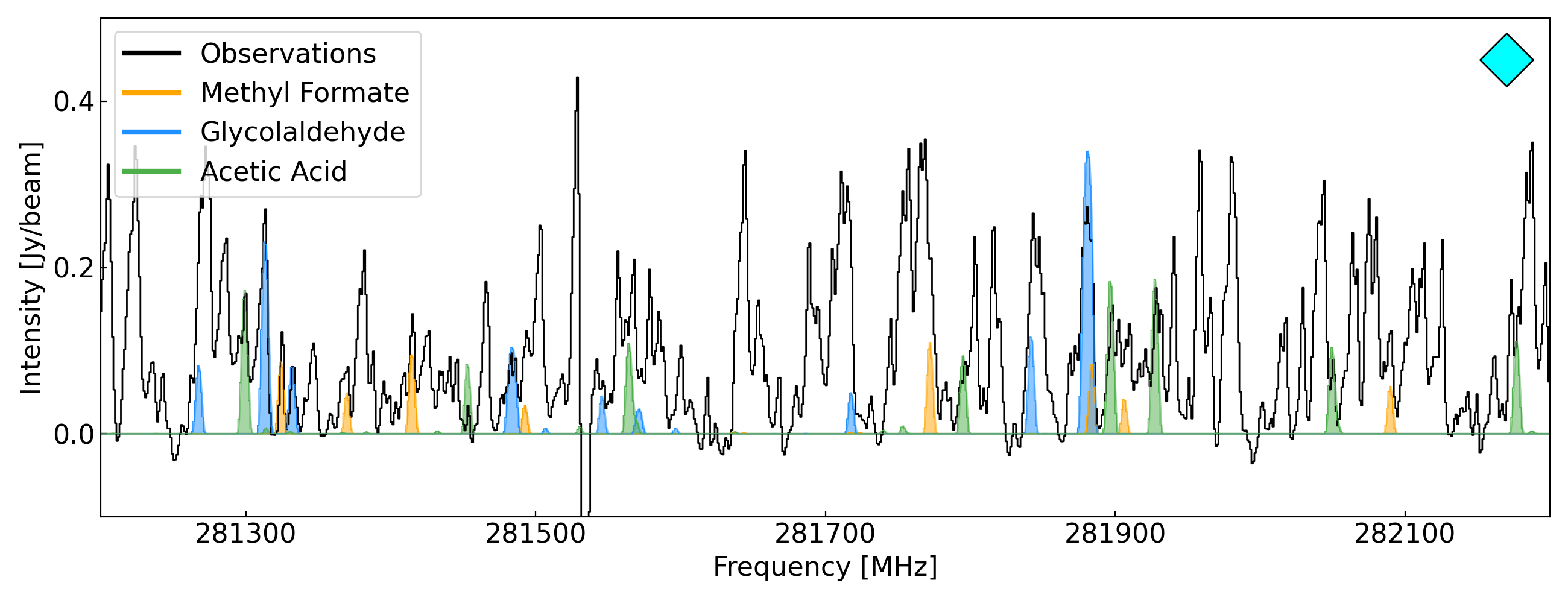}
    \caption{Spectra of MM1 extracted from the positions marked in Figure~\ref{mm1_finder}.}
    \label{mm1spectra2}
\end{figure*}

\clearpage

\begin{figure*}
    \centering
    \includegraphics[width=0.85\textwidth]{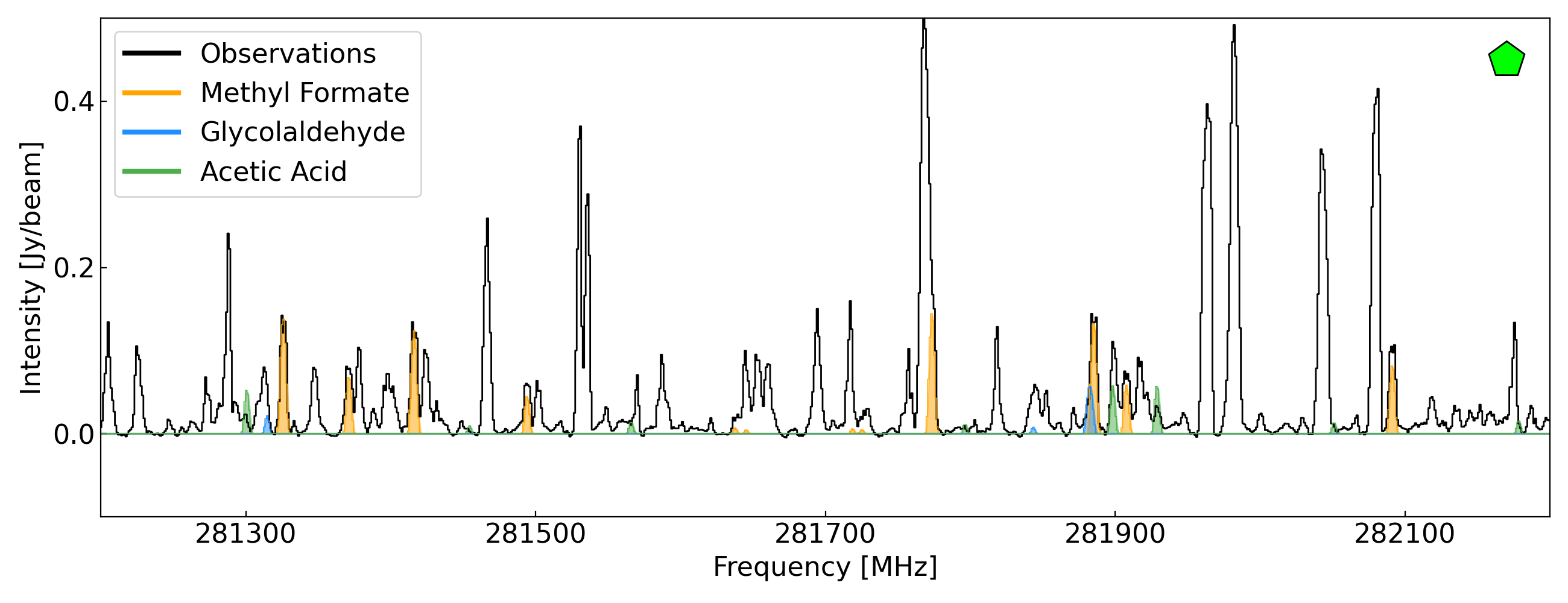}
    \includegraphics[width=0.85\textwidth]{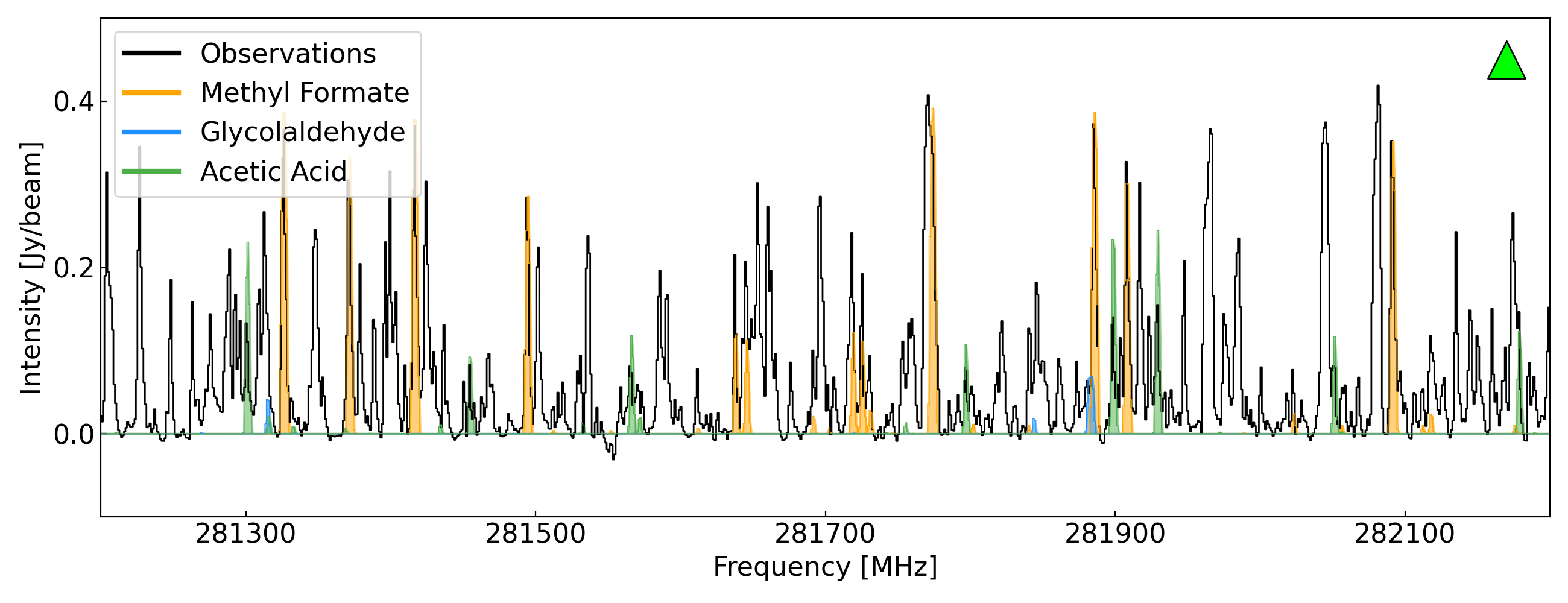}
    \includegraphics[width=0.85\textwidth]{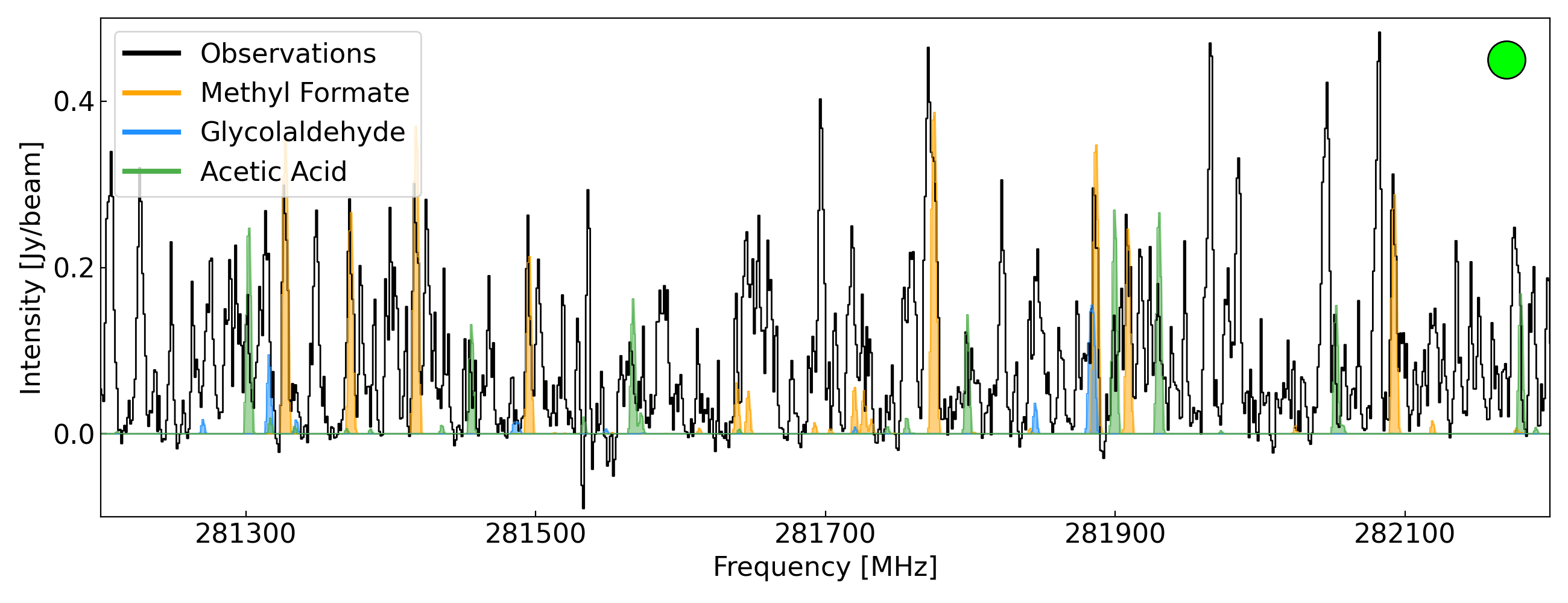}
    \includegraphics[width=0.85\textwidth]{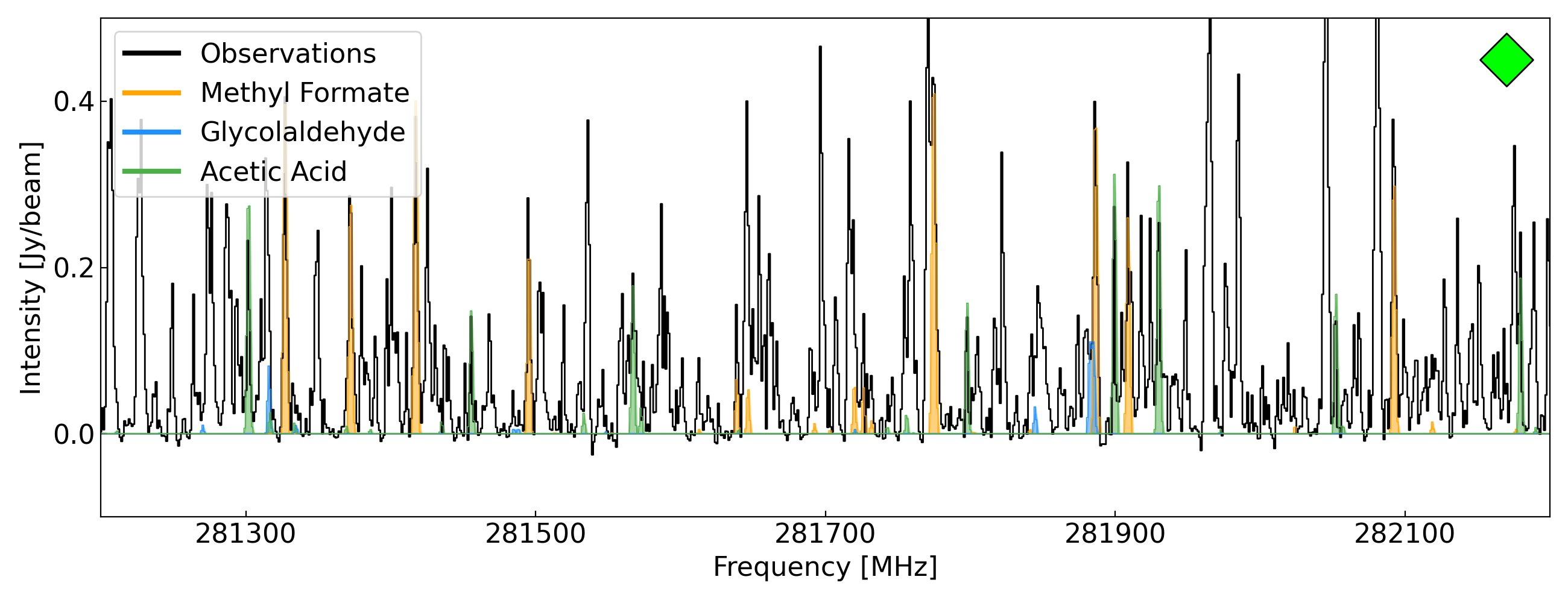}
    \caption{Spectra of MM2 extracted from the positions marked in Figure~\ref{mm2_finder}.}
    \label{mm2spectra}
\end{figure*}

\clearpage

\section{Appendix B}
\label{appendix_b}

\begin{figure}[h!]
    \centering
    \includegraphics[width=0.5\textwidth]{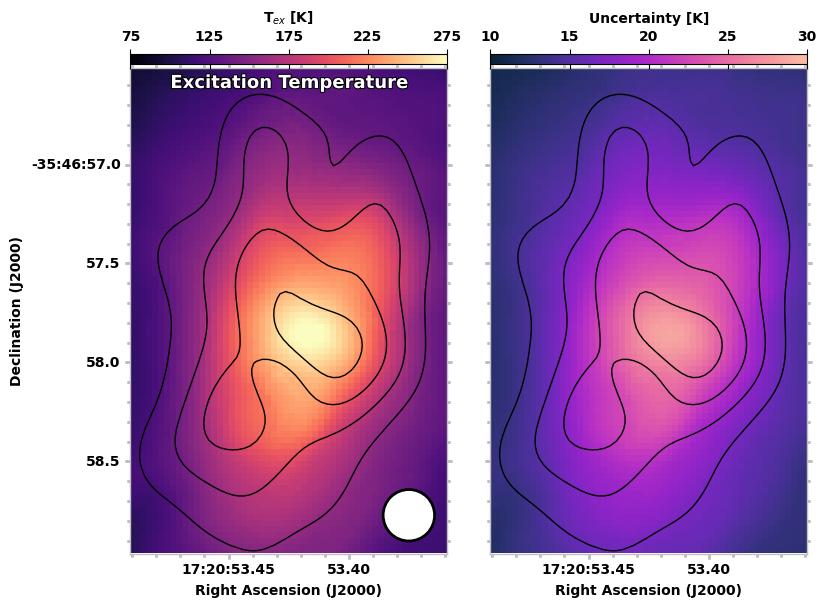}
    \caption{Excitation temperature (left) and excitation temperature uncertainty (right) images produced by the automated fitting routine for NGC 6334I-MM1.}
    \label{fig:b00}
\end{figure}

\begin{figure}[h!]
    \centering
    \includegraphics[width=0.5\textwidth]{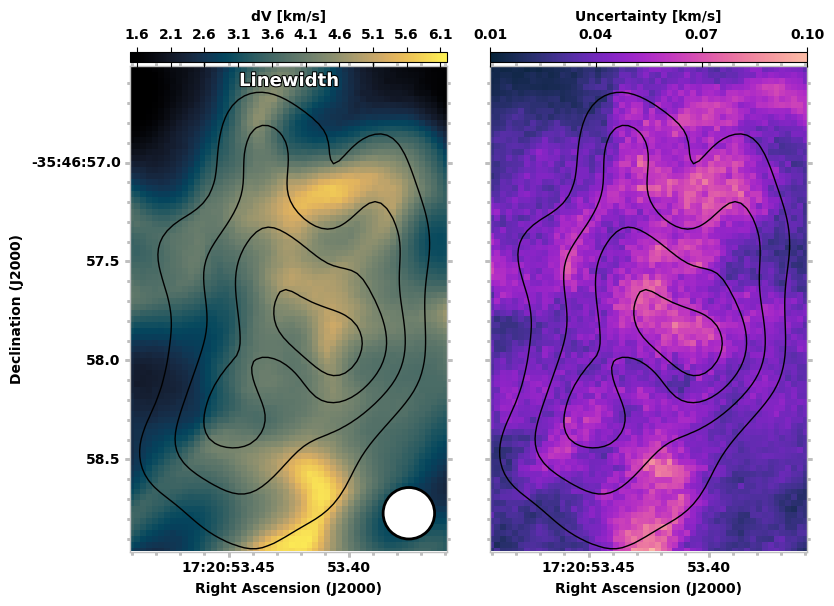}
    \caption{Linewidth (left) and linewidth uncertainty (right) images produced by the automated fitting routine for NGC 6334I-MM1.}
    \label{fig:b01}
\end{figure}

\begin{figure}[h!]
    \centering
    \includegraphics[width=0.5\textwidth]{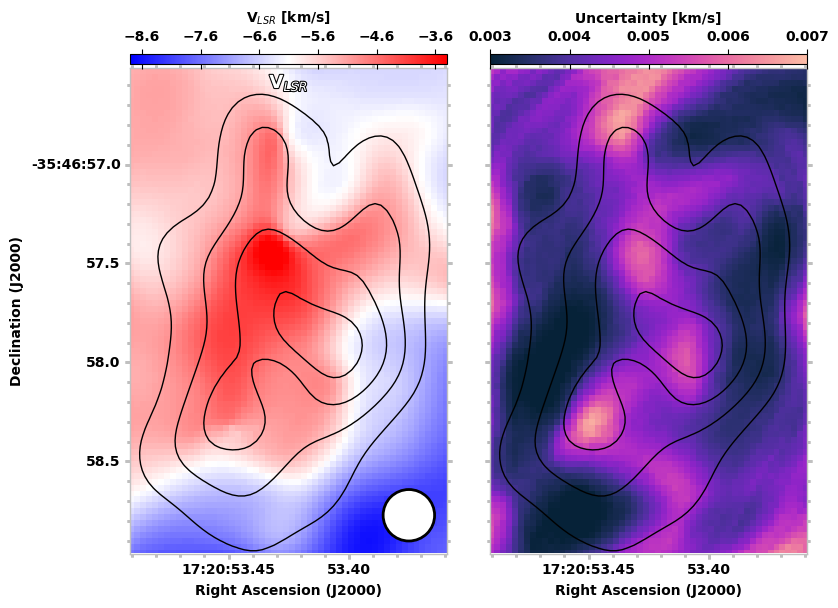}
    \caption{Velocity (left) and velocity uncertainty (right) images produced by the automated fitting routine for NGC 6334I-MM1.}
    \label{fig:b02}
\end{figure}

\begin{figure}[h!]
    \centering
    \includegraphics[width=0.5\textwidth]{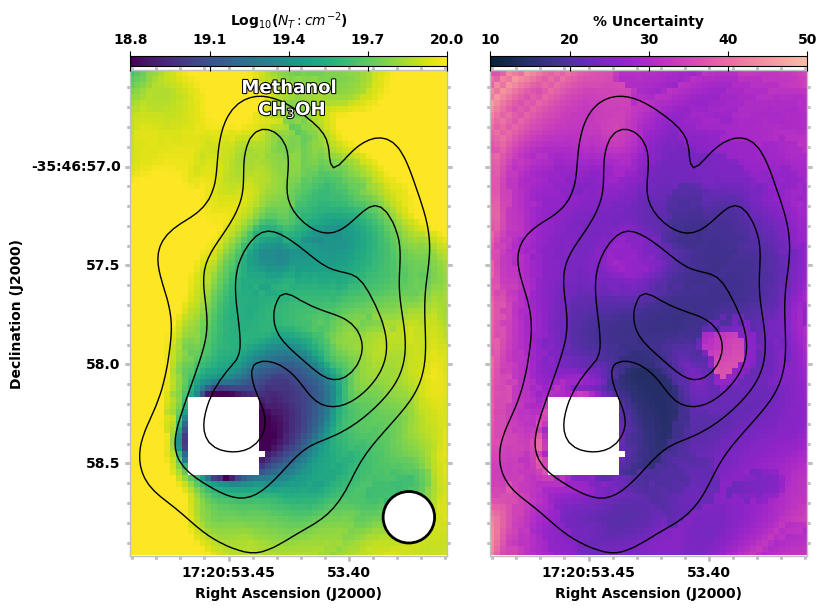}
    \caption{Column density (left) and column density uncertainty (right) images produced by the automated fitting routine for methanol in NGC 6334I-MM1.}
    \label{fig:b1}
\end{figure}

\begin{figure}[h!]
    \centering
    \includegraphics[width=0.5\textwidth]{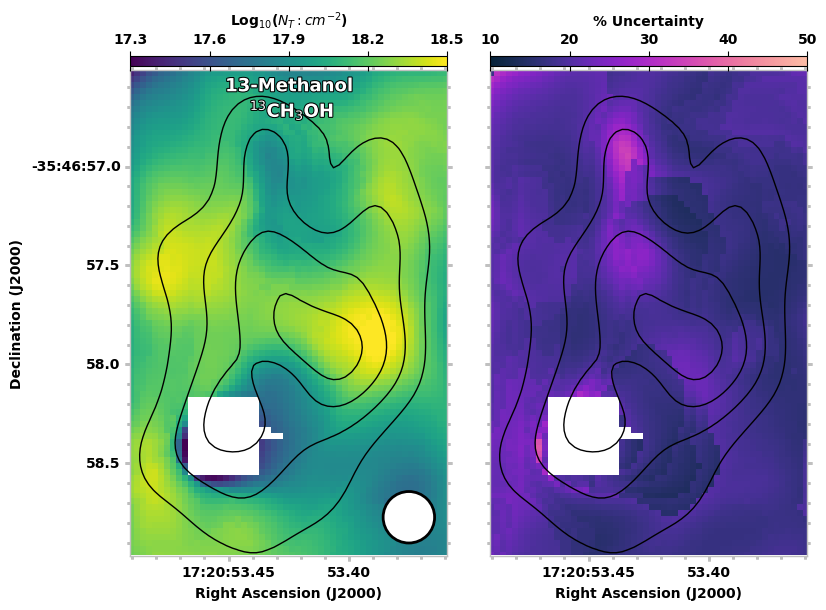}
    \caption{Same as Figure \ref{fig:b1} but for 13-methanol.}
    \label{b2}
\end{figure}

\begin{figure}[h!]
    \centering
    \includegraphics[width=0.5\textwidth]{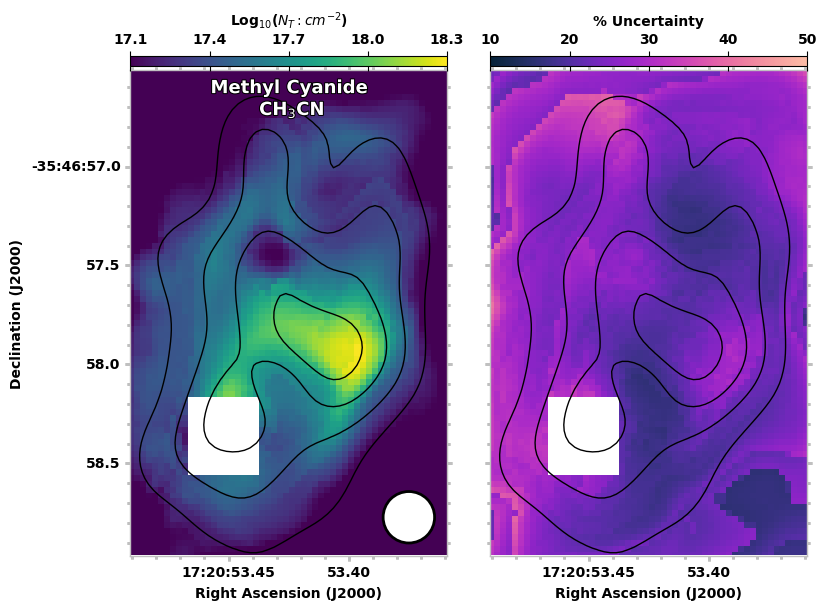}
    \caption{Same as Figure \ref{fig:b1} but for methyl cyanide. See Section \ref{channel_exclusion} for a description of some effects which may be impacting this image.}
    \label{b3}
\end{figure}

\begin{figure}[h!]
    \centering
    \includegraphics[width=0.5\textwidth]{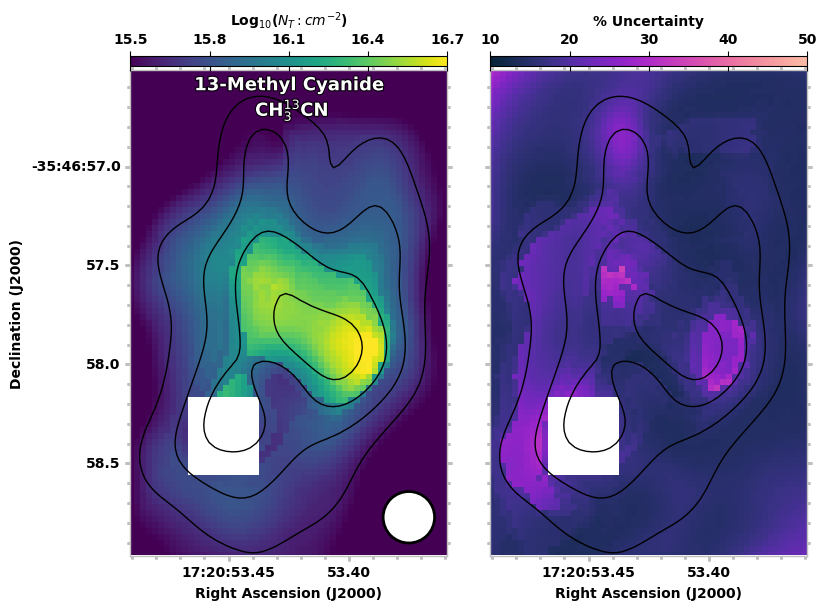}
    \caption{Same as Figure \ref{fig:b1} but for 13-methyl cyanide.}
    \label{b4}
\end{figure}

\begin{figure}[h!]
    \centering
    \includegraphics[width=0.5\textwidth]{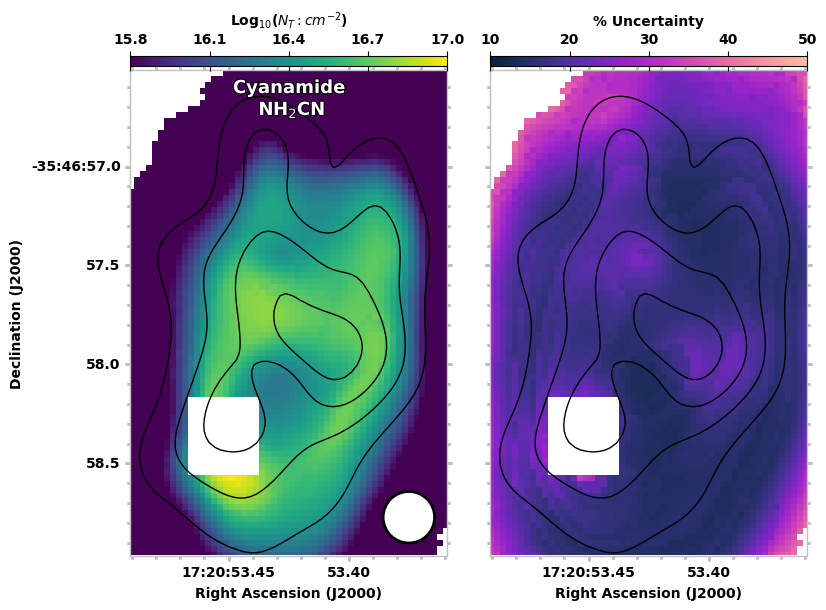}
    \caption{Same as Figure \ref{fig:b1} but for cyanamide. }
    \label{b5}
\end{figure}

\begin{figure}[h!]
    \centering
    \includegraphics[width=0.5\textwidth]{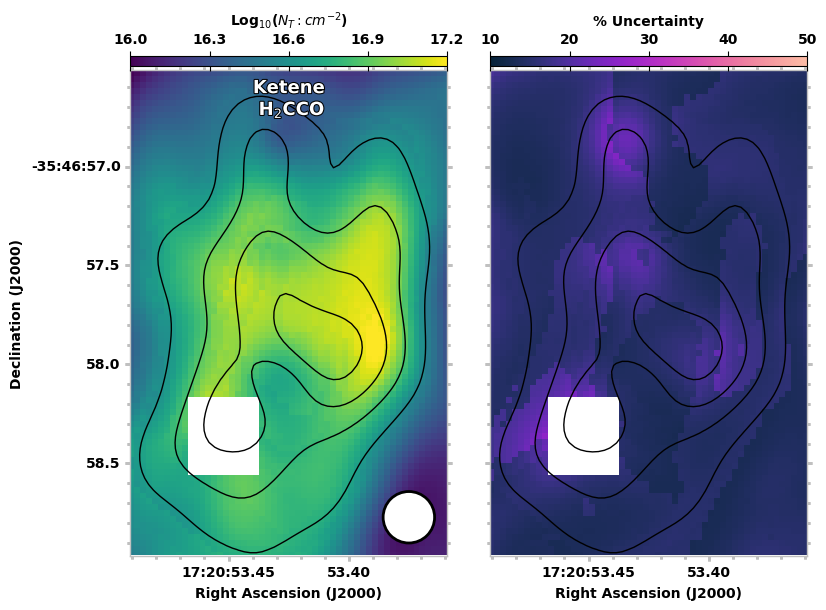}
    \caption{Same as Figure \ref{fig:b1} but for ketene.}
    \label{b6}
\end{figure}

\begin{figure}[h!]
    \centering
    \includegraphics[width=0.5\textwidth]{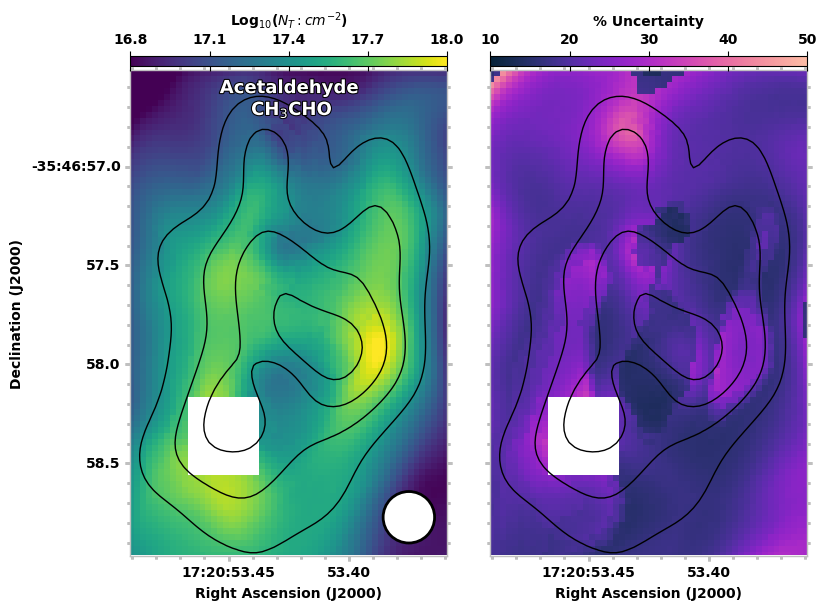}
    \caption{Same as Figure \ref{fig:b1} but for acetaldehyde.}
    \label{b7}
\end{figure}

\begin{figure}[h!]
    \centering
    \includegraphics[width=0.5\textwidth]{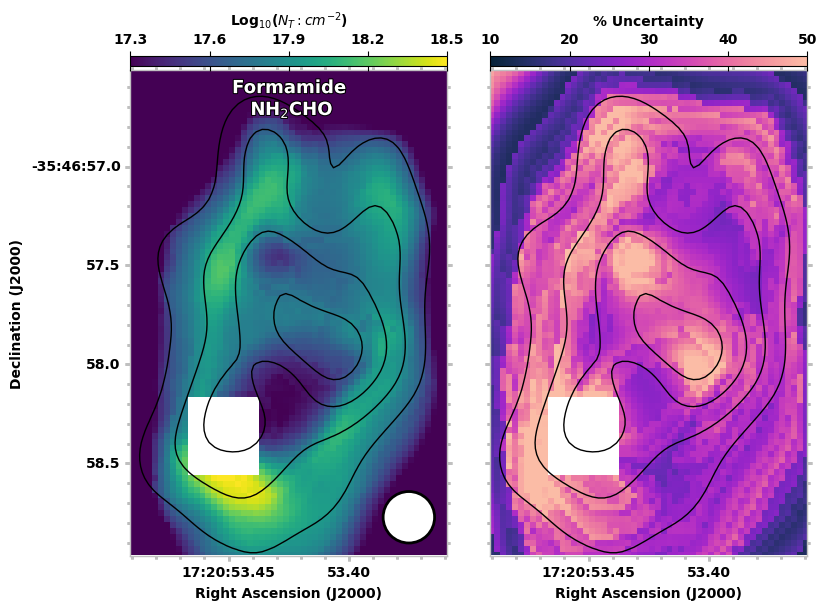}
    \caption{Same as Figure \ref{fig:b1} but for formamide. See Section \ref{channel_exclusion} for a description of some effects which may be impacting this image.}
    \label{b8}
\end{figure}

\begin{figure}[h!]
    \centering
    \includegraphics[width=0.5\textwidth]{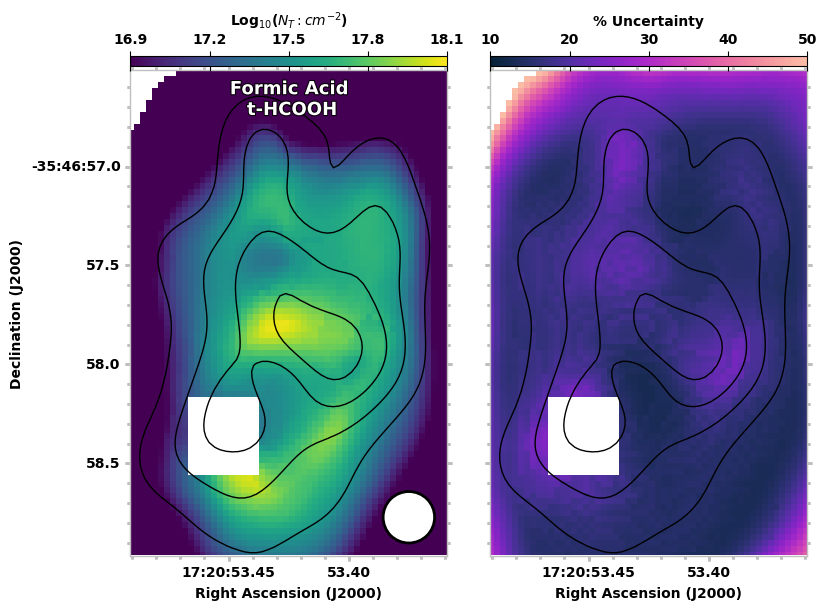}
    \caption{Same as Figure \ref{fig:b1} but for the trans conformer formic acid.}
    \label{b9}
\end{figure}

\begin{figure}[h!]
    \centering
    \includegraphics[width=0.5\textwidth]{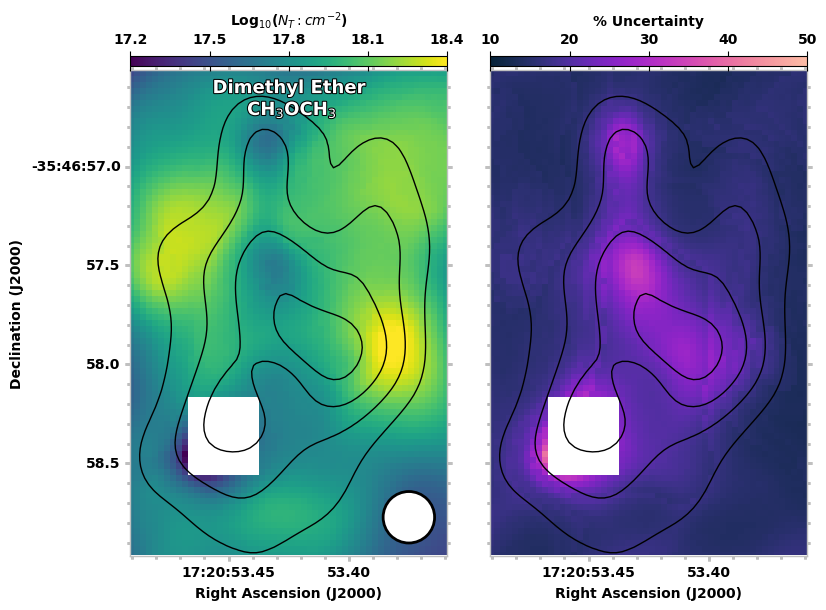}
    \caption{Same as Figure \ref{fig:b1} but for dimethyl ether.}
    \label{b10}
\end{figure}

\begin{figure}[h!]
    \centering
    \includegraphics[width=0.5\textwidth]{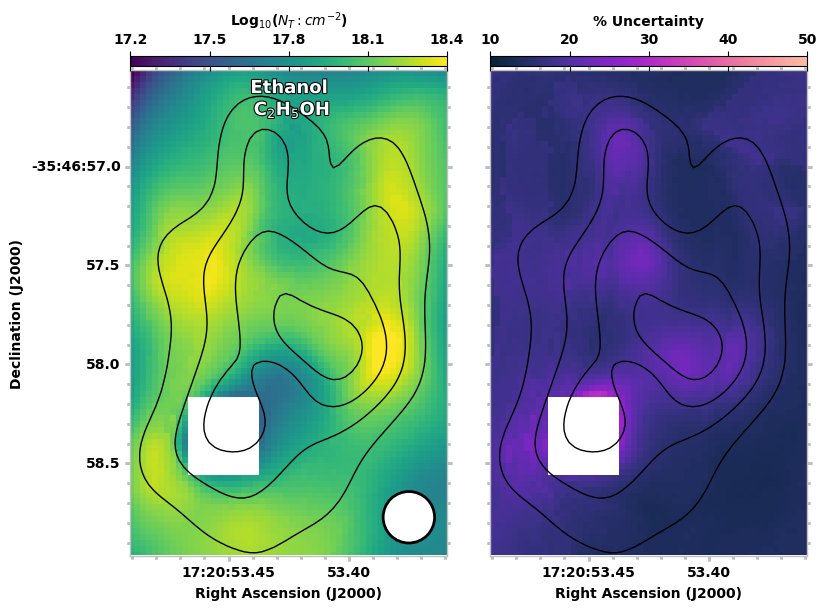}
    \caption{Same as Figure \ref{fig:b1} but for ethanol.}
    \label{b11}
\end{figure}

\begin{figure}[h!]
    \centering
    \includegraphics[width=0.5\textwidth]{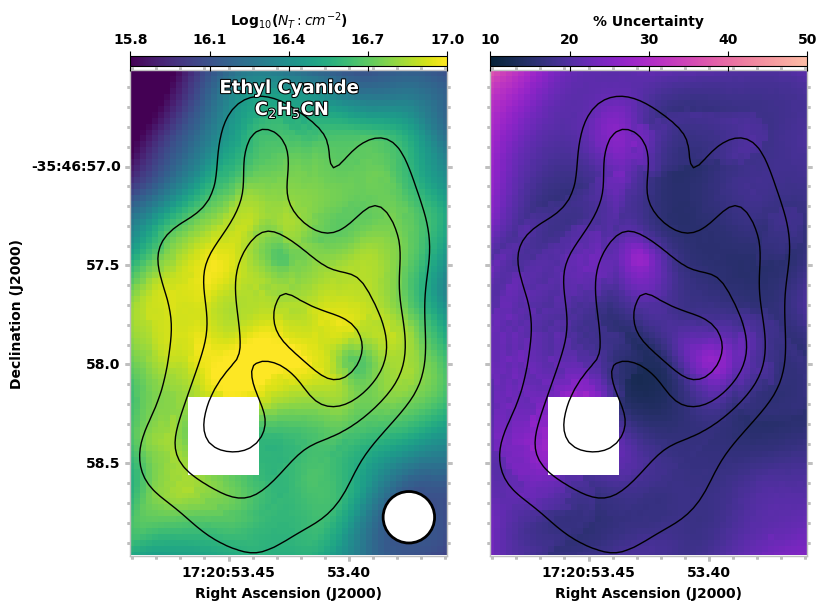}
    \caption{Same as Figure \ref{fig:b1} but for ethyl cyanide.}
    \label{b12}
\end{figure}

\begin{figure}[h!]
    \centering
    \includegraphics[width=0.5\textwidth]{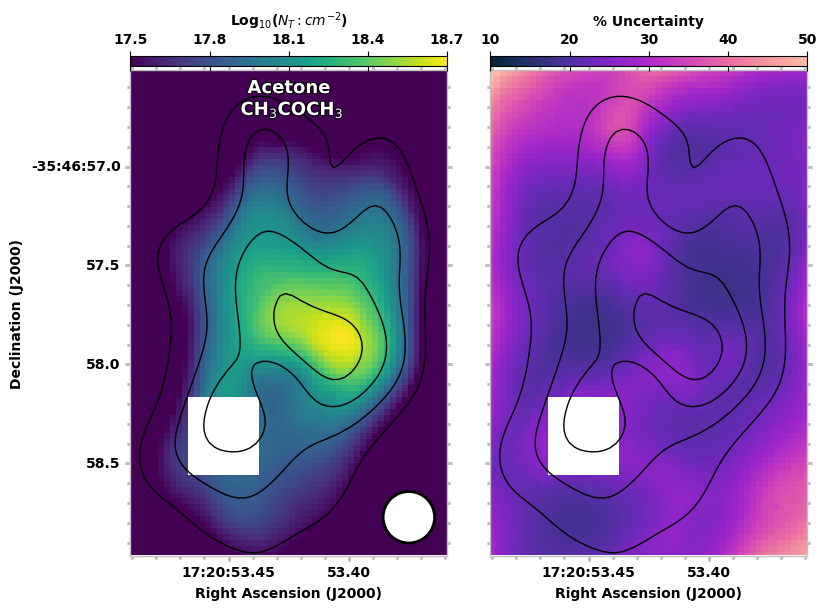}
    \caption{Same as Figure \ref{fig:b1} but for acetone.}
    \label{b13}
\end{figure}

\begin{figure}[h!]
    \centering
    \includegraphics[width=0.5\textwidth]{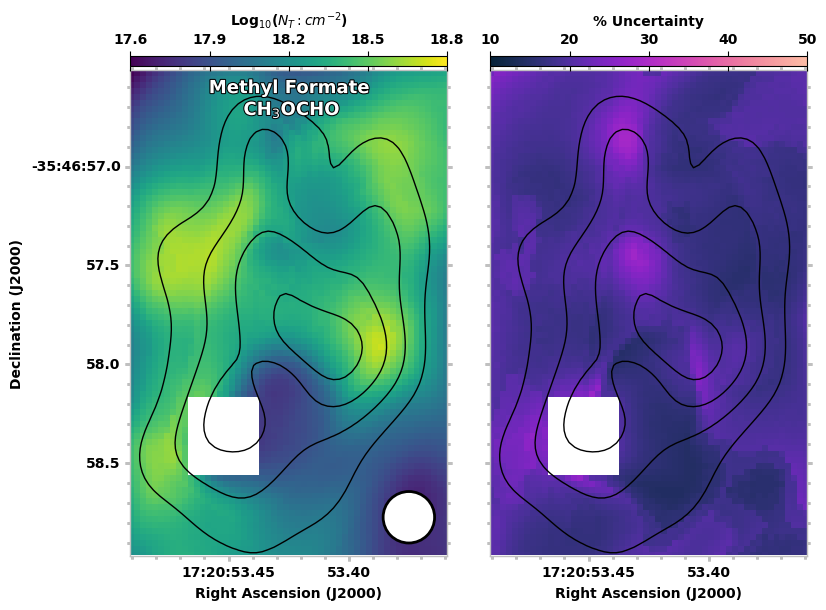}
    \caption{Same as Figure \ref{fig:b1} but for methyl formate.}
    \label{b14}
\end{figure}

\begin{figure}[h!]
    \centering
    \includegraphics[width=0.5\textwidth]{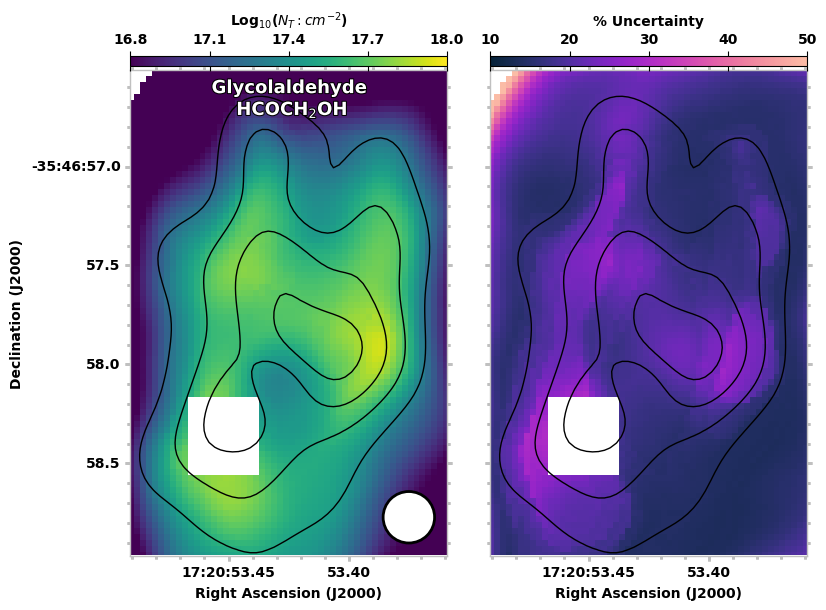}
    \caption{Same as Figure \ref{fig:b1} but for glycolaldehyde.}
    \label{b15}
\end{figure}

\begin{figure}[h!]
    \centering
    \includegraphics[width=0.5\textwidth]{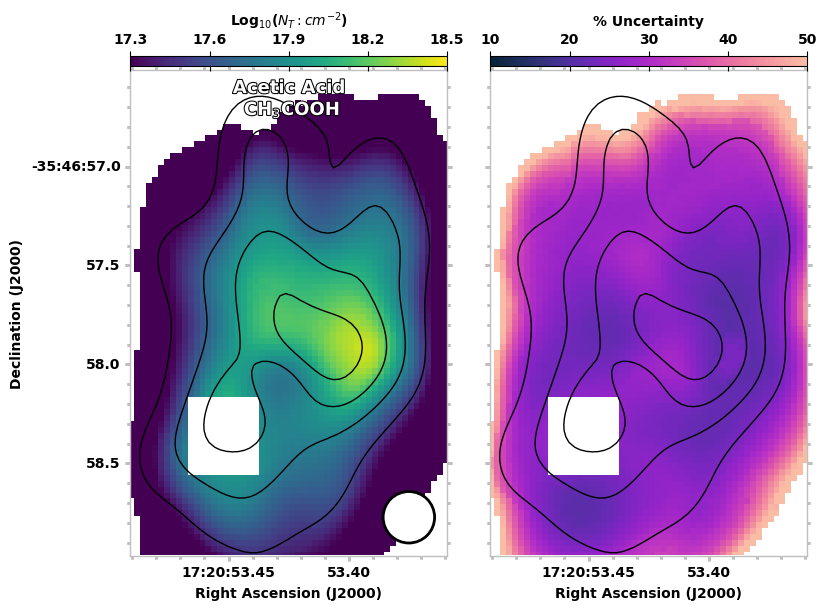}
    \caption{Same as Figure \ref{fig:b1} but for acetic acid.}
    \label{b16}
\end{figure}

\begin{figure}[h!]
    \centering
    \includegraphics[width=0.5\textwidth]{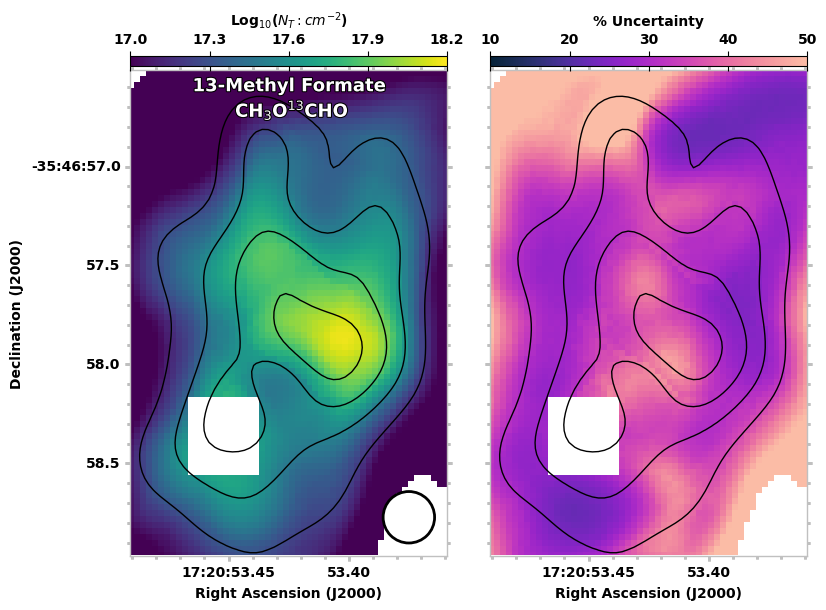}
    \caption{Same as Figure \ref{fig:b1} but for 13-methyl formate.}
    \label{b17}
\end{figure}

\begin{figure}[h!]
    \centering
    \includegraphics[width=0.5\textwidth]{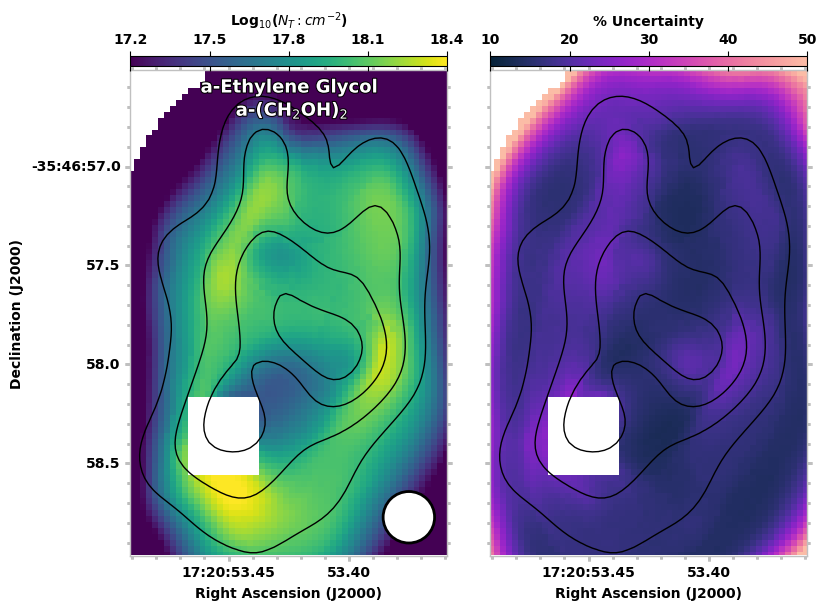}
    \caption{Same as Figure \ref{fig:b1} but for a-ethylene glycol.}
    \label{b18}
\end{figure}

\begin{figure}[h!]
    \centering
    \includegraphics[width=0.5\textwidth]{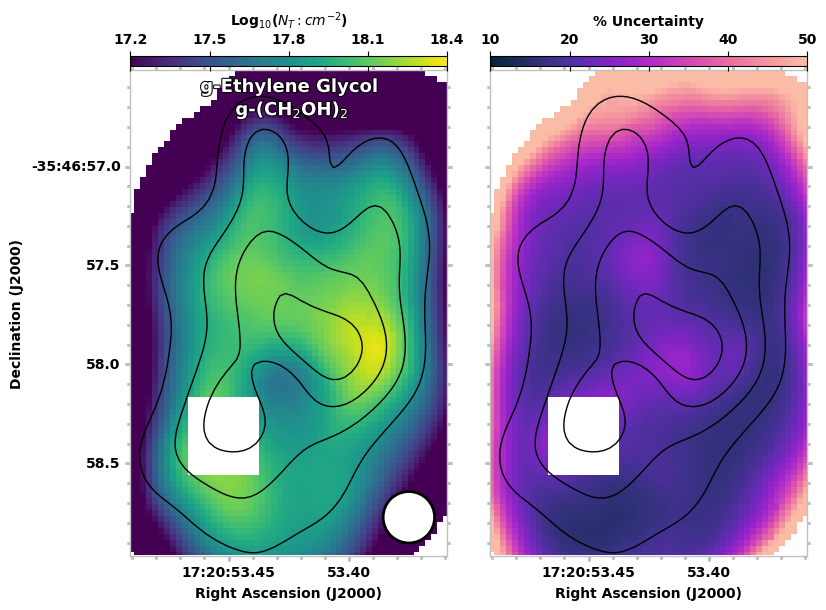}
    \caption{Same as Figure \ref{fig:b1} but for g-ethylene glycol.}
    \label{b19}
\end{figure}

\begin{figure}[h!]
    \centering
    \includegraphics[width=0.5\textwidth]{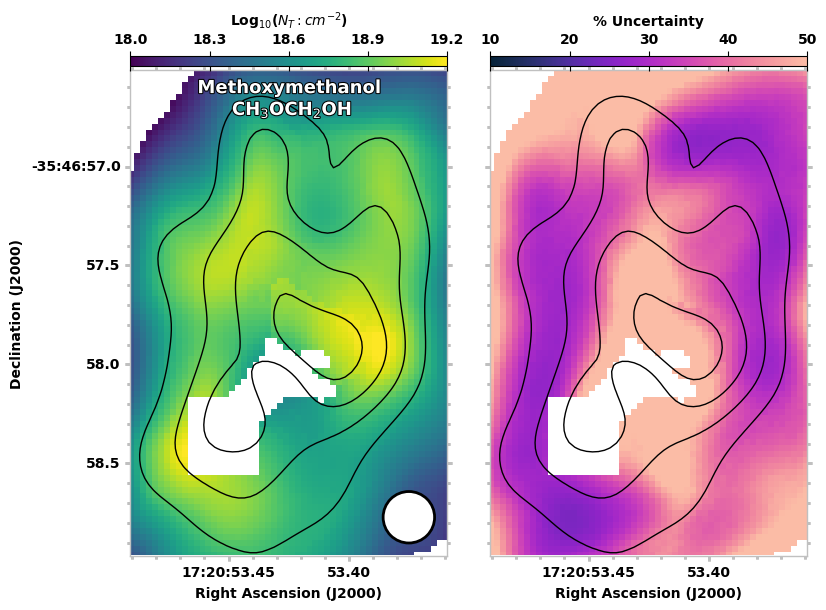}
    \caption{Same as Figure \ref{fig:b1} but for methoxymethanol.}
    \label{b20}
\end{figure}

\begin{figure}[h!]
    \centering
    \includegraphics[width=0.5\textwidth]{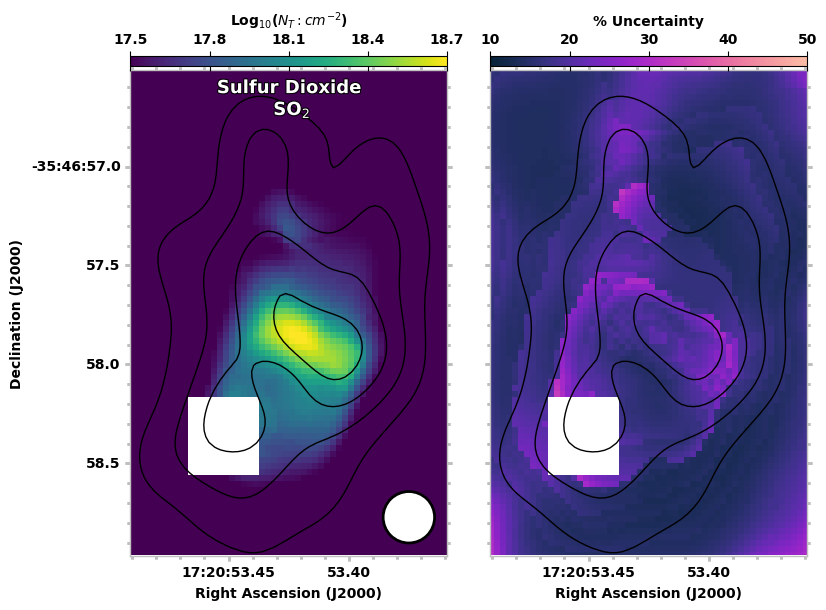}
    \caption{Same as Figure \ref{fig:b1} but for sulfur dioxide.}
    \label{b21}
\end{figure}

\begin{figure}[h!]
    \centering
    \includegraphics[width=0.5\textwidth]{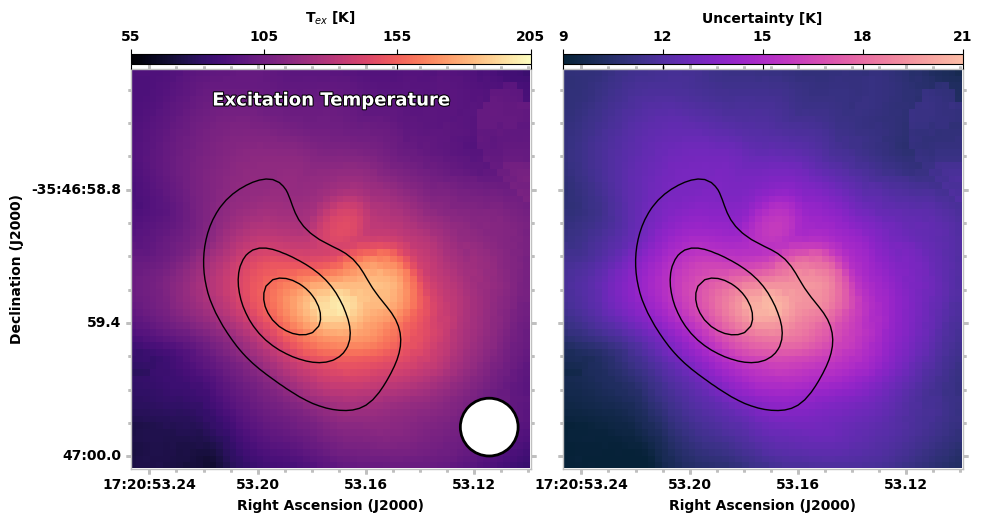}
    \caption{Excitation temperature (left) and excitation temperature uncertainty (right) images produced by the automated fitting routine for NGC 6334I-MM2.}
    \label{fig:b03}
\end{figure}

\begin{figure}[h!]
    \centering
    \includegraphics[width=0.5\textwidth]{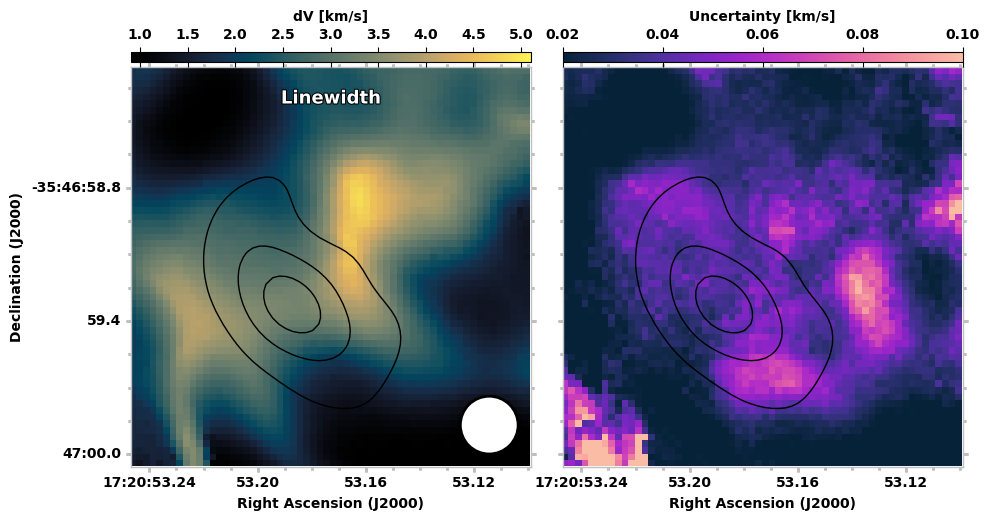}
    \caption{Linewidth (left) and linewidth uncertainty (right) images produced by the automated fitting routine for NGC 6334I-MM2.}
    \label{fig:b04}
\end{figure}

\begin{figure}[h!]
    \centering
    \includegraphics[width=0.5\textwidth]{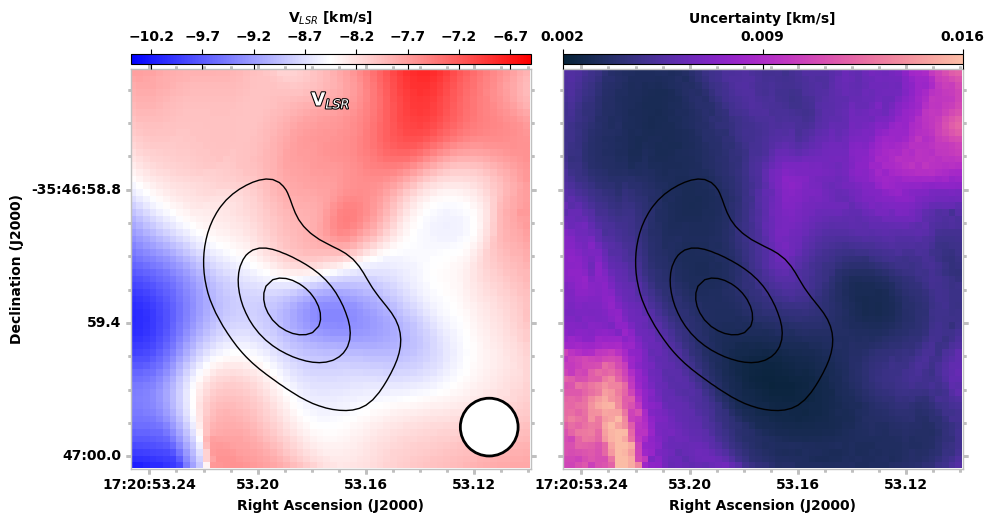}
    \caption{Velocity (left) and velocity uncertainty (right) images produced by the automated fitting routine for NGC 6334I-MM2.}
    \label{fig:b05}
\end{figure}

\begin{figure}[h!]
    \centering
    \includegraphics[width=0.5\textwidth]{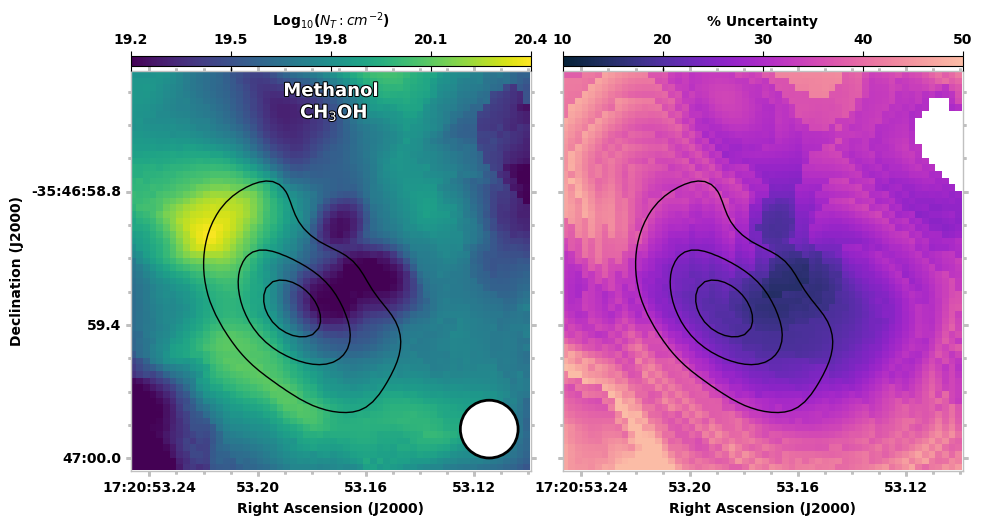}
    \caption{Column density (left) and column density uncertainty (right) images produced by the automated fitting routine for methanol in NGC 6334I-MM2.}
    \label{b22}
\end{figure}

\begin{figure}[h!]
    \centering
    \includegraphics[width=0.5\textwidth]{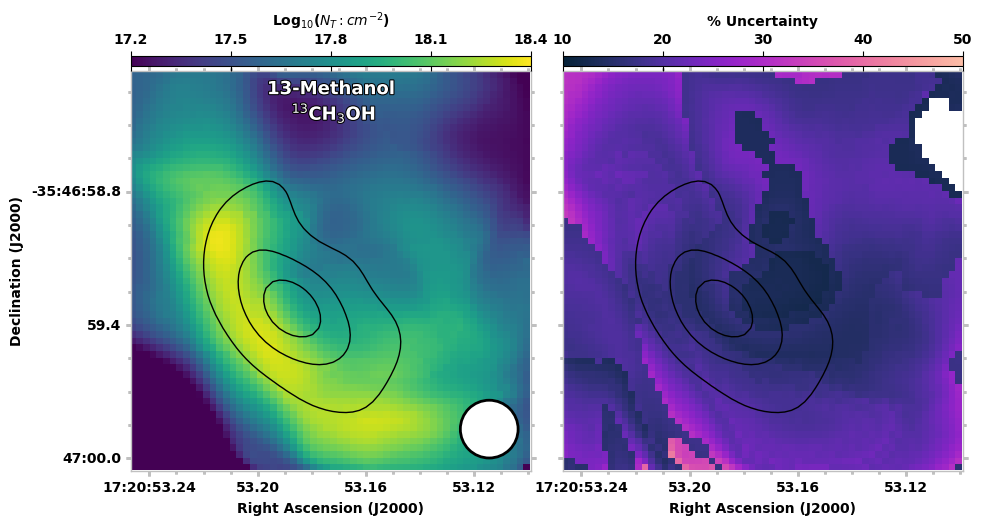}
    \caption{Same as Figure \ref{b22} but for 13-methanol.}
    \label{b23}
\end{figure}

\begin{figure}[h!]
    \centering
    \includegraphics[width=0.5\textwidth]{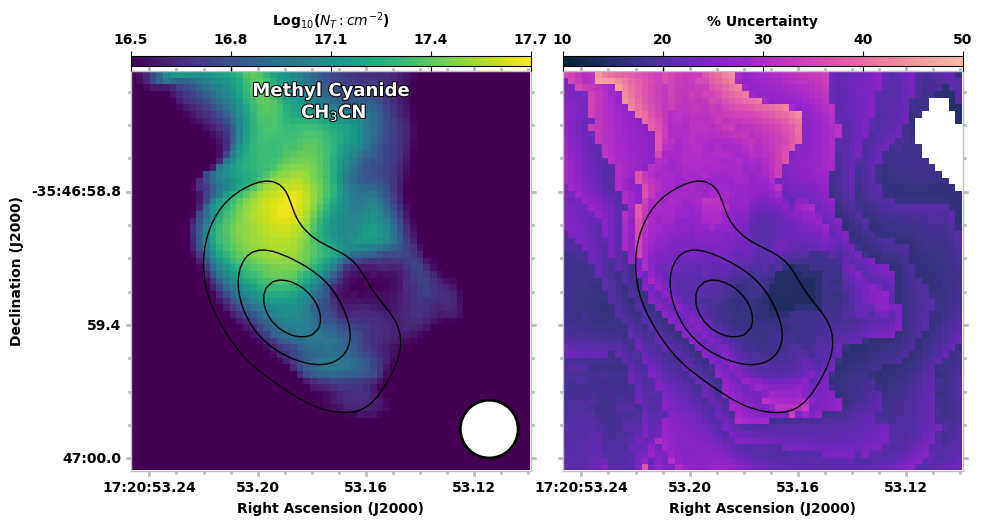}
    \caption{Same as Figure \ref{b22} but for methyl cyanide.}
    \label{b24}
\end{figure}

\begin{figure}[h!]
    \centering
    \includegraphics[width=0.5\textwidth]{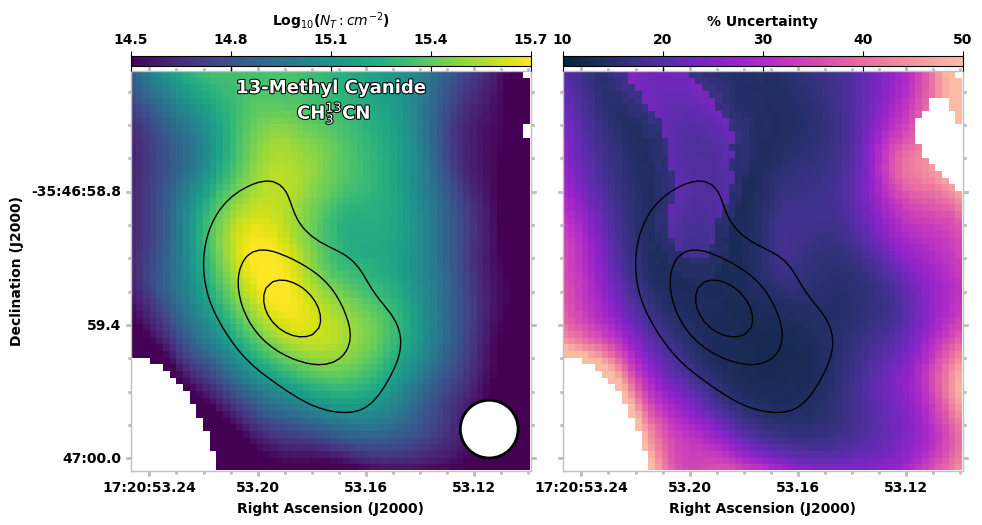}
    \caption{Same as Figure \ref{b22} but for 13-methyl cyanide.}
    \label{b25}
\end{figure}

\begin{figure}[h!]
    \centering
    \includegraphics[width=0.5\textwidth]{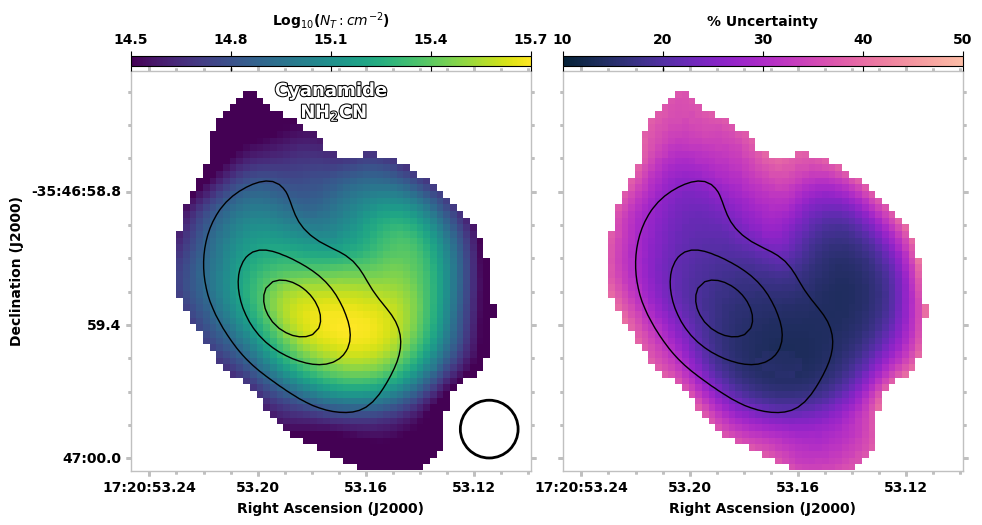}
    \caption{Same as Figure \ref{b22} but for cyanamide.}
    \label{b26}
\end{figure}

\begin{figure}[h!]
    \centering
    \includegraphics[width=0.5\textwidth]{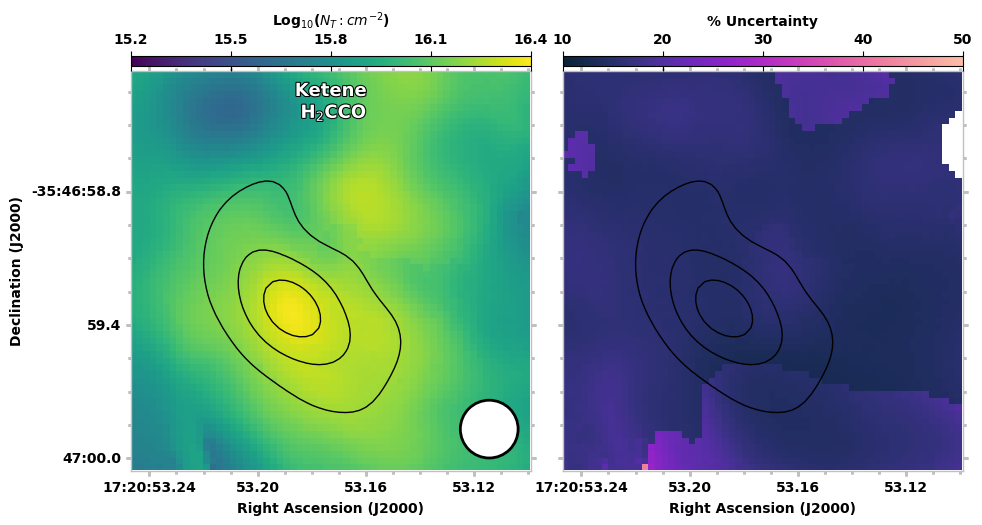}
    \caption{Same as Figure \ref{b22} but for ketene.}
    \label{b27}
\end{figure}

\begin{figure}[h!]
    \centering
    \includegraphics[width=0.5\textwidth]{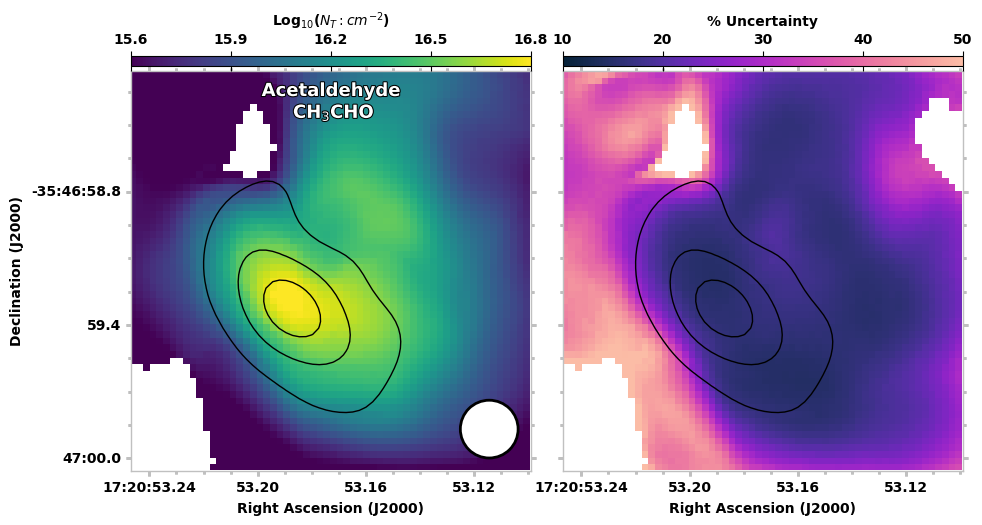}
    \caption{Same as Figure \ref{b22} but for acetaldehyde.}
    \label{b28}
\end{figure}

\begin{figure}[h!]
    \centering
    \includegraphics[width=0.5\textwidth]{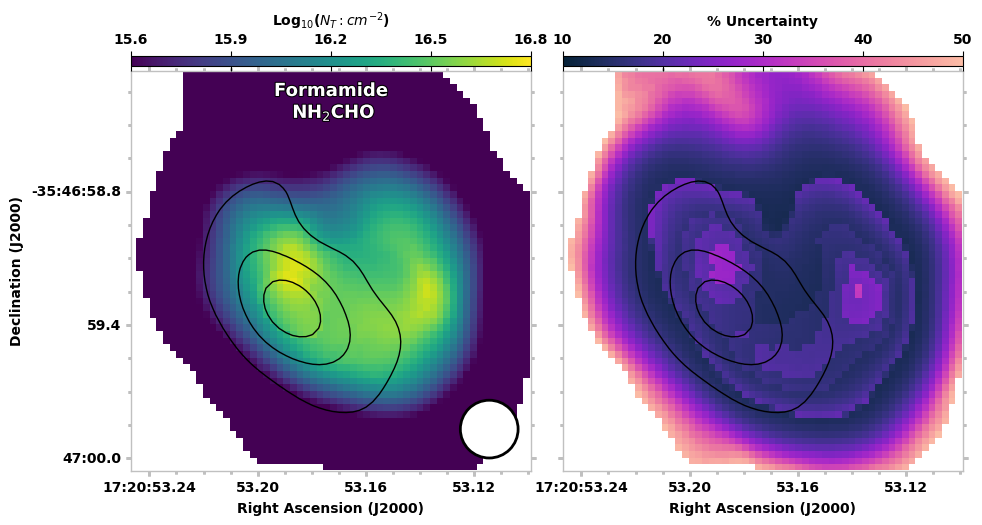}
    \caption{Same as Figure \ref{b22} but for formamide.}
    \label{b29}
\end{figure}

\begin{figure}[h!]
    \centering
    \includegraphics[width=0.5\textwidth]{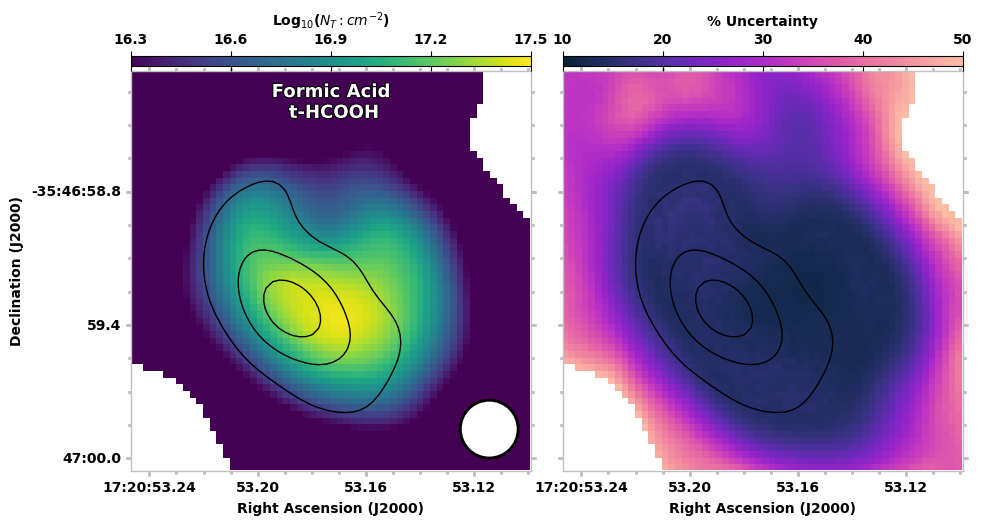}
    \caption{Same as Figure \ref{b22} but for the trans conformer of formic acid.}
    \label{b30}
\end{figure}

\begin{figure}[h!]
    \centering
    \includegraphics[width=0.5\textwidth]{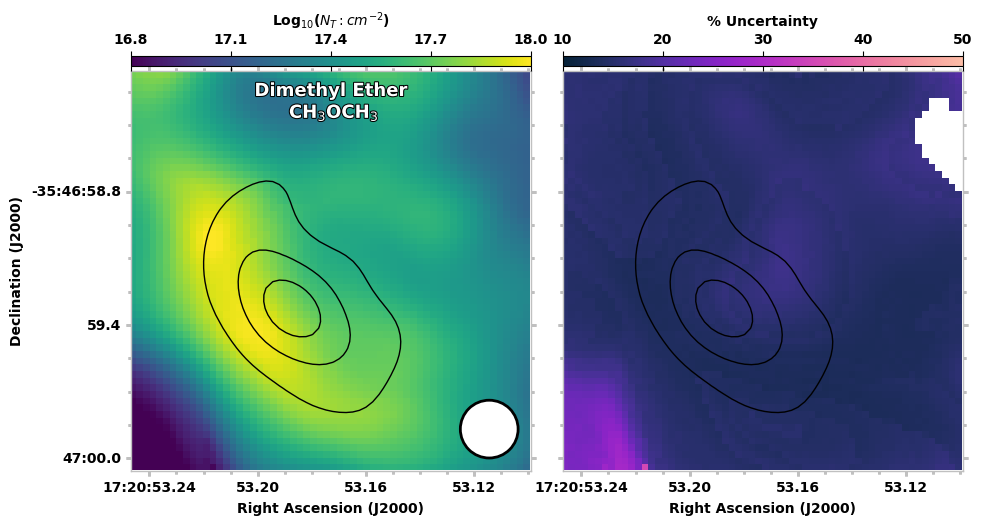}
    \caption{Same as Figure \ref{b22} but for dimethyl ether.}
    \label{b31}
\end{figure}

\begin{figure}[h!]
    \centering
    \includegraphics[width=0.5\textwidth]{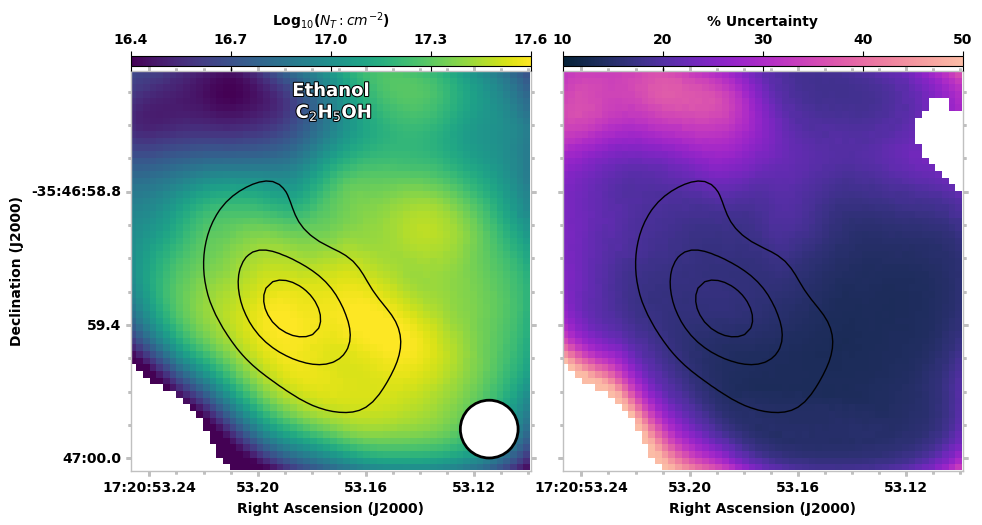}
    \caption{Same as Figure \ref{b22} but for ethanol.}
    \label{b32}
\end{figure}

\begin{figure}[h!]
    \centering
    \includegraphics[width=0.5\textwidth]{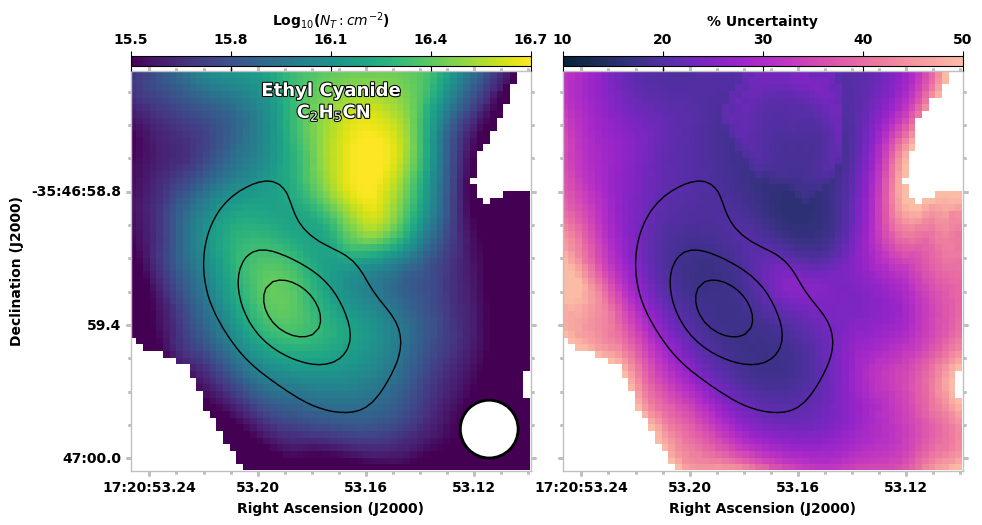}
    \caption{Same as Figure \ref{b22} but for ethyl cyanide.}
    \label{b33}
\end{figure}

\begin{figure}[h!]
    \centering
    \includegraphics[width=0.5\textwidth]{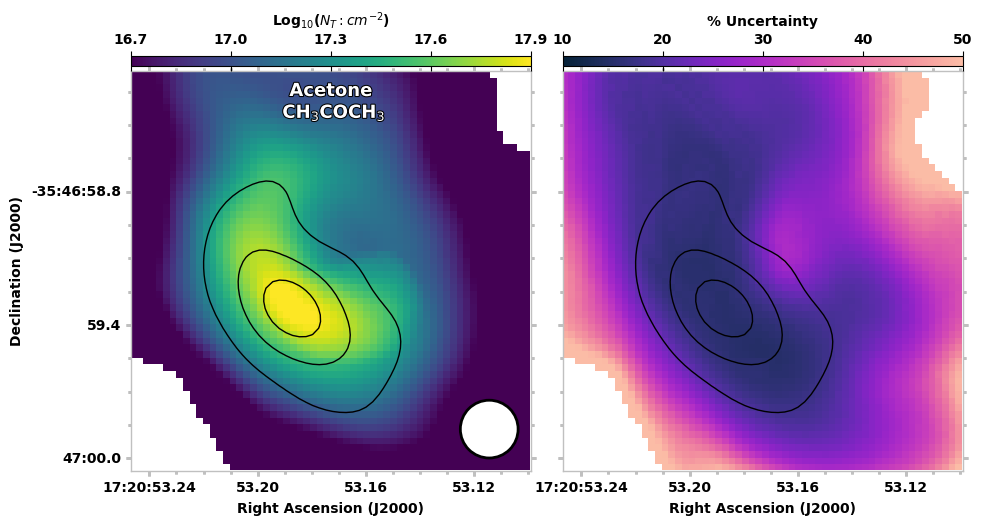}
    \caption{Same as Figure \ref{b22} but for acetone.}
    \label{b34}
\end{figure}

\begin{figure}[h!]
    \centering
    \includegraphics[width=0.5\textwidth]{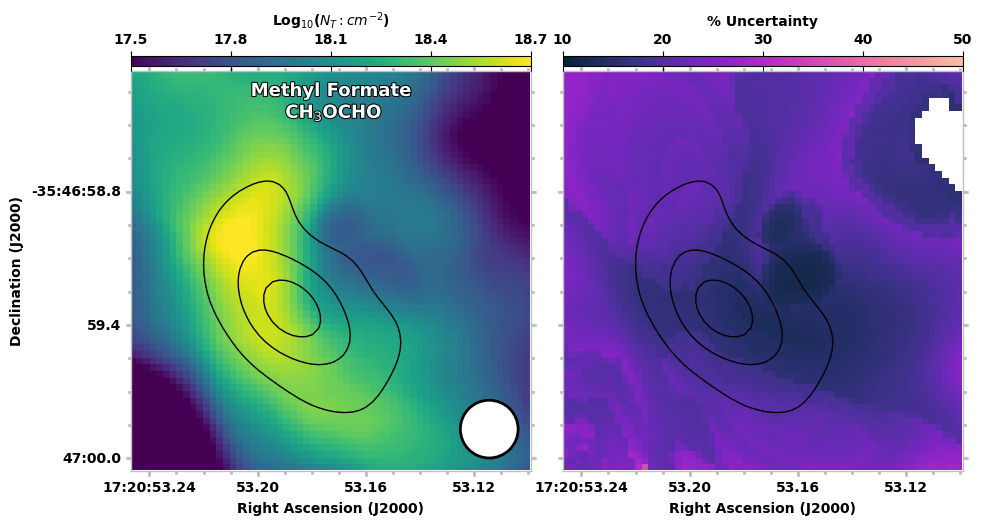}
    \caption{Same as Figure \ref{b22} but for methyl formate.}
    \label{b35}
\end{figure}

\begin{figure}[h!]
    \centering
    \includegraphics[width=0.5\textwidth]{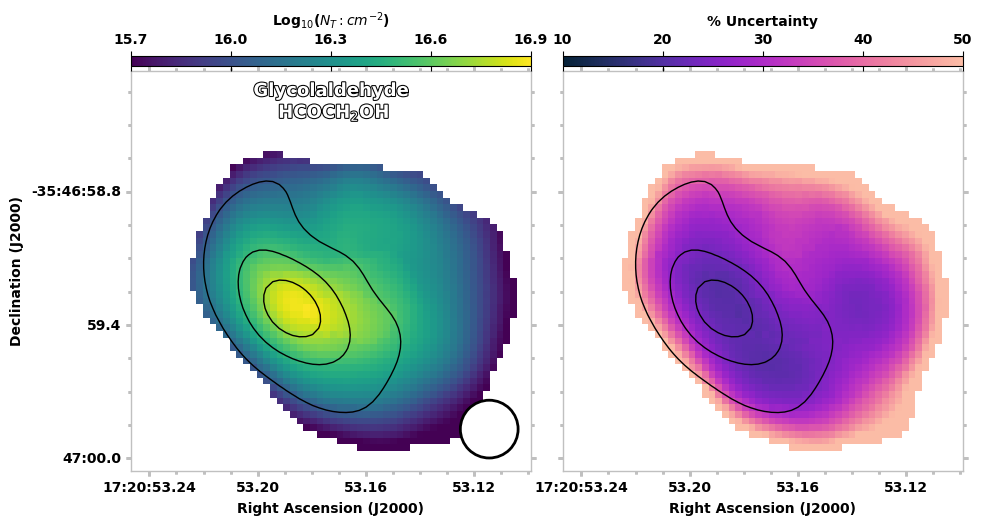}
    \caption{Same as Figure \ref{b22} but for glycolaldehyde. Due to the weak emission of this molecule in this source these values should be treated as an upper limit.}
    \label{b36}
\end{figure}

\begin{figure}[h!]
    \centering
    \includegraphics[width=0.5\textwidth]{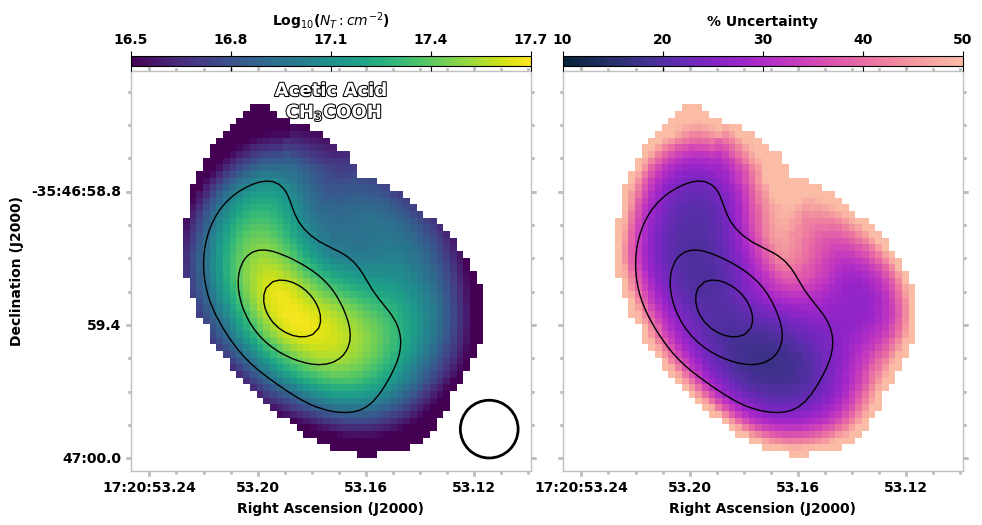}
    \caption{Same as Figure \ref{b22} but for acetic acid.}
    \label{b37}
\end{figure}

\begin{figure}[h!]
    \centering
    \includegraphics[width=0.5\textwidth]{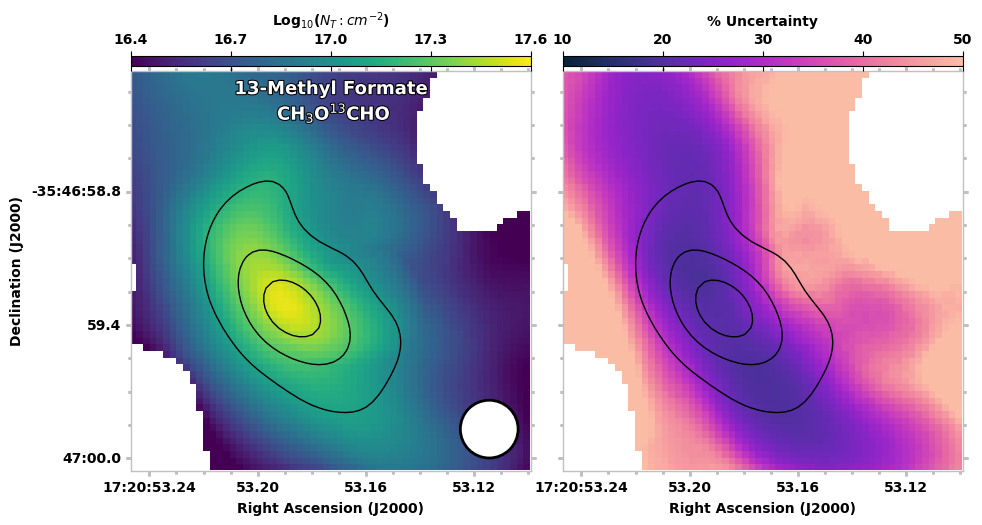}
    \caption{Same as Figure \ref{b22} but for 13-methyl formate.}
    \label{b38}
\end{figure}

\begin{figure}[h!]
    \centering
    \includegraphics[width=0.5\textwidth]{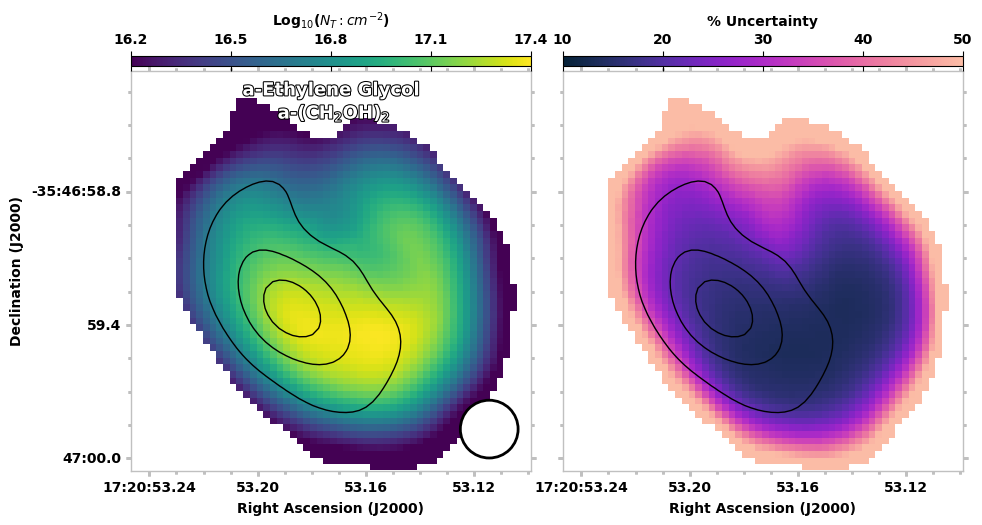}
    \caption{Same as Figure \ref{b22} but for a-ethylene glycol.}
    \label{b39}
\end{figure}

\begin{figure}[h!]
    \centering
    \includegraphics[width=0.5\textwidth]{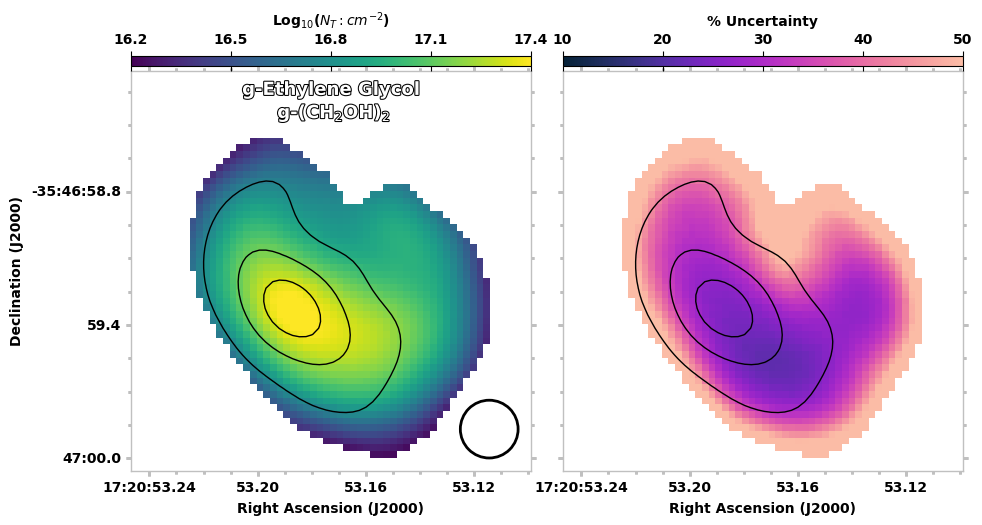}
    \caption{Same as Figure \ref{b22} but for g-ethylene glycol.}
    \label{b40}
\end{figure}

\begin{figure}[h!]
    \centering
    \includegraphics[width=0.5\textwidth]{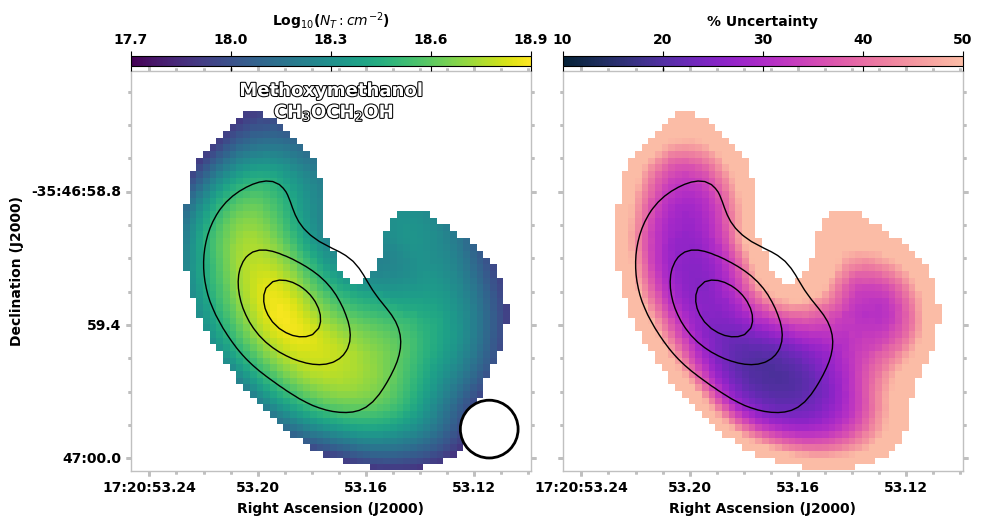}
    \caption{Same as Figure \ref{b22} but for methoxymethanol.}
    \label{b41}
\end{figure}

\begin{figure}[h!]
    \centering
    \includegraphics[width=0.5\textwidth]{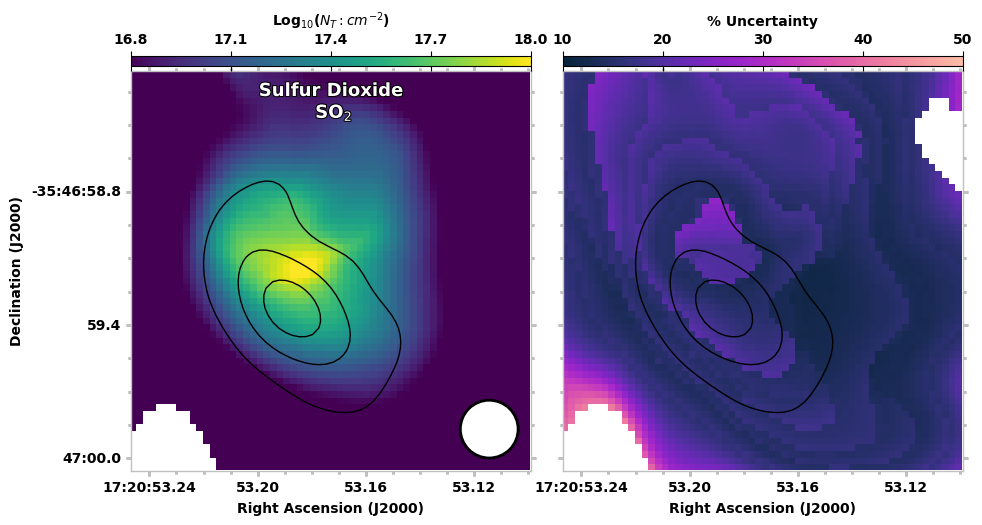}
    \caption{Same as Figure \ref{b22} but for sulfur dioxide.}
    \label{fig:b42}
\end{figure}

Images for each of the parameters derived using the fitting routine along with the associated uncertainties are presented in Figures~\ref{fig:b00}--\ref{fig:b42} for all of the molecules in the emission model. Note that while each column density image spans 1.2 orders of magnitude, the range is unique for each molecule. The column density uncertainty images cover a uniform range for all molecules and are expressed as a percentage of the value for each pixel. The uncertainty images for the other parameters are expressed in physical units. The same masks described in Section \ref{column_densities} have been applied to these images. 

\clearpage

\clearpage

\clearpage

\clearpage

\clearpage

\clearpage

\clearpage

\clearpage

\clearpage

\clearpage

\clearpage

\clearpage

%\end{comment}

\clearpage

\section{Appendix C}
\label{appendix_c}

\begin{figure}[h!]
    \centering
    \includegraphics[width=0.5\textwidth]{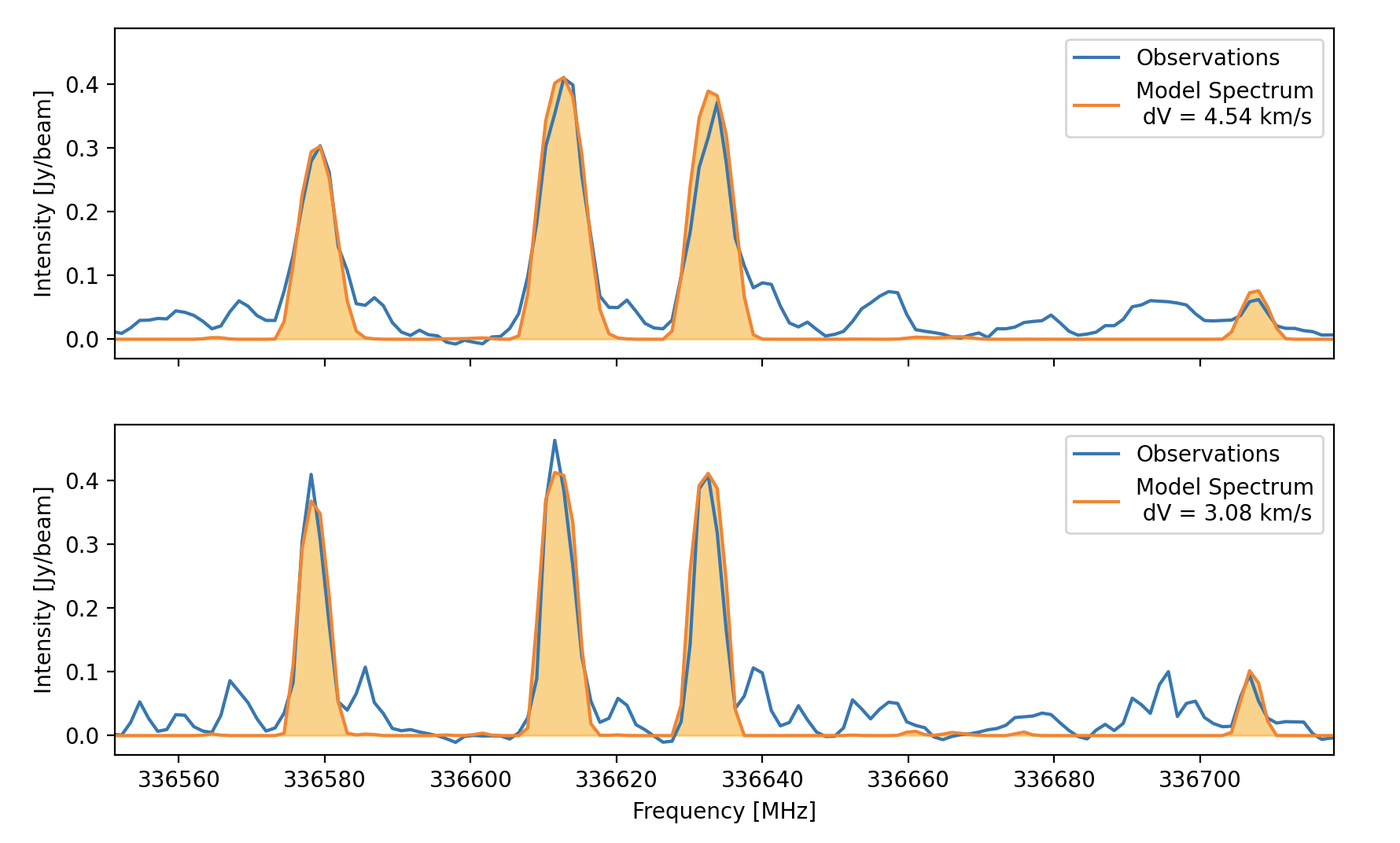}
    \caption{A pair of spectra extracted from the northernmost ``filament" in the linewidth map of Figure \ref{3panel_mm1_univ}. The two spectra were extracted from pixels that were in the center \emph{(top)} and edge \emph{(bottom)} of the feature.}
    \label{dv_image}
\end{figure}

We present two spectra extracted from the northernmost ``filament" in the linewidth map of Figure \ref{3panel_mm1_univ} to showcase the measurable difference in the lineshapes in Figure \ref{dv_image}. We also present a pair of spectra demonstrating how the variation in spectral crowding across the source can impact the measured velocity using the moment map technique.

%We also present a pair of spectra centered on the methyl formate transition used to create the moment maps in Section \ref{vlsr_image} to demonstrate how the two methods arrived at such different values for certain pixels.

\begin{figure}[h!]
    \centering
    \includegraphics[width=0.5\textwidth]{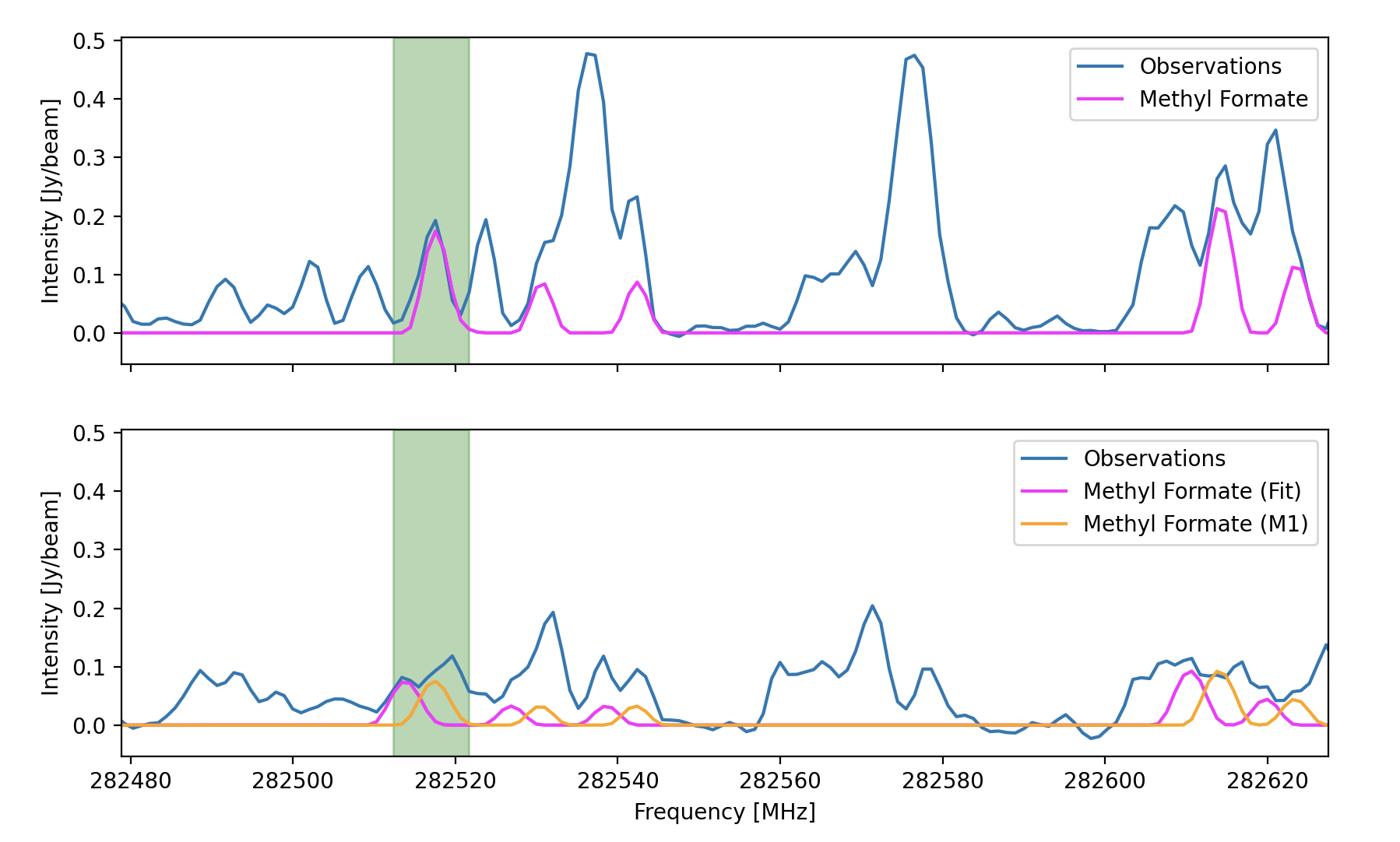}
    \caption{The top spectrum is from one of the pixels that was used in \citet{El-Abd:2019:129} to measure the properties of methyl formate in MM1 with the channels used for the moment 1 map highlighted in green. The bottom spectrum is from a pixel where our method of measuring the velocity disagreed with the moment map by almost 4 \kms\/ with a simulated spectrum at both velocities overlaid. Note how the moment 1 map understandably skews toward the stronger transition in the channel range while our method (correctly) picks out the weaker transition. This speaks to both the difficulty in picking out an appropriate channel range for the entirety of a turbulent region for a moment 1 map as well as the strength of our method as the velocity is not skewed by an individual transition for a molecule.}
    \label{vlsr_image}
\end{figure}

\acknowledgements

This paper makes use of the following ALMA data: \#2015.A.00022.T. ALMA is a partnership of ESO (representing its member states), NSF (USA) and NINS (Japan), together with NRC (Canada) and NSC and ASIAA (Taiwan) and KASI (Republic of Korea), in cooperation with the Republic of Chile. The Joint ALMA Observatory is operated by ESO, AUI/NRAO and NAOJ.   The National Radio Astronomy Observatory is a facility of the National Science Foundation operated under cooperative agreement by Associated Universities, Inc.  This research made use of NASA’s Astrophysics Data System  Bibliographic  Services,  Astropy,  a community-developed core Python package for Astronomy \citep{AstropyCollaboration:2022:167}, and APLpy, an open-source plotting package for Python \citep{aplpy2012}. Support for this work was provided by the NSF through the Grote Reber Fellowship Program administered by Associated Universities, Inc./National Radio Astronomy Observatory. We would like to thank the referee for their comments regarding this work. 
% hosted at http://aplpy.github.com.

\software{
%analysisUtils \citep{au2023},
APLpy \citep{aplpy2012},
Astropy \citep{AstropyCollaboration:2022:167},
CASA \citep{CASA2022},
Jupyter \citep{Kluyver:2016:87},
lmfit \citep{Newville:2014:},
molsim \citep{Lee:2023:},
scipy \citep{Gommers:2023:}
}

\clearpage

% \bibliography{references,other_refs}

\begin{thebibliography}{}
\expandafter\ifx\csname natexlab\endcsname\relax\def\natexlab#1{#1}\fi
\providecommand{\url}[1]{\href{#1}{#1}}

\bibitem[{{Astropy Collaboration} {et~al.}(2022){Astropy Collaboration},
  Price-Whelan, Lim, Earl, Starkman, Bradley, Shupe, Patil, Corrales, Brasseur,
  Nöthe, Donath, Tollerud, Morris, Ginsburg, Vaher, Weaver, Tocknell,
  Jamieson, van Kerkwijk, Robitaille, Merry, Bachetti, Günther, Aldcroft,
  Alvarado-Montes, Archibald, Bódi, Bapat, Barentsen, Bazán, Biswas, Boquien,
  Burke, Cara, Cara, Conroy, Conseil, Craig, Cross, Cruz, D'Eugenio, Dencheva,
  Devillepoix, Dietrich, Eigenbrot, Erben, Ferreira, Foreman-Mackey, Fox,
  Freij, Garg, Geda, Glattly, Gondhalekar, Gordon, Grant, Greenfield, Groener,
  Guest, Gurovich, Handberg, Hart, Hatfield-Dodds, Homeier, Hosseinzadeh,
  Jenness, Jones, Joseph, Kalmbach, Karamehmetoglu, Kałuszyński, Kelley,
  Kern, Kerzendorf, Koch, Kulumani, Lee, Ly, Ma, MacBride, Maljaars, Muna,
  Murphy, Norman, O'Steen, Oman, Pacifici, Pascual, Pascual-Granado, Patil,
  Perren, Pickering, Rastogi, Roulston, Ryan, Rykoff, Sabater, Sakurikar,
  Salgado, Sanghi, Saunders, Savchenko, Schwardt, Seifert-Eckert, Shih, Jain,
  Shukla, Sick, Simpson, Singanamalla, Singer, Singhal, Sinha, Sipőcz,
  Spitler, Stansby, Streicher, Šumak, Swinbank, Taranu, Tewary, Tremblay,
  de~Val-Borro, Van~Kooten, Vasović, Verma, de~Miranda~Cardoso, Williams,
  Wilson, Winkel, Wood-Vasey, Xue, Yoachim, Zhang, Zonca, \& {Astropy Project
  Contributors}}]{AstropyCollaboration:2022:167}
{Astropy Collaboration}, Price-Whelan, A.~M., Lim, P.~L., {et~al.} 2022, The
  Astrophysical Journal, 935, 167, aDS Bibcode: 2022ApJ...935..167A.

\bibitem[{Beuther {et~al.}(2007)Beuther, Walsh, Thorwirth, Zhang, Hunter,
  Megeath, \& Menten}]{Beuther:2007:989}
Beuther, H., Walsh, A.~J., Thorwirth, S., {et~al.} 2007, Astronomy \&
  Astrophysics, 466, 989, arXiv:astro-ph/0702190.

\bibitem[{{Brogan} {et~al.}(2007){Brogan}, {Chandler}, {Hunter}, {Shirley}, \&
  {Sarma}}]{Brogan2007}
{Brogan}, C.~L., {Chandler}, C.~J., {Hunter}, T.~R., {Shirley}, Y.~L., \&
  {Sarma}, A.~P. 2007, \apjl, 660, L133

\bibitem[{{Brogan} {et~al.}(2016){Brogan}, {Hunter}, {Cyganowski}, {Chandler},
  {Friesen}, \& {Indebetouw}}]{Brogan2016}
{Brogan}, C.~L., {Hunter}, T.~R., {Cyganowski}, C.~J., {et~al.} 2016, \apj,
  832, 187

\bibitem[{Brogan {et~al.}(2018)Brogan, Hunter, Cyganowski, Chibueze, Friesen,
  Hirota, MacLeod, McGuire, \& Sobolev}]{Brogan:2018:87}
Brogan, C.~L., Hunter, T.~R., Cyganowski, C.~J., {et~al.} 2018, The
  Astrophysical Journal, 866, 87, arXiv:1809.04178 [astro-ph].

\bibitem[{Bøgelund {et~al.}(2018)Bøgelund, McGuire, Ligterink, Taquet,
  Brogan, Hunter, Pearson, Hogerheijde, \& van Dishoeck}]{Bogelund:2018:A88}
Bøgelund, E.~G., McGuire, B.~A., Ligterink, N. F.~W., {et~al.} 2018, Astronomy
  \& Astrophysics, 615, A88, arXiv:1804.01090 [astro-ph].

\bibitem[{Calcutt {et~al.}(2018)Calcutt, Jørgensen, Müller, Kristensen,
  Coutens, Bourke, Garrod, Persson, van~der Wiel, van Dishoeck, \&
  Wampfler}]{Calcutt:2018:A90}
Calcutt, H., Jørgensen, J.~K., Müller, H. S.~P., {et~al.} 2018, Astronomy \&
  Astrophysics, 616, A90, arXiv:1804.09210 [astro-ph].

\bibitem[{Campbell(2018)}]{Campbell:2018:}
Campbell, S. 2018, Comparison of the generalized centroid with {Gaussian} and
  quadratic peak localization methods,  arXiv, arXiv:1807.08355 [astro-ph].

\bibitem[{{CASA Team} {et~al.}(2022){CASA Team}, {Bean}, {Bhatnagar}, {Castro},
  {Donovan Meyer}, {Emonts}, {Garcia}, {Garwood}, {Golap}, {Gonzalez Villalba},
  {Harris}, {Hayashi}, {Hoskins}, {Hsieh}, {Jagannathan}, {Kawasaki},
  {Keimpema}, {Kettenis}, {Lopez}, {Marvil}, {Masters}, {McNichols},
  {Mehringer}, {Miel}, {Moellenbrock}, {Montesino}, {Nakazato}, {Ott}, {Petry},
  {Pokorny}, {Raba}, {Rau}, {Schiebel}, {Schweighart}, {Sekhar}, {Shimada},
  {Small}, {Steeb}, {Sugimoto}, {Suoranta}, {Tsutsumi}, {van Bemmel},
  {Verkouter}, {Wells}, {Xiong}, {Szomoru}, {Griffith}, {Glendenning}, \&
  {Kern}}]{CASA2022}
{CASA Team}, {Bean}, B., {Bhatnagar}, S., {et~al.} 2022, \pasp, 134, 114501

\bibitem[{{Chibueze} {et~al.}(2014){Chibueze}, {Omodaka}, {Handa}, {Imai},
  {Kurayama}, {Nagayama}, {Sunada}, {Nakano}, {Hirota}, \&
  {Honma}}]{Chibueze2014}
{Chibueze}, J.~O., {Omodaka}, T., {Handa}, T., {et~al.} 2014, \apj, 784, 114

\bibitem[{Cortes {et~al.}(2023)Cortes, Vlahakis, Hales, Carpenter, Dent,
  Kameno, Loomis, Vila~Vilaro, Biggs, Miotello, Rosen, Stoehr, \&
  Saini}]{CortesPaulo:2023:}
Cortes, P., Vlahakis, C., Hales, A., {et~al.} 2023, doi:10.5281/ZENODO.4511521,
  publisher: Zenodo Version Number: Cycle 10; Doc. 10.3; version 1.1.

\bibitem[{{Cunningham} {et~al.}(2023){Cunningham}, {Ginsburg},
  {Galv{\'a}n-Madrid}, {Motte}, {Csengeri}, {Stutz}, {Fern{\'a}ndez-L{\'o}pez},
  {{\'A}lvarez-Guti{\'e}rrez}, {Armante}, {Baug}, {Bonfand}, {Bontemps},
  {Braine}, {Brouillet}, {Busquet}, {D{\'\i}az-Gonz{\'a}lez}, {Di Francesco},
  {Gusdorf}, {Herpin}, {Liu}, {L{\'o}pez-Sepulcre}, {Louvet}, {Lu}, {Maud},
  {Nony}, {Olguin}, {Pouteau}, {Rivera-Soto}, {Sandoval-Garrido}, {Sanhueza},
  {Tatematsu}, {Towner}, \& {Valeille-Manet}}]{Cunningham2023}
{Cunningham}, N., {Ginsburg}, A., {Galv{\'a}n-Madrid}, R., {et~al.} 2023, arXiv
  e-prints, arXiv:2306.14710

\bibitem[{El-Abd {et~al.}(2019)El-Abd, Brogan, Hunter, Willis, Garrod, \&
  McGuire}]{El-Abd:2019:129}
El-Abd, S.~J., Brogan, C.~L., Hunter, T.~R., {et~al.} 2019, The Astrophysical
  Journal, 883, 129, arXiv:1907.13551 [astro-ph].

\bibitem[{Fischer {et~al.}(2022)Fischer, Hillenbrand, Herczeg, Johnstone,
  Kóspál, \& Dunham}]{Fischer:2022:}
Fischer, W.~J., Hillenbrand, L.~A., Herczeg, G.~J., {et~al.} 2022, Accretion
  {Variability} as a {Guide} to {Stellar} {Mass} {Assembly},  arXiv,
  arXiv:2203.11257 [astro-ph].

\bibitem[{Fischer {et~al.}(2019)Fischer, Safron, \& Megeath}]{Fischer:2019:183}
Fischer, W.~J., Safron, E., \& Megeath, S.~T. 2019, The Astrophysical Journal,
  872, 183, aDS Bibcode: 2019ApJ...872..183F.

\bibitem[{Garatti {et~al.}(2017)Garatti, Stecklum, Lopez, Eislöffel, Ray,
  Sanna, Cesaroni, Walmsley, Oudmaijer, de~Wit, Moscadelli, Greiner, Krabbe,
  Fischer, Klein, \& Ibañez}]{Garatti:2017:276}
Garatti, A. C.~o., Stecklum, B., Lopez, R.~G., {et~al.} 2017, Nature Physics,
  13, 276, arXiv:1704.02628 [astro-ph].

\bibitem[{Ginsburg \& Mirocha(2011)}]{Ginsburg:2011:ascl:1109.001}
Ginsburg, A., \& Mirocha, J. 2011, Astrophysics Source Code Library,
  ascl:1109.001, aDS Bibcode: 2011ascl.soft09001G.

\bibitem[{Ginsburg {et~al.}(2022)Ginsburg, Sokolov, de~Val-Borro, Rosolowsky,
  Pineda, Sipőcz, \& Henshaw}]{Ginsburg:2022:291}
Ginsburg, A., Sokolov, V., de~Val-Borro, M., {et~al.} 2022, The Astronomical
  Journal, 163, 291, arXiv:2205.04987 [astro-ph].}

\bibitem[{Gommers {et~al.}(2023)Gommers, Virtanen, Burovski, Haberland,
  Weckesser, Oliphant, Reddy, Cournapeau, {Alexbrc}, Nelson, Peterson, Roy,
  Wilson, {Ilhan Polat}, {Endolith}, Mayorov, Van Der~Walt, Brett, Laxalde,
  Larson, Millman, Sakai, Lars, {Peterbell10}, Van~Mulbregt, {CJ Carey},
  {Eric-Jones}, McKibben, Kern, \& Kai}]{Gommers:2023:}
Gommers, R., Virtanen, P., Burovski, E., {et~al.} 2023, scipy/scipy: {SciPy}
  1.11.1,  Zenodo, doi:10.5281/ZENODO.8092679.

\bibitem[{Herbst \& Van~Dishoeck(2009)}]{Herbst:2009:427}
Herbst, E., \& Van~Dishoeck, E.~F. 2009, Annual Review of Astronomy and
  Astrophysics, 47, 427.

\bibitem[{Hollis {et~al.}(2004)Hollis, Jewell, Lovas, \&
  Remijan}]{Hollis:2004:L45}
Hollis, J.~M., Jewell, P.~R., Lovas, F.~J., \& Remijan, A. 2004, The
  Astrophysical Journal, 613, L45, aDS Bibcode: 2004ApJ...613L..45H.

\bibitem[{Hunter {et~al.}(2017)Hunter, Brogan, MacLeod, Cyganowski, Chandler,
  Chibueze, Friesen, Indebetouw, Thesner, \& Young}]{Hunter:2017:L29}
Hunter, T.~R., Brogan, C.~L., MacLeod, G., {et~al.} 2017, The Astrophysical
  Journal, 837, L29, aDS Bibcode: 2017ApJ...837L..29H.

\bibitem[{Hunter {et~al.}(2021)Hunter, Brogan, De~Buizer, Towner, Dowell,
  MacLeod, Stecklum, Cyganowski, El-Abd, \& McGuire}]{Hunter:2021:L17}
Hunter, T.~R., Brogan, C.~L., De~Buizer, J.~M., {et~al.} 2021, The
  Astrophysical Journal Letters, 912, L17, arXiv:2104.05187 [astro-ph].

\bibitem[{{Hunter} {et~al.}(2023){Hunter}, {Indebetouw}, {Brogan}, {Berry},
  {Chang}, {Francke}, {Geers}, {G{\'o}mez}, {Hibbard}, {Humphreys}, {Kent},
  {Kepley}, {Kunneriath}, {Lipnicky}, {Loomis}, {Mason}, {Masters}, {Maud},
  {Muders}, {Sabater}, {Sugimoto}, {Sz{\H{u}}cs}, {Vasiliev}, {Videla},
  {Villard}, {Williams}, {Xue}, \& {Yoon}}]{Hunter2023}
{Hunter}, T.~R., {Indebetouw}, R., {Brogan}, C.~L., {et~al.} 2023, arXiv
  e-prints, arXiv:2306.07420

\bibitem[{Jorgensen {et~al.}(2020)Jorgensen, Belloche, \&
  Garrod}]{Jorgensen:2020:727}
Jorgensen, J.~K., Belloche, A., \& Garrod, R.~T. 2020, Annual Review of
  Astronomy and Astrophysics, 58, 727, arXiv:2006.07071 [astro-ph].

\bibitem[{Kluyver {et~al.}(2016)Kluyver, Ragan-Kelley, Pérez, Granger,
  Bussonnier, Frederic, Kelley, Hamrick, Grout, Corlay, Ivanov, Avila, Abdalla,
  Willing, \& {Jupyter Development Team}}]{Kluyver:2016:87}
Kluyver, T., Ragan-Kelley, B., Pérez, F., {et~al.} 2016, Jupyter
  {Notebooks}—a publishing format for reproducible computational workflows,
  pages: 87-90 Publication Title: IOS Press ADS Bibcode: 2016ppap.book...87K,
  doi:10.3233/978-1-61499-649-1-87.

\bibitem[{Lee {et~al.}(2023)Lee, Loomis, Xue, El-Abd, \& McGuire}]{Lee:2023:}
Lee, K. L.~K., Loomis, R.~A., Xue, C., El-Abd, S., \& McGuire, B.~A. 2023,
  molsim,  Zenodo, doi:10.5281/ZENODO.8118192.

\bibitem[{{MacLeod} {et~al.}(2018){MacLeod}, {Smits}, {Goedhart}, {Hunter},
  {Brogan}, {Chibueze}, {van den Heever}, {Thesner}, {Banda}, \&
  {Paulsen}}]{Macleod2018}
{MacLeod}, G.~C., {Smits}, D.~P., {Goedhart}, S., {et~al.} 2018, \mnras, 478,
  1077

\bibitem[{Maret {et~al.}(2011)Maret, Hily-Blant, Pety, Bardeau, \&
  Reynier}]{Maret:2011:A47}
Maret, S., Hily-Blant, P., Pety, J., Bardeau, S., \& Reynier, E. 2011,
  Astronomy and Astrophysics, 526, A47, aDS Bibcode: 2011A\&A...526A..47M.


\bibitem[{Martín {et~al.}(2019)Martín, Martín-Pintado, Blanco-Sánchez,
  Rivilla, Rodríguez-Franco, \& Rico-Villas}]{Martin:2019:A159}
Martín, S., Martín-Pintado, J., Blanco-Sánchez, C., {et~al.} 2019, Astronomy
  \& Astrophysics, 631, A159.

\bibitem[{McGuire {et~al.}(2017)McGuire, Shingledecker, Willis, Burkhardt,
  El-Abd, Motiyenko, Brogan, Hunter, Margulès, Guillemin, Garrod, Herbst, \&
  Remijan}]{McGuire:2017:L46}
McGuire, B.~A., Shingledecker, C.~N., Willis, E.~R., {et~al.} 2017, The
  Astrophysical Journal, 851, L46, arXiv:1712.03256 [astro-ph].

\bibitem[{{McGuire} {et~al.}(2018){McGuire}, {Brogan}, {Hunter}, {Remijan},
  {Blake}, {Burkhardt}, {Carroll}, {van Dishoeck}, {Garrod}, {Linnartz},
  {Shingledecker}, \& {Willis}}]{McGuire2018}
{McGuire}, B.~A., {Brogan}, C.~L., {Hunter}, T.~R., {et~al.} 2018, \apjl, 863,
  L35

\bibitem[{McGuire {et~al.}(2020)McGuire, Burkhardt, Loomis, Shingledecker, Lee,
  Charnley, Cordiner, Herbst, Kalenskii, Momjian, Willis, Xue, Remijan, \&
  McCarthy}]{McGuire:2020:L10}
McGuire, B.~A., Burkhardt, A.~M., Loomis, R.~A., {et~al.} 2020, The
  Astrophysical Journal, 900, L10, arXiv:2008.12349 [astro-ph].

\bibitem[{Meyer {et~al.}(2021)Meyer, Vorobyov, Elbakyan, Eislöffel, Sobolev,
  \& Stöhr}]{Meyer:2021:4448}
Meyer, D. M.~A., Vorobyov, E.~I., Elbakyan, V.~G., {et~al.} 2021, Monthly
  Notices of the Royal Astronomical Society, 500, 4448, aDS Bibcode:
  2021MNRAS.500.4448M.

\bibitem[{Möller {et~al.}(2017)Möller, Endres, \& Schilke}]{Moller:2017:A7}
Möller, T., Endres, C., \& Schilke, P. 2017, Astronomy \& Astrophysics, 598,
  A7.

\bibitem[{Müller {et~al.}(2005)Müller, Schlöder, Stutzki, \&
  Winnewisser}]{Muller:2005:215}
Müller, H.~S., Schlöder, F., Stutzki, J., \& Winnewisser, G. 2005, Journal of
  Molecular Structure, 742, 215.

\bibitem[{Newville {et~al.}(2014)Newville, Stensitzki, Allen, \&
  Ingargiola}]{Newville:2014:}
Newville, M., Stensitzki, T., Allen, D.~B., \& Ingargiola, A. 2014, {LMFIT}:
  {Non}-{Linear} {Least}-{Square} {Minimization} and {Curve}-{Fitting} for
  {Python},  Zenodo, doi:10.5281/ZENODO.11813.

\bibitem[{Pickett {et~al.}(1998)Pickett, Poynter, Cohen, Delitsky, Pearson, \&
  Müller}]{Pickett:1998:883}
Pickett, H.~M., Poynter, R.~L., Cohen, E.~A., {et~al.} 1998, Journal of
  Quantitative Spectroscopy and Radiative Transfer, 60, 883, aDS Bibcode:
  1998JQSRT..60..883P.

\bibitem[{{Qiu} {et~al.}(2011){Qiu}, {Wyrowski}, {Menten}, {G{\"u}sten},
  {Leurini}, \& {Leinz}}]{Qiu2011}
{Qiu}, K., {Wyrowski}, F., {Menten}, K.~M., {et~al.} 2011, \apjl, 743, L25

\bibitem[{{Reid} {et~al.}(2014){Reid}, {Menten}, {Brunthaler}, {Zheng}, {Dame},
  {Xu}, {Wu}, {Zhang}, {Sanna}, {Sato}, {Hachisuka}, {Choi}, {Immer},
  {Moscadelli}, {Rygl}, \& {Bartkiewicz}}]{Reid2014}
{Reid}, M.~J., {Menten}, K.~M., {Brunthaler}, A., {et~al.} 2014, \apj, 783, 130

\bibitem[{Remijan {et~al.}(2022)Remijan, Xue, Margulès, Belloche, Motiyenko,
  Carder, Codella, Balucani, Brogan, Ceccarelli, Hunter, Maris, Melandri,
  Siebert, \& McGuire}]{Remijan:2022:A85}
Remijan, A., Xue, C., Margulès, L., {et~al.} 2022, Astronomy \& Astrophysics,
  658, A85, arXiv:2112.03356 [astro-ph].

\bibitem[{Rivilla {et~al.}(2013)Rivilla, Martin-Pintado, Jimenez-Serra, \&
  Rodriguez-Franco}]{Rivilla:2013:A48}
Rivilla, V.~M., Martin-Pintado, J., Jimenez-Serra, I., \& Rodriguez-Franco, A.
  2013, Astronomy \& Astrophysics, 554, A48, arXiv:1302.2763 [astro-ph].

\bibitem[{{Robitaille} \& {Bressert}(2012)}]{aplpy2012}
{Robitaille}, T., \& {Bressert}, E. 2012, {APLpy: Astronomical Plotting Library
  in Python}, , , ascl:1208.017

\bibitem[{Schuessler {et~al.}(2022)Schuessler, Remijan, Xue, Carder, Scolati,
  \& McGuire}]{Schuessler:2022:102}
Schuessler, C., Remijan, A., Xue, C., {et~al.} 2022, The Astrophysical Journal,
  941, 102, arXiv:2208.05823 [astro-ph].

\bibitem[{Turner(1991)}]{Turner:1991:617}
Turner, B.~E. 1991, The Astrophysical Journal Supplement Series, 76, 617, aDS
  Bibcode: 1991ApJS...76..617T.

\bibitem[{Vastel {et~al.}(2015)Vastel, Bottinelli, Caux, Glorian, \&
  Boiziot}]{Vastel:2015:313}
Vastel, C., Bottinelli, S., Caux, E., Glorian, J.~M., \& Boiziot, M. 2015,
  {CASSIS}: a tool to visualize and analyse instrumental and synthetic
  spectra., conference Name: SF2A-2015: Proceedings of the Annual meeting of
  the French Society of Astronomy and Astrophysics Pages: 313-316 ADS Bibcode:
  2015sf2a.conf..313V.

\bibitem[{Wilkins {et~al.}(2022)Wilkins, Carroll, \& Blake}]{Wilkins:2022:4}
Wilkins, O.~H., Carroll, P.~B., \& Blake, G.~A. 2022, The Astrophysical
  Journal, 924, 4, aDS Bibcode: 2022ApJ...924....4W.

\bibitem[{Zernickel {et~al.}(2012)Zernickel, Schilke, Schmiedeke, Lis, Brogan,
  Ceccarelli, Comito, Emprechtinger, Hunter, \& Möller}]{Zernickel:2012:A87}
Zernickel, A., Schilke, P., Schmiedeke, A., {et~al.} 2012, Astronomy \&
  Astrophysics, 546, A87, arXiv:1208.5516 [astro-ph].

\bibitem[{Zhu {et~al.}(1997)Zhu, Byrd, Lu, \& Nocedal}]{Zhu:1997:550}
Zhu, C., Byrd, R.~H., Lu, P., \& Nocedal, J. 1997, ACM Transactions on
  Mathematical Software, 23, 550.

\end{thebibliography}
% \bibliographystyle{aasjournal}

\end{document}